\documentstyle[eqsecnum,aps,prc,floats,graphicx,tighten]{revtex}

\def\nuc#1#2{${}^{#1}$#2}
\def\mnuc#1#2{{}^{#1}{\protect\text{#2}}}
\def\degreesC{$\,^{\circ}$C}

\begin{document}
\draft  

\twocolumn[\hsize\textwidth\columnwidth\hsize\csname@twocolumnfalse\endcsname

\begin{flushright}
{\footnotesize astro-ph/9907113} \\
{\footnotesize Phys.\ Rev. C {\bf 60}, 055801 (1999)}
\end{flushright}

\title{Measurement of the solar neutrino capture rate with gallium metal}

\author{J.\,N.\,Abdurashitov, V.\,N.\,Gavrin, S.\,V.\,Girin,
        V.\,V.\,Gorbachev, T.\,V.\,Ibragimova, A.\,V.\,Kalikhov,
        N.\,G.\,Khairnasov, T.\,V.\,Knodel, I.\,N.\,Mirmov, A.\,A.\,Shikhin,
        E.\,P.\,Veretenkin, V.\,M.\,Vermul, V.\,E.\,Yants, and
        G.\,T.\,Zatsepin}
\address{\it Institute for Nuclear Research, Russian Academy of Sciences,
         RU-117312 Moscow, Russia}

\author{T.\,J.\,Bowles, W.\,A.\,Teasdale, and D.\,L.\,Wark\footnotemark}
\address{\it Los Alamos National Laboratory, Los Alamos, New Mexico 87545}

\author{M.\,L.\,Cherry}
\address{\it Louisiana State University, Baton Rouge, Louisiana 70803}

\author{J.\,S.\,Nico}
\address{National Institute of Standards and Technology, Gaithersburg,
         Maryland 20899}

\author{B.\,T.\,Cleveland, R.\,Davis, Jr., K.\,Lande, and P.\,S.\,Wildenhain}
\address{\it University of Pennsylvania, Philadelphia, Pennsylvania 19104}

\author{S.\,R.\,Elliott and J.\,F.\,Wilkerson}
\address{\it University of Washington, Seattle, Washington 98195}

\author{(SAGE Collaboration)}

\date{Received 29 April 1999}
\maketitle

\begin{abstract} The solar neutrino capture rate measured by the
Russian-American Gallium Experiment (SAGE) on metallic gallium during the
period January 1990 through December 1997 is $67.2_{-7.0 -3.0}^{+7.2 +3.5}$
SNU, where the uncertainties are statistical and systematic, respectively.
This represents only about half of the predicted standard solar model rate of
129 SNU.  All the experimental procedures, including extraction of germanium
from gallium, counting of \nuc{71}{Ge}, and data analysis, are discussed in
detail.

\end{abstract}

] 

\section{Introduction}\label{intro}

     The Sun produces its energy by the nuclear fusion of four protons into
an $\alpha$ particle, chains of reactions that yield two positrons and two
neutrinos.  Since these low-energy neutrinos are weakly interacting, it was
assumed that they traverse the Sun and reach the Earth without change.
\footnotetext[1]{Present address: Department of Particle and Nuclear Physics,
Oxford University, Keble Road, Oxford OX1 3RH, UK.}
Measurement of the neutrino energy spectrum should thus give information
about the conditions under which the nuclear reactions take place in the Sun.
All solar neutrino experiments, however, have observed considerably fewer
neutrinos than are predicted by detailed models of the physical processes in
the Sun that are based on the nuclear reaction chains.  As a result of this
neutrino deficit, the assumption that the neutrinos are unchanged during
their passage from the Sun to the Earth is now seriously questioned.  For
such transformations to occur neutrinos must have mass, a hypothesis of
far-reaching consequences.

     The experimental study of solar neutrinos is now over 30 years old.  The
first experiment, a radiochemical detector based on chlorine
\cite{DAV95,DAV98}, observed a capture rate of $2.55 \pm 0.17 \pm 0.18$ SNU,
where 1 SNU = 1 interaction/s in a target that contains $10^{36}$ atoms of
the neutrino absorbing isotope.  Although different standard solar models
(SSM's) predict somewhat different rates for the chlorine experiment (for
example, $7.7^{+1.2}_{-1.0}$ SNU \cite{BAH95,BAH98} and 7.2 SNU
\cite{TUR98}), all such models predict a rate significantly higher than
observed.

\begin{table*}
\caption{Predicted solar neutrino fluxes {\protect \cite{BAH98}} and their
contribution to the Ga capture rate.}
\label{theoretical_fluxes}
\begin{tabular}{lcccd}
         & $\nu$   & $\nu$ energy & $\nu$ flux           & Ga capture   \\
Reaction &  branch & (MeV)        & (cm$^{-2}$ s$^{-1}$) & rate (SNU)   \\
\hline
$p + p {\rightarrow} d + e^+ + \nu$                         & $pp$   &
0--0.42     & $(5.94 \pm 0.06) \times 10^{10}$              & 69.6      \\
$^7\text{Be} + e^- {\rightarrow} ^8\text{B} + \nu$          & $^7$Be &
0.38, 0.86  & $(4.80 \pm 0.43) \times 10^{9}$               & 34.4      \\
$^8\text{B} {\rightarrow} ^8\text{Be}^* + e^+ + \nu$        & $^8$B  &
0--14.1     & $(5.15^{+0.98}_{-0.72})\times 10^{6}$         & 12.4      \\
CNO reactions                                               & CNO    &
0--1.73     & $(1.1 \pm 0.2) \times 10^{8}$                 & 9.8       \\
$p + e^- + p {\rightarrow} d + \nu$                         & $pep$  &
1.44        & $(1.39 \pm 0.01) \times 10^{8}$               & 2.8       \\
\end{tabular}
\end{table*}

     For 20 years, until about 1985, the chlorine experiment was the only
measurement.  This experiment is primarily sensitive to high-energy
\nuc{8}{B} neutrinos with a $\sim20$\% contribution from other sources,
mainly \nuc{7}{Be}.  The flux of \nuc{8}{B} neutrinos is very dependent on
the central temperature of the Sun ($T_{\odot}^{24}$\cite{BAH96}).  As a
result many models were suggested that would slightly suppress $T_{\odot}$
and hence decrease the \nuc{8}{B} flux significantly.  (See
Ref.\ \cite{BAH89} for a description of a large number of such models.)  Most
of these models, however, run into difficulty with some other measured aspect
of the Sun.  An alternative solution to this discrepancy could be neutrino
oscillations.  The Cl experiment operates on the inverse beta decay reaction
and thus is only sensitive to electron neutrinos.  If the neutrinos were to
change flavor on their trip from the solar core to the Earth, the Cl
experiment would not observe them.

     In the mid 1980s, the Kamioka nucleon decay experiment (Kamiokande)
began to measure the solar neutrino flux.  This large water Cherenkov
detector was originally designed to look for high-energy signals from proton
decay.  After great effort, the energy threshold was reduced to a level to
permit a sensitivity to recoil electrons from \nuc{8}{B} solar neutrino
interactions.  The path of the recoil electrons is in the direction of the
initial neutrino trajectory, and thus this experiment demonstrated for the
first time that neutrinos were coming from the Sun.  The measured flux
\cite{FUK96} of $(2.80 \pm 0.19 \pm 0.33) \times 10^6$/(cm$^2$ s) was less
than half of the solar model prediction, and the solar neutrino problem was
thus confirmed by a second experiment.

     Because the high-energy solar neutrino flux was suppressed, it became
very important to also determine the flux of low-energy neutrinos produced in
the dominant proton-proton ($pp$) reaction.  Exotic hypotheses aside, the
rate of the $pp$ reaction is directly related to the solar luminosity and is
insensitive to alterations in the solar model.  In the early 1990s the
Russian-American Gallium Experiment (SAGE) and then the Gallium Experiment
(GALLEX) began to publish results.  These experiments are based on the
neutrino capture reaction \nuc{71}{Ga}$(\nu_e,e^-)$\nuc{71}{Ge} \cite{KUZ65}
and have the very low threshold of 233 keV \cite{TOI96}.  They are thus
sensitive to low-energy $pp$ neutrinos, whose end point energy is 423 keV
\cite{BAH97}, and provide the only feasible means at present to measure
low-energy solar neutrinos.  The SAGE result \cite{GAV98} of $66.9 ^{+7.1 +
5.4}_{-6.8 -5.7}$ SNU with a target of Ga metal and the GALLEX result
\cite{HAM99} of $77.5 \pm 6.2 ^{+4.3}_{-4.7}$ SNU with a target of GaCl$_3$
are both well below the SSM prediction from the Bahcall-Pinsonneault solar
model \cite{BAH98} of $129^{+8}_{-6}$ SNU.  The insensitivity of Ga to the
solar model is seen in the capture rate calculation from the model of Brun,
Turck-Chi\`eze, and Morel \cite{TUR98} of 127.2 SNU.  The contributions of
the components of the solar neutrino flux to the \nuc{71}{Ga} capture rate
are given in Table~\ref{theoretical_fluxes}.

     With the four experiments having three different thresholds, one can
deduce some information concerning the $\nu$ energy spectral distribution.
If one fits the data from all experiments with the neutrino fluxes as free
parameters, the best fit is when the $\nu_{\text{Be}}$ flux is greatly
reduced whereas there is an appreciable $\nu_{\text{B}}$ flux
\cite{HEE97,Smirnov98,BAH982}.  This is an apparent paradox as it is
difficult to form \nuc{8}{B} in the Sun without forming \nuc{7}{Be}.

     In 1996 Super-Kamiokande began to take data.  This 50 kton water
Cherenkov detector is the first high-count-rate solar neutrino experiment.
The present result \cite{Super-K} for the \nuc{8}{B} neutrino flux, assuming
that neutrino transformations do not occur, is $(2.44 \pm 0.05 _{-0.07}
^{+0.09}) \times 10^6$/(cm$^2$ s), in agreement with its predecessor.

     The purpose of this paper is to summarize all of the SAGE data for the
last eight years.  It is organized in the same way as the SAGE experiment is
carried out: after presenting some general aspects of the experiment, we
consider the chemical extraction of Ge from Ga and the subsequent Ge
purification.  Then we present how the Ge is counted, how \nuc{71}{Ge} events
are identified, and how the data are analyzed to give the solar neutrino
capture rate.  Finally, we consider the sources of systematic uncertainty,
give the overall results, and conclude with the implications for solar and
neutrino physics.

     In an attempt to make the material understandable to the general reader,
but still useful to the specialist, each of these subjects is first discussed
in a general way, followed by subsections that give more detail.  The reader
who wants a general overview need only read the beginning of each section.
The reader who desires more information regarding a particular subject should
read the appropriate subsection.

\section{SAGE Overview}\label{overview}

     In this section we give some general information on the location of the
experiment, its physical characteristics, and the division of the SAGE data
into three experimental periods.

\subsection{Baksan Neutrino Observatory}\label{bno}

     The SAGE experiment is situated in a specially built underground
laboratory at the Baksan Neutrino Observatory (BNO) of the Institute for
Nuclear Research of the Russian Academy of Sciences in the northern Caucasus
mountains.  The main chamber of the laboratory is 60~m long, 10~m wide, and
12~m high.  It is located 3.5 km from the entrance of a horizontal adit
excavated into the side of Mount Andyrchi and has an overhead shielding of
4700 meters of water equivalent.  To reduce neutron and gamma backgrounds
from the rock, the laboratory is entirely lined with 60 cm of
low-radioactivity concrete with an outer 6 mm steel shell.  All aspects of
the experiment are in this underground area, with additional rooms devoted to
chemistry, counting, and a low-background solid-state Ge detector.  Other
facilities for subsidiary measurements are in a general laboratory building
outside the adit.

\subsection{Extraction history}\label{extrac_hist}

     The data from SAGE span nearly a decade during which the experiment
evolved a great deal.  As a result, the data can be naturally divided into
several periods characterized by different experimental conditions.
Extractions on approximately 30 tons of Ga began in 1988; by late 1989
backgrounds were low enough to begin solar neutrino measurements.  The data
period referred to as SAGE~I began in January 1990 and ended in May 1992
\cite{ABD94}.  In the summer of 1991, the extraction mass was increased to
nearly 60 tons.  The SAGE~I data were taken without digitized wave forms and
the $L$ peak could not be analyzed because of high electronic noise at low
energy.  (The decay modes of \nuc{71}{Ge} are described below in
Sec.~\ref{counting}.)  The solar neutrino capture rate determined from this
data were published in Ref.\ \cite{NIC94}.

     Within a few months after SAGE~I was completed, the experiment was
greatly improved with respect to electronic noise.  The following period of
data, from September 1992 to December 1994, is referred to as SAGE~II.  It is
distinguished by recording of the counter wave form in most runs which makes
possible analysis of events in the low-energy $L$ peak.

\begin{table}
\caption{Definition of the various segments of SAGE data.}
\label{Data_Assignment_Table}
\begin{tabular}{lcc}

Designation & Included extractions          & Comments                   \\
\hline
SAGE I      & Jan. 90 $\rightarrow$ May  92 & Rise time from ADP         \\
SAGE II.1   & Sep. 92 $\rightarrow$ Oct. 93 & Rise time from wave form,  \\
            &                               & begin to use $L$ peak      \\
SAGE II.2   & Nov. 93 $\rightarrow$ June 94 & Ga theft period            \\
SAGE II.3   & July 94 $\rightarrow$ Dec. 94 & Before Cr experiment       \\
SAGE III.1  & Jan. 95 $\rightarrow$ June 95 & Some extractions           \\
            &                               & during Cr experiment       \\
SAGE III.2  & July 95 $\rightarrow$ present & After Cr experiment        \\
\end{tabular}
\end{table}

     During SAGE~II, there was a period (which we call SAGE~II.2) in which 2
tons of gallium, approximately 3.6\% of the total mass, was stolen from the
experiment.  The gallium was apparently removed in small quantities from
November 1993 to June 1994.  During this period a prototype gravity wave
laser interferometer at BNO detected unapproved transport of materials from
underground.  After discovery of the theft, all of the gallium was cleaned,
additional security controls for access to the gallium were instituted, and
SAGE resumed operation.  As this period of time has some uncertainty with
respect to experimental control, it is singled out for separate treatment,
and is not included in our best estimate for the neutrino capture rate.

     An experiment using a 517 kCi $^{51}$Cr neutrino source
\cite{ABD96,ABD98} began in late December of 1994 and continued until May
1995.  We refer to all data after January 1995 as SAGE~III, with a special
designation of SAGE~III.1 for solar neutrino extractions during the Cr
experiment.

     Table \ref{Data_Assignment_Table} summarizes the data period
designations.  The exposure times and other data for all runs of SAGE that
are potentially useful for solar neutrino capture rate determination are
given in Table~\ref{run_parameter_table}.

\begin{table*}
\squeezetable
\caption{Parameters for all 67 runs that are potentially useful for solar
neutrino measurement.  The efficiency values include the reduction for the
energy cut (usually 0.9815) and for the rise time cut (usually 0.95).
$L$-peak efficiencies are only given for those extractions for which wave
form data were available.  Peak ratio values, whose uncertainty is
approximately $\pm 0.02$, are only listed for those extractions for which
\nuc{109}{Cd} calibrations were made.  This ratio is 1.0 for unpolymerized
counters.}
\label{run_parameter_table}
\begin{tabular}{l c d c c c c c c c c c c}

           & Mean       & Exposure  & Ga \\
Exposure   & exposure   & time      & mass   & Extraction & Counter &
Pressure & Percent & Operating & Counting & $K$-peak   & $L$-peak   & Peak
\\
 date      & date       & (days)    & (tons) & efficiency & name    & (mm Hg)
& GeH$_4$ & voltage   & system   & efficiency & efficiency & ratio  \\
\hline
Jan. 90   & 1990.040 &  42.0 & 28.67 & 0.78 & Ni 1   & 604 & 28.0 & 1230 & 2
& 0.333 &       &      \\
Feb. 90   & 1990.139 &  30.0 & 28.59 & 0.79 & LA12   & 635 & 53.0 & 1450 & 5
& 0.249 &       &      \\
Mar. 90   & 1990.218 &  26.0 & 28.51 & 0.81 & Ni 1   & 640 & 25.0 & 1238 & 2
& 0.343 &       &      \\
Apr. 90   & 1990.285 &  19.0 & 28.40 & 0.76 & LA24   & 850 & 30.0 & 1430 & 5
& 0.335 &       &      \\
July 90   & 1990.540 &  21.0 & 21.01 & 0.78 & Ni 1   & 524 & 19.3 & 1130 & 2
& 0.327 &       &      \\
June 91   & 1991.463 &  53.0 & 27.43 & 0.82 & LA74   & 715 & 28.0 & 1300 & 2
& 0.334 &       &      \\
July 91   & 1991.539 &  23.0 & 27.37 & 0.66 & LA77   & 710 & 24.0 & 1300 & 3
& 0.320 &       &      \\
Aug. 91   & 1991.622 &  26.3 & 49.33 & 0.78 & RD2    & 570 & 34.0 & 1700 & 5
& 0.250 &       &      \\
Sep. 91   & 1991.707 &  27.0 & 56.55 & 0.78 & LA40   & 935 & 40.0 & 1630 & 2
& 0.338 &       &      \\
Nov. 91   & 1991.872 &  26.0 & 56.32 & 0.81 & LA46   & 108 & 30.0 & 1746 & 3
& 0.339 &       &      \\
Dec. 91   & 1991.948 &  26.8 & 56.24 & 0.79 & LA51   & 870 & 27.0 & 1394 & 2
& 0.336 &       &      \\
Feb. 92-1 & 1992.138 &  24.5 & 43.03 & 0.80 & LA71   & 666 & 12.0 & 1110 & 2
& 0.322 &       &      \\
Feb. 92-2 & 1992.138 &  24.5 & 13.04 & 0.80 & LA50   & 640 & 30.0 & 1165 & 2
& 0.305 &       &      \\
Mar. 92   & 1992.214 &  20.9 & 55.96 & 0.78 & LA46   & 810 & 20.5 & 1292 & 2
& 0.316 &       &      \\
Apr. 92   & 1992.284 &  23.5 & 55.85 & 0.83 & LA51   & 815 & 23.0 & 1386 & 2
& 0.333 &       &      \\
May  92   & 1992.383 &  27.5 & 55.72 & 0.67 & LA95   & 675 & 69.0 & 1620 & 2
& 0.282 &       &      \\
Sep. 92   & 1992.700 & 116.8 & 55.60 & 0.53 & LA110  & 720 & 21.0 & 1311 & 3
& 0.338 & 0.322 &      \\
Oct. 92   & 1992.790 &  27.2 & 55.48 & 0.83 & LA111  & 725 & 25.0 & 1391 & 3
& 0.341 & 0.327 &      \\
Nov. 92   & 1992.871 &  26.7 & 55.38 & 0.81 & LA105  & 730 & 23.0 & 1351 & 3
& 0.315 & 0.297 &      \\
Dec. 92   & 1992.945 &  24.3 & 55.26 & 0.85 & LA116  & 740 & 26.0 & 1406 & 3
& 0.325 & 0.315 & 1.04 \\
Jan. 93   & 1993.039 &  32.3 & 55.14 & 0.76 & LA110  & 770 & 25.0 & 1412 & 3
& 0.342 & 0.314 &      \\
Feb. 93   & 1993.115 &  23.0 & 55.03 & 0.79 & LA107  & 730 & 24.0 & 1336 & 6
& 0.315 &       &      \\
Apr. 93   & 1993.281 &  26.6 & 48.22 & 0.83 & LA111* & 710 & 23.0 & 1352 & 3
& 0.322 &       &      \\
May  93   & 1993.364 &  30.9 & 48.17 & 0.82 & LA116  & 705 & 16.0 & 1210 & 3
& 0.327 &       & 1.04 \\
June 93   & 1993.454 &  30.4 & 54.66 & 0.80 & LA110  & 740 & 24.0 & 1352 & 3
& 0.338 & 0.313 &      \\
July 93   & 1993.537 &  27.9 & 40.44 & 0.80 & LA111  & 675 & 22.0 & 1266 & 3
& 0.353 &       &      \\
Aug. 93-1 & 1993.631 &  34.0 & 40.36 & 0.79 & LA107  & 680 & 12.0 & 1210 & 6
& 0.317 &       & 1.00 \\
Aug. 93-2 & 1993.628 &  63.8 & 14.09 & 0.51 & A9     & 765 & 12.0 & 1130 & 6
& 0.322 &       & 1.20 \\
Oct. 93-1 & 1993.749 &  13.0 & 14.06 & 0.79 & A12    & 750 & 14.0 & 1224 & 6
& 0.333 &       & 1.00 \\
Oct. 93-2 & 1993.800 &  34.7 & 14.10 & 0.80 & LA111* & 710 & 15.0 & 1162 & 3
& 0.328 & 0.309 & 1.03 \\
Oct. 93-3 & 1993.812 &  24.6 & 14.02 & 0.84 & LA116  & 665 & 14.0 & 1184 & 3
& 0.323 & 0.299 & 1.04 \\
Nov. 93-1 & 1993.855 &  14.0 & 14.07 & 0.87 & LA119  & 665 & 13.0 & 1113 & 3
& 0.321 & 0.316 & 1.08 \\
Nov. 93-2 & 1993.844 &  53.4 & 26.16 & 0.52 & LA110  & 675 &  9.0 & 1094 & 3
& 0.340 & 0.326 & 1.00 \\
Dec. 93-1 & 1993.936 &  30.5 & 26.13 & 0.78 & A19    & 760 & 12.0 & 1287 & 3
& 0.336 &       & 1.08 \\
Dec. 93-2 & 1993.939 &  39.9 & 28.05 & 0.80 & LA111  & 690 & 12.0 & 1230 & 3
& 0.345 & 0.331 & 1.02 \\
Jan. 94-1 & 1994.048 &  42.2 & 26.67 & 0.82 & LA107  & 760 & 12.0 & 1196 & 6
& 0.328 &       & 1.00 \\
Jan. 94-2 & 1994.051 &  41.1 & 27.44 & 0.80 & LA111* & 750 & 12.5 & 1065 & 3
& 0.308 &       & 1.04 \\
Feb. 94   & 1994.137 &  28.0 & 54.01 & 0.64 & LA116  & 600 & 15.0 & 1090 & 3
& 0.312 & 0.326 & 1.04 \\
Mar. 94   & 1994.218 &  31.0 & 53.94 & 0.78 & LA105  & 625 & 10.0 & 1190 & 3
& 0.309 & 0.311 & 1.00 \\
Apr. 94   & 1994.283 &  22.5 & 53.88 & 0.73 & LA110  & 685 & 27.0 & 1331 & 3
& 0.328 & 0.335 & 1.00 \\
May  94-3 & 1994.374 &  32.9 & 26.99 & 0.85 & LA111  & 610 & 17.0 & 1215 & 3
& 0.329 & 0.343 & 1.00 \\
July 94   & 1994.551 &  31.3 & 50.60 & 0.80 & LA107  & 620 & 22.0 & 1236 & 3
& 0.301 & 0.269 & 1.00 \\
Aug. 94   & 1994.634 &  31.0 & 50.55 & 0.80 & LA105  & 655 & 13.0 & 1196 & 3
& 0.312 & 0.307 & 1.05 \\
Sep. 94-1 & 1994.722 &  33.2 & 37.21 & 0.76 & A13    & 695 & 18.0 & 1270 & 3
& 0.334 & 0.319 & 1.07 \\
Oct. 94   & 1994.799 &  28.8 & 50.45 & 0.76 & A19    & 695 & 25.0 & 1375 & 3
& 0.334 &       & 1.06 \\
Nov. 94   & 1994.886 &  31.0 & 50.40 & 0.79 & LA113  & 685 & 28.5 & 1383 & 3
& 0.306 & 0.314 & 1.05 \\
Dec. 94   & 1994.951 &  21.0 & 13.14 & 0.80 & A12*   & 610 & 16.5 & 1184 & 6
& 0.310 &       & 1.02 \\
Mar. 95   & 1995.209 &  42.5 & 24.03 & 0.92 & A28    & 690 & 18.5 & 1222 & 6
& 0.321 &       & 1.00 \\
July 95   & 1995.538 &  19.9 & 50.06 & 0.86 & LA107  & 635 & 30.0 & 1333 & 3
& 0.298 & 0.317 & 1.01 \\
Aug. 95   & 1995.658 &  46.7 & 50.00 & 0.70 & A12    & 710 & 17.0 & 1260 & 3
& 0.325 & 0.312 & 1.01 \\
Sep. 95   & 1995.742 &  28.8 & 49.95 & 0.67 & LA46   & 645 & 37.0 & 1382 & 3
& 0.283 & 0.294 & 1.02 \\
Oct. 95   & 1995.807 &  18.7 & 49.83 & 0.49 & A19    & 680 & 18.5 & 1248 & 3
& 0.319 & 0.294 & 1.08 \\
Nov. 95   & 1995.875 &  25.8 & 49.76 & 0.89 & A9     & 685 & 33.0 & 1429 & 3
& 0.310 & 0.294 & 1.21 \\
Dec. 95-2 & 1995.962 &  32.7 & 41.47 & 0.73 & LA113  & 725 & 18.5 & 1271 & 3
& 0.319 & 0.278 & 1.00 \\
Jan. 96   & 1996.045 &  29.7 & 49.64 & 0.77 & A12    & 715 & 24.0 & 1340 & 3
& 0.321 & 0.310 & 1.00 \\
May  96   & 1996.347 &  49.9 & 49.47 & 0.75 & LA116  & 685 & 21.5 & 1295 & 3
& 0.320 & 0.319 & 1.00 \\
Aug. 96   & 1996.615 &  45.0 & 49.26 & 0.77 & A13    & 675 & 23.0 & 1332 & 3
& 0.327 & 0.330 & 1.09 \\
Oct. 96   & 1996.749 &  45.8 & 49.15 & 0.83 & LA116  & 635 & 15.0 & 1185 & 3
& 0.318 & 0.319 & 1.02 \\
Nov. 96   & 1996.882 &  48.7 & 49.09 & 0.78 & A12    & 720 & 21.5 & 1308 & 3
& 0.323 & 0.306 & 1.00 \\
Jan. 97   & 1997.019 &  49.8 & 49.04 & 0.85 & LA113  & 700 & 29.0 & 1372 & 3
& 0.308 & 0.295 & 1.00 \\
Mar. 97   & 1997.151 &  44.9 & 48.93 & 0.93 & A13    & 650 & 23.5 & 1339 & 3
& 0.323 & 0.335 & 1.08 \\
Apr. 97   & 1997.277 &  42.9 & 48.83 & 0.90 & LA116  & 670 & 29.0 & 1360 & 3
& 0.313 & 0.320 & 1.02 \\
June 97   & 1997.403 &  45.6 & 48.78 & 0.87 & A12    & 675 & 24.5 & 1320 & 3
& 0.314 & 0.314 & 1.00 \\
July 97   & 1997.537 &  45.9 & 48.67 & 0.91 & LA51   & 690 & 15.5 & 1242 & 3
& 0.321 & 0.312 & 1.03 \\
Sep. 97   & 1997.671 &  46.4 & 48.56 & 0.75 & A13    & 650 & 25.0 & 1318 & 3
& 0.322 & 0.335 & 1.04 \\
Oct. 97   & 1997.803 &  45.0 & 48.45 & 0.83 & LA116  & 635 & 23.5 & 1318 & 3
& 0.328 & 0.327 & 1.03 \\
Dec. 97   & 1997.940 &  47.0 & 48.34 & 0.88 & A12    & 710 & 27.0 & 1382 & 3
& 0.318 & 0.306 & 1.00
\end{tabular}
\normalsize
\end{table*}

\section{Extraction of G\lowercase{e} from G\lowercase{a}}

     The extraction and concentration of germanium in the SAGE experiment
consists of the following steps.
\begin{enumerate}
\item{Ge is extracted from the Ga metal into an aqueous solution by an
      oxidation reaction.}
\item{The aqueous solution is concentrated.}
      \begin{enumerate}
      \item{Vacuum evaporation reduces the volume of aqueous solution by a
            factor of 8.}
      \item{Ge is swept from this solution as volatile GeCl$_4$ by a gas flow
            and trapped in 1~l of de-ionized water.}
      \item{A solvent extraction is made from the water which concentrates
            the Ge into a volume of 100~ml.}
      \end{enumerate}
\item{The gas GeH$_4$ is synthesized, purified, and put into a proportional
      counter.}
\end{enumerate}
\noindent The average extraction efficiency from the Ga metal to GeH$_4$ was
77\% before 1997 and 87\% thereafter.  Each of these steps will now be
briefly described and this section concludes with a description of the
evidence that the extraction procedure does indeed remove germanium with high
efficiency.

\subsection{Chemical extraction procedure}\label{chem_extr_proc}

\subsubsection{Extraction of Ge from metal Ga}

     A procedure for the extraction of Ge from metallic Ga was first
investigated at Brookhaven National Laboratory \cite{BAH78}.  It is based on
the selective oxidation of Ge in liquid Ga metal by a weakly acidic
H$_2$O$_2$ solution.  This method was developed and fully tested in a 7.5-ton
pilot installation at the Institute for Nuclear Research \cite{BAR84}.  The
final procedure extracts Ge with high efficiency and dissolves only a small
amount of Ga \cite{IS90,Vere91}.

\begin{figure}
\begin{center}
\includegraphics*[width=3.375in]{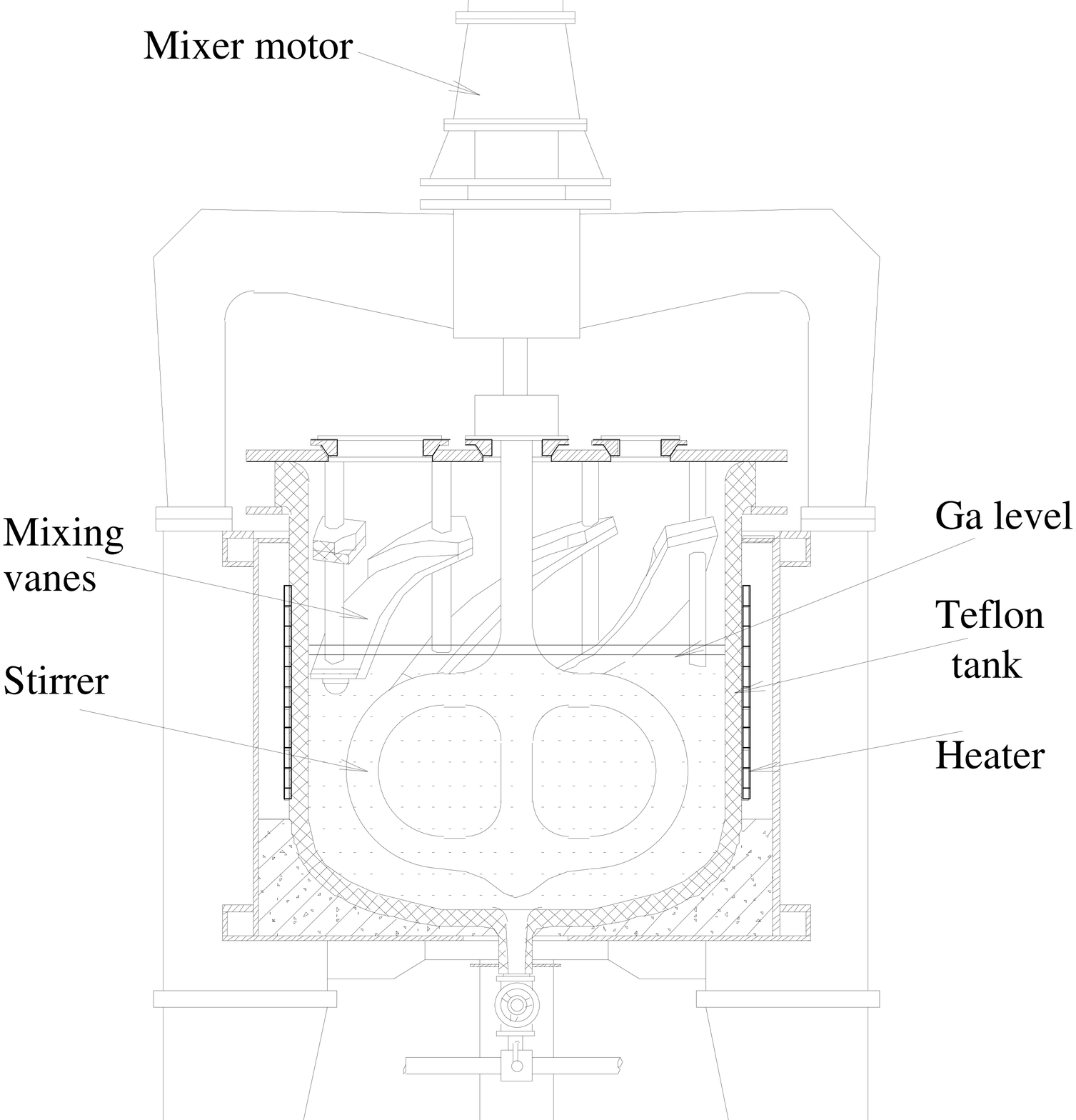}
\end{center}
\caption{Chemical reactor for extraction of Ge from Ga.}
\label{reactor}
\end{figure}

     The Ga at BNO is contained in chemical reactors, each of which is able
to extract from as much as 8 tons of Ga.  The reactor (Fig.\ \ref{reactor})
is a 2-m$^3$ Teflon tank with $\sim 40$-mm-thick walls to which band heaters
are attached.  The Teflon tank is placed inside a secondary stainless steel
tank.  The Ga can be stirred with a motor that can turn an internal mixer at
up to 80 rpm.  A specially designed set of vanes are attached to the inside
cover of the reactor that serve to completely disperse the extraction
reagents (density 1.0 kg/l) throughout the liquid Ga (density 6.1 kg/l).  The
vanes are made from Teflon and the stirrer and cover are Teflon lined.  A
glass viewport in the reactor cover enables one to see the extremely vigorous
mixing action.  Ten such reactors are installed at BNO which are connected
with a system of heated Teflon tubing and a Teflon pump that can transfer Ga
between reactors.  A system of glass-Teflon dosing pumps can put a measured
volume of reagents into any reactor, and a vacuum suction device made from
Teflon, glass, and zirconium extends to the Ga surface to remove the
reagents.  The filling of a reactor with reagents and the stirring are
controlled by an automated system.

     Each measurement of the solar neutrino flux begins by adding to the Ga
approximately 700 $\mu$g of stable Ge carrier (distributed equally among all
of the reactors) in the form of a solid Ga-Ge alloy with known Ge content
$(\sim 2 \times 10^{-4}$ mass \%).  The reactor contents are then stirred so
as to thoroughly disperse the carrier throughout the Ga metal.  After a
typical exposure interval of 4--6 weeks, the Ge carrier and any additional Ge
atoms produced by solar neutrinos or other processes are chemically extracted
from the Ga.

     The efficiency of Ge extraction depends on a number of parameters.
Since the efficiency falls rapidly as the Ga temperature increases, we begin
to extract with the Ga at 30.0--30.5\degreesC, just slightly above its
freezing temperature (29.8\degreesC).  The efficiency increases with an
increase in the amount of oxidizing agent (H$_2$O$_2$), but this has the
detrimental effect of dissolving more Ga.  The efficiency also depends on the
volume of aqueous phase which defines the time of later concentration of Ge,
the most time consuming part of the entire extraction process.  Taking into
account all of these factors, a procedure was developed which extracts about
85\% of the Ge and dissolves only 0.1\% of the Ga.

     The extraction solution for a reactor containing 7.5 tons of Ga consists
of 200~l of de-ionized water, 5~l of 7~M HCl,\footnote{The symbol~M stands
for the amount of substance concentration in moles per liter.} and 16~l of a
30\% solution of H$_2$O$_2$.  All components of this solution are purified so
their Ge content is negligible.  Immediately after the reagents are added,
reactor stirring starts at a speed of 70 rpm.  As the mixture is intensively
stirred, the gallium turns into fine droplets which are covered with a Ga
oxide film.  This film prevents fusion of the droplets and holds the Ga as an
emulsion \cite{Vere87,Rowley}.  The dissolved Ge in the Ga migrates to the
surface of the droplets where it is oxidized and incorporated into the oxide
film.  Because of the highly exothermic oxidation reaction, the Ga
temperature rapidly rises.  After approximately 25 min, the H$_2$O$_2$ has
been consumed; the Ga temperature plateaus and the emulsion spontaneously
breaks down.  To dissolve the oxide containing Ge, the extraction procedure
is finished by adding 45~l of 7~M HCl (cooled to $-15$\degreesC) and stirring
for 1--2 min.  The Ga temperature at the end of this extraction process is
$\sim 50$\degreesC.

     The extraction solution is immediately decanted and sent to the first
step of concentration, which is evaporation.  The Ga in each reactor is then
washed by adding 20~l of 0.5~M HCl.  This solution is stirred with the liquid
Ga for about 1 min, is decanted out, and is added to the previous extraction
solution.  Finally, to prevent oxidation of the Ga during the interval
between extractions, a solution of 0.5~M HCl is added to the reactor and left
there until the next extraction.

\subsubsection{Vacuum evaporation of extraction solutions}

     Extraction is made sequentially from one reactor to the next.  All the
extraction solutions, whose total volume is 2200~l for 60 tons of gallium,
are combined at the evaporation step, which is carried out in a glass
recirculation apparatus with a steam-heated active volume of 70~l.  As the
evaporation proceeds, the acidity of the evaporated solution increases.  Ge
is volatile from concentrated chloride solutions, so the evaporation is
stopped when the volume of solution reaches 250--270~l, before loss of Ge can
begin.  The average time for evaporation is 15 h.

\subsubsection{Sweeping}

     The next step is based on the volatility of GeCl$_4$ from a concentrated
solution of HCl.  The evaporated extraction solution, which contains 250~g of
Ga/l in the form of chloride, is transferred to glass vessels with a volume
of 200~l.  These vessels are part of a sealed gas flow system.  The HCl
concentration is raised to 9~M by adding purified 12~M HCl and an air flow at
1.0 m$^3$/h is initiated.  Ge is swept as GeCl$_4$ from this 50\degreesC\
acid solution through a counter-current scrubber where the GeCl$_4$ is
absorbed in a 1.0~l volume of de-ionized H$_2$O.  The amount of Ge remaining
in the solution $C(t)$ falls exponentially: $C(t) = C(0) \exp[-1.84 V(t)]$
where $V$ is the volume of sweep gas in m$^3$.  The duration of sweeping is
usually 2.5 h which gives 99\% Ge extraction efficiency.  At the end of the
sweep the acidity of the absorber solution is in the range of 4.0~M to 4.2M,
which excludes loss of Ge.

\subsubsection{Solvent extraction}

     A solvent extraction is then carried out to further concentrate the Ge.
This procedure is based on the high distribution coefficient of Ge between an
acidic water solution and an organic solvent, such as CCl$_4$.  To achieve an
optimal acidity (8.5~M), the appropriate amount of purified 12~M HCl is added
to the solution obtained from sweeping.  The Ge is first extracted into
CCl$_4$ and then is back extracted into low-tritium H$_2$O.  This process is
repeated 3 times.  To remove the residual CCl$_4$, a very small amount of
hexane is added to the organic phase at the last step of the final back
extraction.  The final traces of hexane are removed by heating the solution
at 90\degreesC\ for 40 min.  This results in the Ge being concentrated in a
volume of 100~ml of low-tritium H$_2$O.

\subsubsection{Germane synthesis}

     The final step of the extraction process is to synthesize germane
(GeH$_4$) which is used as a 20\%--30\% fraction of the counting gas in a
proportional counter.  NaOH is added to the 100-ml water solution to adjust
the $p$H to the range of 8--9, and the solution is placed in a reaction flask
on a high-vacuum glass apparatus.  Any air is swept out of the solution and
the connecting piping with a He flow and 2~g of low-tritium NaBH$_4$
dissolved in 40~ml of low-tritium H$_2$O is added.  The mixture is then
heated to 70\degreesC, at which temperature the Ge is reduced by the NaBH$_4$
to make GeH$_4$.  The H$_2$ generated by the reaction and the flowing He
sweep the GeH$_4$ onto a Chromosorb 102 gas chromatography column at
$-196$\degreesC\ where it is trapped.  When the reaction is finished the
column temperature is raised to $-35$\degreesC\ and the GeH$_4$ is eluted
with He carrier gas.  It is then frozen on another Chromosorb 102 trap at
$-196$\degreesC\ where most of the He is pumped away.  The GeH$_4$ is then
transferred with a mercury-filled Toepler pump to a glass bulb at
$-196$\degreesC\ where any residual He is pumped away.  The Toepler pump is
used again to transfer the GeH$_4$ to a calibrated stem, where the GeH$_4$
volume is measured.  During this transfer the temperature of the bulb is held
at $-142$\degreesC\ so as to minimize Rn.  A measured quantity of old
low-background Xe is added and this gas mixture is inserted into a miniature
proportional counter.  The counter has been evacuated at $10^{-6}$ torr and
baked at 100\degreesC\ for at least 6 h.

\subsubsection{Modified procedures for SAGE III}

\begin{figure}
\begin{center}
\includegraphics[width=3.375in]{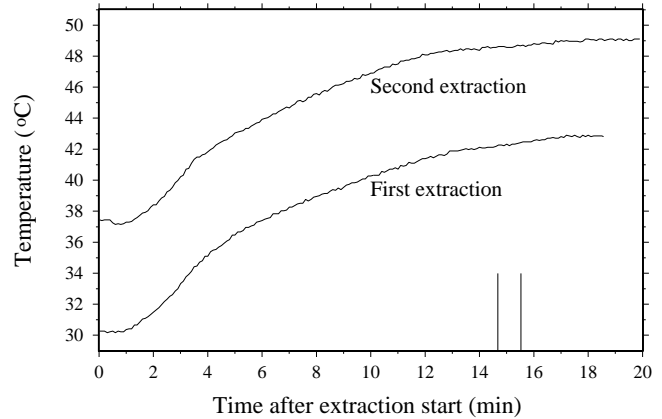}
\end{center}
\caption{Ga temperature using two extraction procedure begun in 1997.   The
extraction reagents are added 1 min before time zero.  Extraction begins when
the mixer is started.  The two vertical lines about 15 min after the start of
extraction are when the HCl is added to dissolve the oxide containing Ge.}
\label{reacextr}
\end{figure}

\paragraph{Extraction from Ga.}
     At the beginning of 1997, the extraction procedure was modified to a
two-step extraction process.  In the first step the volume of reagents added
to each reactor is reduced from the values given previously by a factor of 2.
The remaining steps in the removal of Ge from the Ga proceed the same as
previously described, but only require about 15 min because of the reduced
H$_2$O$_2$ volume.  This first step extracts about 75\% of the Ge from the
Ga, dissolves 0.05\% of the Ga, and raises the Ga temperature to about
40\degreesC\ (Fig.\ \ref{reacextr}, lower curve).  After the first extraction
from each reactor, a second extraction is carried out in the same order using
the same volume of reagents as in the first extraction.  By the time the
second extraction begins the Ga has cooled to 37\degreesC\ and an additional
drop of 1.5--2\degreesC\ occurs when the new reagents are added.  Since the
initial Ga temperature is now elevated, the efficiency of the second
extraction is less than the first, and averages 70\%.  Again 0.05\% of the Ga
is dissolved and the final Ga temperature is 49\degreesC\ at the end of this
second extraction (Fig.\ \ref{reacextr}, upper curve).  This modified
procedure results in a total efficiency of Ge removal from the Ga in excess
of 90\%, but both procedures dissolve the same total amount of Ga (0.1\%).

\paragraph{Evaporation of extraction solutions.}
     The vacuum evaporation was modified at the beginning of SAGE~III.
Instead of stopping the distillation before the Ge volatilizes, the Ge is
allowed to evaporate, at which time collection of Ge in the condensate is
begun.  Evaporation is continued until all the Ge has been transferred to the
condenser.  The condensate is then further evaporated until its acidity is
4.5~M.  This solution, whose volume is about 130 l, is transferred to the
sweeping apparatus, 12~M hydrochloric acid is added to obtain 9~M acidity,
and the Ge is swept out in the same manner as for SAGE~I and II.  An
important advantage of this new method is that the solution that results from
sweeping is pure 9~M HCl, free from Ga or Ge, so it can be used in later
extractions.  These chemical technology modifications in SAGE~III increase
the efficiency of Ge extraction by 6\%--7\%, decrease the average duration of
concentration by 3--4 h, and reduce the consumption of concentrated HCl by
2.5 times.

\subsection {Chemical extraction efficiency}
\label{Chemical_extraction_efficiency}

     The total efficiency of extraction of Ge is given by the ratio of the Ge
content of the synthesized germane to the Ge present in the reactors at the
beginning of the exposure interval.  As a check, the amount of extracted Ge
is also determined by atomic absorption analysis of a small fraction of the
solution used in the GeH$_4$ synthesis.  The extraction efficiency prior to
1997 was typically 80\%.  The modified extraction procedure initiated in 1997
gives about a 10\% higher overall efficiency.  The extraction efficiency for
each run is given in Table~\ref{run_parameter_table}.

     Since each extraction leaves 10\%--20\% of the carrier Ge still present
in the Ga, it is customary to make a second extraction within a few days
after the first.  This second extraction removes most of the residual Ge so
that the Ge content of each reactor is well known after the carrier Ge is
added.  Occasionally a third extraction is made to totally deplete the Ge
content.  The extracts from these additional extractions are usually
processed in the same manner as for the solar neutrino extraction, including
counting of the synthesized GeH$_4$.

\subsection{Tests of the extraction efficiency}\label{extr_effs}

     The Ga experiment relies on the ability to extract a few tens of atoms
of \nuc{71}{Ge} from $5 \times 10^{29}$ atoms of Ga.  To measure the
efficiency of extraction, about 700 $\mu$g of stable Ge carrier is added to
the Ga at the beginning of each exposure, but even after this addition, the
separation factor of Ge from Ga is still 1 atom in $10^{11}$.  In such a
situation one can legitimately question how well the extraction efficiency is
known.  We have performed auxiliary measurements to verify that this
efficiency is well established, and briefly describe these tests in this
section.

\subsubsection{\nuc{51}{Cr} experiment}\label{cr_expt}
     The most direct experiment of this type involved the irradiation of Ga
with the 747-keV neutrinos from an artificial source of \nuc{51}{Cr}
\cite{ABD96,ABD98}.  Eight exposures of 13 tons of Ga were made to a 517 kCi
\nuc{51}{Cr} source.  The \nuc{71}{Ge} atoms were extracted by our usual
chemical procedure and their number determined by counting.  The ratio $R$ of
the measured neutrino capture cross section \cite{ABD96,ABD98} to the
theoretically calculated cross sections of Bahcall \cite{BAH97} and Haxton
\cite{HAX98} was

\begin{eqnarray}
R & \equiv & \frac{\sigma_{\text{measured}}}
                  {\sigma_{\text{theoretical}}} \\
  & =      & \left\{ \begin{array}{l l}
                     0.95 \pm 0.12 \text{ (expt)}\ ^{+0.035}_{-0.027}
                              \text{ (theor)}  & \text{ (Bahcall)}, \\
                     0.87 \pm 0.11 \text{ (expt)}\ \pm 0.09
                               \text{ (theor)} & \text{ (Haxton)}.
                     \end{array}
             \right.
       \nonumber
\end{eqnarray}

\noindent With either of these theoretical cross sections, $R$ is consistent
with 1.0, which implies that the extraction efficiency of \nuc{71}{Ge} atoms
produced in Ga by the neutrinos from \nuc{51}{Cr} is the same as that of
natural Ge carrier.

\subsubsection{Ga(n,${\gamma}$) experiment}
     To test the possibility that atomic excitations might tie up
\nuc{71}{Ge} in a chemical form from which it would not be efficiently
extracted, the radioactive isotopes \nuc{70}{Ga} and \nuc{72}{Ga}, which beta
decay to \nuc{70}{Ge} and \nuc{72}{Ge}, were produced in liquid gallium by
neutron irradiation.  The Ge isotopes were extracted from the Ga using our
standard procedure.  The number of Ge atoms produced was determined by mass
spectroscopic measurements \cite{ABD94} and was found to be consistent with
the number expected based on the known neutron flux and capture cross
section, thus suggesting that chemical traps are not present.

\subsubsection{Removal of \nuc{68}{Ge}}
     Further evidence that the extraction efficiency is well understood came
from monitoring the initial removal from the Ga of cosmogenically produced
\nuc{68}{Ge}.  This nuclide was generated in the Ga as it resided outside the
laboratory exposed to cosmic rays.  When the Ga was brought underground, the
reduction in the \nuc{68}{Ge} content in the initial extractions was the same
as for the Ge carrier.

\subsubsection{\nuc{71}{Ge} carrier}\label{Ge_carrier}
     A special Ge carrier was produced which contained a known number of
\nuc{71}{Ge} atoms.  This carrier was added to a reactor holding 7 tons of
Ga, three successive extractions were carried out, and the number of
\nuc{71}{Ge} atoms in each extraction was determined by counting.  The
results \cite{Abazov91} verified that the extraction efficiencies of the
natural Ge carrier and \nuc{71}{Ge} track each other very closely.

\section {Counting of $^{\bf71}$G\lowercase{e}}\label{counting}

\subsection{General overview}

     Once the \nuc{71}{Ge} is isolated internally in the proportional
counter, its decay must be identified.  \nuc{71}{Ge} decays solely by
electron capture to the ground state of $^{71}$Ga with a half-life of 11.43
days \cite{HAM85}.  The probabilities of $K$, $L$, and $M$ capture are 88\%,
10.3\%, and 1.7\%, respectively \cite{GEN71}.  $K$ capture gives Auger
electrons with an energy of 10.367 keV (41.5\% of all decays), 9.2-keV x rays
accompanied by 1.2-keV Auger electrons from the subsequent $M$--$L$
transition (41.2\% of all decays), and 10.26-keV x rays accompanied by
0.12-keV Auger electrons (5.3\% of all decays).  $L$ and $M$ capture give
essentially only Auger electrons with energies of 1.2 keV and 0.12 keV,
respectively \cite{BRO86}.  The proportional counter observes the Auger
electrons and, with considerably less efficiency, the x rays emitted during
the relaxation of the atomic electron shell.  As a result, about the same
fraction of events occur in the $L$ and $K$ peaks.

     These low-energy Auger electrons and x rays produce a nearly pointlike
ionization in the counter gas.  This ionization will arrive at the anode wire
of the proportional counter as a unit, resulting in a fast rise time for the
pulse.  In contrast, although a typical $\beta$ particle produced by a
background process may also lose 1--15 keV in the counter gas, it will leave
an extended trail of ionization.  This ionization will arrive at the anode
wire distributed in time according to its radial extent in the counter, which
usually gives a pulse with a slower rise time than for a \nuc{71}{Ge} event.
The identification of true \nuc{71}{Ge} events and the rejection of
background events are thus greatly facilitated by using a two parameter
analysis: a candidate \nuc{71}{Ge} event must not only fall within the
appropriate energy region, but must also have a rise time consistent with
pointlike ionization.

     To properly determine the background rate it is necessary to count each
sample for a long time after any \nuc{71}{Ge} has decayed.  We endeavor to
begin to count as soon as possible after extraction and to continue counting
for at least 160 days.  Since the number of high-quality low-background
counters and of available counting channels is limited, runs are occasionally
ended before the desired counting duration is met to permit another run to
begin.  Further, since many counters are measured in a common system,
counting time is frequently lost for calibration or for counter installation
or removal.

     This section continues with a discussion of how the proportional
counters are made, how their counting efficiency is determined, and how they
are calibrated, and concludes with a description of the counting electronics.

\subsection{Proportional counters}\label{prop_cntrs}

     The design and construction of the proportional counters are based on
the experience gained in the Cl experiment.  They are made only from
materials that are radioactively clean, are assembled in a clean environment,
are only exposed to high levels of radioactivity during efficiency
measurement, and are always counted for background before use in a solar
neutrino extraction.

\begin{figure}
\begin{center}
\includegraphics[width=3.375in]{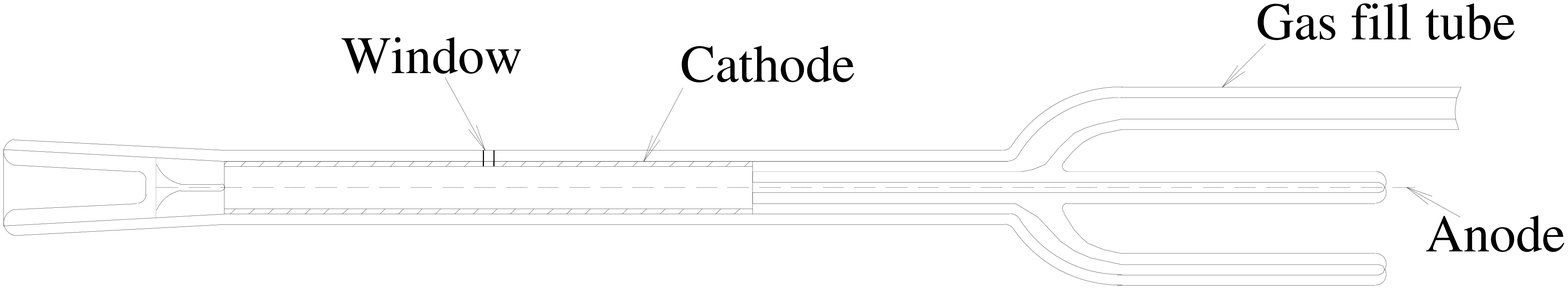}
\end{center}
\caption{Schematic view of a proportional counter.}
\label{counter_view}
\end{figure}

     Although several different types of counters were used at the beginning
of the experiment, all counters used since the extraction of September 1991
are of a common type, shown in Fig.\ \ref{counter_view}.  The counter bodies
are fabricated by a glassblower from Heraeus Amersil transparent synthetic
fused silica (Suprasil).  The main body is 10 cm long with an 8-mm outer
diameter.  One end is open for insertion of the cathode but can be sealed
with a flared plug.  Three tubes are attached to the other end --- one tube
with 2-mm inner diameter is used for insertion of the filling gas; the other
two tubes are capillaries for the cathode and anode electrical feedthroughs.
A 2-mm hole is made in the counter body near its center over which a very
thin piece of blown silica is sealed.  There is a corresponding hole in the
cathode at this position so that x rays from external sources can pass
through this window into the counter gas for calibration.

     The main body of the counter is large enough to hold a snug-fitting
zone-refined iron cathode sleeve, whose dimensions are approximately 5 mm
diameter, 5 cm length, and 1/3 mm thickness.  The cathodes are individually
machined to fit each counter body, making sure that there is sufficient space
between the cathode and the body to permit the counter to be heated to at
least 100\degreesC\ for bakeout of impurities.  The iron is drilled and cut
to length using only new tools and ultrapure hexane as lubricant.

     The major component of the counters, Suprasil, has a total metallic
impurity content of $\leq 1$ ppm by weight and OH and equivalent H$_2$O
contents of $\sim 10^3$ ppm.  The cathode material typically has less than 1
ppm metallic impurities, except for copper which is present at $\sim 7$ ppm.

     The first step in counter fabrication is a helium leak test of the seals
and thin calibration window of the counter body.  All parts of the counter
are then thoroughly cleaned: the silica parts are soaked overnight in aqua
regia, etched briefly in hydrofluoric acid, thoroughly washed in high-purity
water, and dried in an oven at slightly above 100\degreesC.  The cathodes are
washed in hexane in an ultrasonic bath, baked, and dried under vacuum for
approximately 24 h at 500\degreesC.  After cleaning, all counter parts are
handled only with gloves and clean tools.

     The final steps of counter fabrication take place inside a laminar flow
clean bench.  Under a microscope, a 25-$\mu$m wire of high-purity tungsten is
spot welded to the cathode and then threaded through a thin capillary to the
outside of the counter, where an external lead pin is connected.  Again under
a microscope, a 12.5-$\mu$m tungsten anode wire is threaded through the
second capillary, through the center of the cathode sleeve, and welded to a
50-$\mu$m tungsten spring wire held in place at the end of the counter by the
Suprasil end plug.  With the anode and cathode wires held taut in the
capillaries, the electrical connections to the external leads are made with a
small dab of conducting epoxy injected into the end of the capillary with a
hypodermic needle.  With the wires still held taut, the quartz end plug is
gently welded in place by a glassblower, and then (with the counter filled
with $\sim 0.1$ atm hydrogen to prevent oxidation of the thin wires), the
capillaries are heated and sealed around the cathode and anode wires.  The
counters are then tested for gas tightness, evacuated and baked for $\geq 72$
h, purged, and filled for testing with P-10 counter gas (90\% argon, 10\%
methane).

     Counters are tested at the time of fabrication for stability, gain, and
resolution.  Counter background rates are measured at Baksan and are in the
range of 0.1/day (0.07/day) in the \nuc{71}{Ge} $L$-peak ($K$-peak) candidate
regions.

\subsection{Measurement of proportional counter efficiency}
\label{counter_efficiency}

     This section gives a general description of the methods of counter
efficiency measurement, shows how these methods are applied to determine the
$L$- and $K$-peak efficiencies of several typical counters, and presents how
the counting efficiency of the solar neutrino extractions is determined.

\subsubsection{Measurement methods}

     Two different techniques and three different isotopes are employed:
\nuc{37}{Ar} to measure volume efficiency, and \nuc{69}{Ge} and \nuc{71}{Ge}
to measure the $L$- and $K$-peak efficiencies.

     The first method uses \nuc{37}{Ar} to measure the volume efficiency,
which we define as the probability that the decay of a radioactive atom in
the gas phase in the volume of a proportional counter will produce a
detectable pulse.  The \nuc{37}{Ar} source is produced by the $(n,\alpha)$
reaction on \nuc{40}{Ca} using fast neutrons from the research breeder
reactor of the Institute of Physics and Power Engineering in Obninsk.  The
extracted \nuc{37}{Ar} is purified on a Ti getter and then mixed with 90\% Ar
plus 10\% CH$_4$.  A small sample of this mixture is placed into the counter
under test, the counter high voltage is set so that the \nuc{37}{Ar} $L$ peak
is at least one-quarter scale on the energy analog-to-digital converter
(ADC), and an energy spectrum is measured.  The gas sample is then
transferred with very high efficiency $E_{\text{transfer}}$ ($>99.5$\%) to a
counter that was specially constructed for these measurements.  It is 20 cm
in length with an internal diameter of 4 mm.  It has a deposited carbon film
cathode, shaped ends to minimize end effects, a volume of 2.5 cm$^3$, and a
volume efficiency of $(99.5\pm 0.2)$\%.  Additional Ar--10\% CH$_4$ is added
to bring the pressure in this standardization counter to about the same value
as in the test counter, and another energy spectrum is measured under similar
conditions to that of the test counter.  To find the position of the $L$
peak, these two spectra are fit to a Gaussian plus a constant background.  An
energy threshold is then set at one-third of the peak value, equivalent to
about 80 eV, and the total number of counts above this threshold determined
by summation.  This gives the count rates in the counter under test,
$R_{\text{test}}$, and in the standardization counter, $R_{\text{standard}}$.
After making minor corrections for background rates, the volume efficiency of
the test counter is given by $\epsilon_v = 0.995 R_{\text{test}}
E_{\text{transfer}} D /R_{\text{standard}}$ where $D$ is the decay factor of
the \nuc{37}{Ar} between the times of measurement of the two spectra. Because
of the high and well-known transfer efficiency and standardization counter
efficiency, the total estimated uncertainty in the volume efficiency of the
test counter using this method is only 0.005, or approximately 0.6\%.

     The second counter efficiency measurement method uses \nuc{69}{Ge}.  A
brief description is given here; for more details see
Ref.\ \cite{Abdurashitov and Gusev and Yants 95}.  The \nuc{69}{Ge} source is
made by the $(p,n)$ reaction on 99\%-enriched \nuc{69}{Ga} with 7-MeV protons
from the cyclotron of the Nuclear Physics Institute of Moscow State
University.  \nuc{69}{Ge} is extracted from the gallium target, synthesized
into \nuc{69}{Ge}H$_4$, and added to a normal GeH$_4$-Xe counter filling.
\nuc{69}{Ge} decays both by electron capture (64\%) and by positron emission
(36\%).  About 40\% of the electron capture decays go to an excited state of
the daughter \nuc{69}{Ga} which emits a coincident 1106-keV gamma ray. The
measurements are made by placing a proportional counter with a \nuc{69}{Ge}
filling on the axis of and 10 cm to 12 cm distant from a large Ge
semiconductor detector which observes the gamma rays.  Energy spectra are
taken of the events produced by electron capture decays of \nuc{69}{Ge} by
gating the signal from the proportional counter with the output of a single
channel analyzer set on the 1106-keV gamma ray.  The $K$, $L$, and volume
efficiencies are defined as the ratio of the number of counts in the $K$
peak, $L$ peak, and total spectrum, respectively, to the number of 1106-keV
gammas detected by the large germanium detector.  In these calculations,
small corrections are made to the raw number of observed events because of
random coincidences and background in the Ge detector.  The uncertainty in
this measurement method is mainly from the partial detection of $M$-peak
events.  The $M$ peak in Ge is at $\sim 120$ eV, a higher energy than in Ar,
but it contains a much larger fraction of the total number of decays (7\%
compared to 1.4\% in Ar).  Even though part of the $M$ peak is detected in
these Ge spectra, a substantial correction for the missing fraction of events
below threshold energy is still required.  The estimated uncertainty in the
peak efficiency is thus slightly less than 2.5\% (or 0.008 in absolute
efficiency) and 1.7\% (0.015 in absolute efficiency) in the volume
efficiency.

     The final measurement method uses \nuc{71}{Ge} produced by neutron
irradiation of \nuc{70}{Ge}.  After extraction and purification of the Ge,
\nuc{71}{Ge}H$_4$ is synthesized, and mixed with Xe-GeH$_4$.  Measurements of
the volume efficiency are then made using a similar technique to that
described for \nuc{37}{Ar}. In addition, the $L$- and $K$-peak efficiencies
are determined by integration over the peaks.  The uncertainty in this
measurement method is about the same as for the \nuc{69}{Ge} method.

\subsubsection{Application and test}

\begin{table}
\caption{Volume efficiency measurements using the three techniques described
in the text.}
\label{volume_efficiency}
\begin{tabular}{lccc}

Counter & \multicolumn{3}{c}{Volume efficiency measured with}       \\
\cline{2-4}
\vspace{-0.8em} \\  
name    & \nuc{37}{Ar}      & \nuc{69}{Ge}      & \nuc{71}{Ge}      \\
\hline
LA51    & 0.887 $\pm$ 0.005 &                   &                   \\
LA88    & 0.876 $\pm$ 0.005 & 0.854 $\pm$ 0.015 & 0.879 $\pm$ 0.015 \\
LA105   & 0.872 $\pm$ 0.005 &                   &                   \\
LA107   & 0.874 $\pm$ 0.005 &                   &                   \\
LA110   & 0.933 $\pm$ 0.005 &                   &                   \\
LA111   & 0.948 $\pm$ 0.005 &                   &                   \\
LA111*  & 0.897 $\pm$ 0.005 & 0.895 $\pm$ 0.015 & 0.908 $\pm$ 0.015 \\
LA113   & 0.875 $\pm$ 0.005 &                   &                   \\
LA114   & 0.892 $\pm$ 0.005 & 0.918 $\pm$ 0.015 & 0.913 $\pm$ 0.015 \\
LA116   & 0.901 $\pm$ 0.005 &                   &                   \\
A8      & 0.868 $\pm$ 0.005 & 0.867 $\pm$ 0.015 &                   \\
A13     & 0.928 $\pm$ 0.005 &                   &                   \\
A28     & 0.893 $\pm$ 0.005 &                   &                   \\
A31     & 0.872 $\pm$ 0.005 &                   &                   \\
\hline
Average & 0.894 $\pm$ 0.025 &                   &                   \\
\end{tabular}
\end{table}

\begin{table*}
\caption{Comparison of measured counting efficiency in a 2 FWHM wide energy
window centered on the $L$ and $K$ peaks with the efficiency calculated from
the efficiency formula, Eq.~(\protect \ref{efficiency_formula}), using volume
efficiency measured with \nuc{37}{Ar}.  The uncertainty in the measured
efficiency is $\pm$ 0.008 and the uncertainty in the calculated efficiency is
estimated to be $\pm$ 1.5\%.}
\label{efficiency_comparison}
\begin{tabular}{l c c d d d d d}

        &         &          & GeH$_4$    & \multicolumn{4}{c}{Counting
efficiency}                       \\ \cline{5-8}
Counter & Isotope & Pressure & fraction   & \multicolumn{2}{c}{in $K$ peak} &
\multicolumn{2}{c}{in $L$ peak} \\ \cline{5-6} \cline{7-8}
name    &   used  & (mm Hg)  &(volume \%) & Measured        & Calculated  &
Measured        & Calculated  \\
\hline
LA88    & \nuc{69}{Ge} & 640 & 10.6 & 0.326 & 0.329 & 0.313 & 0.327 \\
LA88    & \nuc{71}{Ge} & 735 & 9.5  & 0.332 & 0.345 & 0.322 & 0.316 \\
LA111*  & \nuc{69}{Ge} & 710 & 15.  & 0.347 & 0.345 & 0.334 & 0.330 \\
LA111*  & \nuc{71}{Ge} & 735 & 9.5  & 0.358 & 0.353 & 0.321 & 0.324 \\
LA114   & \nuc{69}{Ge} & 745 &  8.  & 0.355 & 0.354 & 0.316 & 0.320 \\
LA114   & \nuc{71}{Ge} & 909 & 17.4 & 0.379 & 0.369 & 0.320 & 0.310 \\
A8      & \nuc{69}{Ge} & 800 & 12.  & 0.348 & 0.349 & 0.310 & 0.308

\end{tabular}
\end{table*}

     Table~\ref{volume_efficiency} gives the measured volume efficiencies for
14 counters using the measurement methods based on \nuc{37}{Ar},
\nuc{69}{Ge}, and \nuc{71}{Ge}.  For those counters that were measured with
more than one isotope, the agreement is very good and distributed in the
expected statistical manner.

     The efficiencies in the $L$ and $K$ peaks for four counters measured
with the \nuc{69}{Ge} coincidence method and for three counters measured with
\nuc{71}{Ge} are given in Table~\ref{efficiency_comparison}.  Because
different gas compositions and pressures were used in these counter fillings,
these measurements can only be compared if one has a procedure for correcting
the efficiency for the gas filling.  Since our solar neutrino runs also have
different counter fillings, such a correction procedure is also essential for
determining the counting efficiency for normal extractions.

     The \nuc{71}{Ge} counting efficiency $\epsilon(P,G)$, before the
application of energy or rise time cuts, can be written in the general form
     \begin{equation}
     \label{efficiency_formula}
\epsilon(P,G) = \epsilon_v (1 - f_D) E(P,G),
     \end{equation}
\noindent where $\epsilon_v$ is the volume efficiency, $G$ is the fraction of
the counting gas that is GeH$_4$, $P$ is the total counter pressure in
standard atmospheres, and $f_D$ is the fraction of peak events that lie
outside the $\pm 1$ full width at half maximum (FWHM) energy window,
determined empirically for our counters from \nuc{71}{Ge} and \nuc{69}{Ge}
spectra to be 0.063 for the $L$ peak and 0.202 for the $K$ peak.  Monte Carlo
simulations, based on our standard counter geometry, were made to determine
the dependence of the efficiency on $P$ and $G$ \cite{Kouzes89}.  Fits to
these calculations with a polynomial function give $E(P,G) = A(G) + B(G) P +
C(G) P^2$, where $A(G) = A_0 + A_1 G$, $B(G) = B_0 + B_1 G$, and $C(G) = C_0
+ C_1 G$.  This equation applies to both the $L$ and $K$ peaks with different
constants in the expressions for $A$, $B$, and $C$. For the $L$ peak the
constants are $A_0 = 51.0$, $A_1 = 5.51$, $B_0 = -15.7$, $B_1 = 1.58$, $C_0 =
3.0$, and $C_1 = 0.000113$, and for the $K$ peak the constants are $A_0 =
29.7$, $A_1 = -8.27$, $B_0 = 28.4$, $B_1 = -5.02$, $C_0 = -6.22$, and $C_1 =
2.27$.  Over the range of counter fillings for usual extractions, the
estimated uncertainty in $E$ from the Monte Carlo calculations is $\pm 1$\%.

     With the aid of this efficiency formula it is now possible to compare
the measurements in Table~\ref{efficiency_comparison}.  Calculated
efficiencies for these counters in the $L$ and $K$ peaks, using the volume
efficiency measured with \nuc{37}{Ar} and Eq.\ (\ref{efficiency_formula}),
are given in columns 6 and 8 of Table~\ref{efficiency_comparison}.  The total
uncertainty in the calculated efficiencies is estimated to be 1.5\%,
consisting of 0.6\% from uncertainty in $\epsilon_v$, 1.0\% from uncertainty
in $f_D$, and 1.0\% from the uncertainty in the Monte Carlo simulations.  The
calculated efficiencies agree with the values measured with \nuc{69}{Ge} and
\nuc{71}{Ge} within the errors of calculation and measurement.

\subsubsection{Counting efficiency for solar neutrino extractions}

     The counters used during the course of the experiment are listed in
Table~\ref{run_parameter_table}.  The counter type used for the majority of
extractions is indicated by the designation ``LA'' or ``A''.  The second type
was used for three extractions during 1990 and is indicated by ``Ni''; the
final type was used only for the August 1991 extraction and is indicated by
the designation ``RD''.

     The counting efficiency used for each extraction is calculated by
Eq.\ (\ref{efficiency_formula}) and is given in
Table~\ref{run_parameter_table}.  The volume efficiency of most counters has
been directly measured with \nuc{37}{Ar}; if a counter's volume efficiency
has not been measured, it is assumed to equal the average of all measured
counters.  Because the analysis reported in this section resulted in new
counter efficiencies for SAGE~I, these revised efficiencies are given in this
table and are used in any combined fits which include SAGE~I data.

\subsection{Counter calibration}
\label{ext_cals}

     Immediately after filling counters are calibrated through their side
window with the 5.9-keV x rays from an \nuc{55}{Fe} source.  They are
recalibrated with \nuc{55}{Fe} after about 3 days of operation, and then
again approximately every 2 weeks until counting ends.  This usually gives
more than ten \nuc{55}{Fe} calibrations, with at least four during the first
month of counting while the \nuc{71}{Ge} is decaying.  In addition, beginning
with SAGE~II, calibrations are usually made with a \nuc{109}{Cd} source
whenever an \nuc{55}{Fe} calibration is done.  The 22-keV Ag x rays that
follow \nuc{109}{Cd} decay pass through the counter window and fluoresce the
Fe cathode, giving the $K$ x-ray peak from Fe at 6.4 keV.  Although these x
rays originate near the counter window, they are absorbed throughout the
counter volume, and thus give the average counter response.  Beginning with
the February 1993 extraction, a \nuc{109}{Cd}+Se source was periodically
used.  The Cd x rays fluoresce a Se target whose $L$ and $K$ x rays enter the
counter through its side window and give peaks at 1.4 keV and 11.208 keV.

     The energies of the peaks from these calibration sources are summarized
in Table \ref{x_ray_energies}.  These various calibration lines have been
used to check the linearity of the energy and amplitude of the differentiated
pulse (ADP) counting channels and to determine offsets.  There are also Xe
escape peaks with the \nuc{109}{Cd} and \nuc{109}{Cd}+Se sources, but these
lines are usually weak and not useful for energy scale determination.

\begin{table}
\caption{Summary of the external calibration source x-ray energies.}
\label{x_ray_energies}
\begin{tabular}{dcc}

Energy (keV) &   Source         & Origin                         \\
\hline
\\ [-10pt]  
1.4          & \nuc{109}{Cd}+Se & Se $L$ x ray through window    \\
1.625        & \nuc{55}{Fe}     & Xe escape peak through window  \\
5.895        & \nuc{55}{Fe}     & Mn $K$ x ray through window    \\
6.4          & \nuc{109}{Cd}    & $K$ x ray from Fe cathode      \\
11.208       & \nuc{109}{Cd}+Se & Se $K$ x ray through window    \\
\end{tabular}
\end{table}

     The typical counter resolution measured with an \nuc{55}{Fe} source is
in the range of 20\%--23\%.  Scaling the resolution by the square root of the
energy, this implies resolutions in the \nuc{71}{Ge} $L$ and $K$ peaks of
45\%--50\% and 15\%--17\%, respectively, values that are observed in
\nuc{71}{Ge}-filled counters operated at low voltage.

     Many calibrations are done on each counter.  With each calibration a
small fraction of the GeH$_4$ molecules are broken into fragments which can
be deposited on the anode wire near the counter window.  This process, which
we call ``polymerization'', gradually increases the anode diameter, reduces
the electric field, and gives a depression of the apparent energy measured
with an \nuc{55}{Fe} source or a \nuc{109}{Cd}+Se source.  Polymerization
(see, e.g., \cite{Vavra86}) occurs most readily at high count rates, so we
maintain the rate below 10 events/s during calibration.  A check for the
presence of polymerization is made by comparing the peak positions of the
5.895-keV line from the \nuc{55}{Fe} source (which provides events only at
the counter window) and the 6.4-keV line from the \nuc{109}{Cd} source (which
provides events over a much larger fraction of the counter volume).  If the
counter anode is not polymerized near the window and the energy channel is
linear, the ratio of peak positions will be 6.4/5.895 = 1.086.  For each
extraction the ratio of the 6.4-keV to 5.895-keV peak positions averaged over
all calibrations is given in Table~\ref{run_parameter_table} relative to the
unpolymerized value of 1.086.  Most counters show little or no evidence of
polymerization.  For polymerized counters the peak ratio is greater than 1.00
and is used to correct the energy scale derived from each \nuc{55}{Fe}
calibration.

\subsection{Linearity of counter gain}
\label{gain_linearity}

     Calibrations with the \nuc{109}{Cd}+Se source have been used to check
the predicted position and resolution of the \nuc{71}{Ge} $K$ peak from an
\nuc{55}{Fe} calibration.  Since the 11.208-keV peak energy with the
\nuc{109}{Cd}+Se source is very close to the 10.367-keV energy of
\nuc{71}{Ge} $K$-peak events, this method has the advantage that very little
extrapolation of the peak position in energy is needed.  Some departures from
linearity are present in the region of the \nuc{71}{Ge} $K$ peak.

     Measurements have been made as a function of GeH$_4$ fraction $G$,
counter pressure $P$, and operating voltage $V$.  The ratio of the peak
positions is equal to the ratio of the energies (11.2/5.9) up to a critical
voltage $V_{\text{crit}} = 10.5 G + 0.6 P + 588$.  Above this critical
voltage, the location of the \nuc{71}{Ge} $K$ peak $[P_K(\mnuc{71}{Ge})]$ can
be inferred from the location of the \nuc{55}{Fe} peak $[P(\mnuc{55}{Fe})]$
using the formula

\begin{equation}
\label{Gain_Scaling_Formula}
\frac{P_K(\mnuc{71}{Ge})}{P(\mnuc{55}{Fe})} = \frac{10.367}{5.895}
[1 - (4.5G + 2.78)(V - V_{\text{crit}}) \times 10^{-6}],
\end{equation}

\noindent where $G$ is expressed in percent, $P$ is in mm Hg, and $V$ is in
volts.  The typical correction due to the nonlinearity of the gain is a
reduction in the predicted \nuc{71}{Ge} peak position of 2\%.

     This set of experiments also measured the resolution of the peaks from
\nuc{109}{Cd}+Se and from \nuc{55}{Fe}.  Below a critical voltage, the ratio
of the resolutions was equal to the expected value of $\sqrt{5.9/11.2}$, but
above this voltage, given by $V_{\text{crit}} = 6 G + P/3 + 824$, the
\nuc{109}{Cd}+Se resolution was wider than predicted from the \nuc{55}{Fe}
resolution.  From these measurements the relationship between the
\nuc{71}{Ge} $K$-peak resolution $[R_K(\mnuc{71}{Ge})]$ and the \nuc{55}{Fe}
resolution $[R(\mnuc{55}{Fe})]$ was found to be

\begin{equation}
\label{Resolution_Scaling_Formula}
\frac{R_K(\mnuc{71}{Ge})}{R(\mnuc{55}{Fe})} = \sqrt{\frac{5.895}{10.367}}
[ 1 + 1.5 \times 10^{-3} (V - V_{\text{crit}}) ].
\end{equation}

\noindent Note that the value for $V_{\text{crit}}$ for the resolution
correction is not the same as for the gain correction.  The typical
correction for the $K$ peak results in an increase in the predicted
\nuc{71}{Ge} resolution of 15\%.

     The correction to the gain and resolution predicted by these empirical
formulas is accurate to about 30\%.  The nonlinearity in gain and resolution
is only present at the higher energies.  No corrections are required for the
\nuc{71}{Ge} $L$ peak because the critical voltages are much higher than for
the $K$ peak.

\subsection{Electronic systems}\label{elec_sys}

     As indicated in Table~\ref{run_parameter_table}, SAGE has used several
different counting systems as the experiment progressed.  Most runs of SAGE~I
were counted in what we call system 2.  Since the fall of 1992, during
SAGE~II and III, most first extractions were counted in system 3.  System 6
measured a few first extractions, but most were from a low mass of Ga.  The
major specifications of these various counting systems are given in
Table~\ref{sys_specs}.   Since this article focuses on SAGE~II and III, we
will mainly consider counting system 3 in the following.  Some additional
information concerning system 6 and previous systems is given in
Appendix~\ref{other_counting_systems}.

\begin{table}
\caption{Specifications of counting systems 2, 6, and 3.}
\label{sys_specs}
\begin{tabular}{lccc}
Specification                         & Sys.\ 2    & Sys.\ 6 & Sys.\ 3  \\
\hline
Number of channels                    & 7          & 7       & 8        \\
Number of channels with NaI           & 5          & 6       & 8        \\
Counter dynamic range (keV)           & 0.4--13    & 0.5--18 & 0.3--18  \\
NaI dynamic range (keV)               & 50--3000   & 50--3000& 50--3000 \\
Max.\ counting rate (s$^{-1}$)        & 5          & 1000    & 1.5      \\
NaI coincidence window ($\mu$s)       & 8          & 4       & 5.2      \\
Energy time constant ($\mu$s)         & 1          & NA      & NA       \\
ADP time constant (ns)                & 10         & 10--500 & 10       \\
NaI time constant ($\mu$s)            & 1          & 0.5     & 1        \\
Bandwidth, $-3$ dB (MHz)              & 90         & 45      & 90       \\
Rise time, 10\%--90\% (ns)            & 3.5        & 8       & 4        \\
Noise, peak to peak (mV)              & $< 10$     & $< 12$  & $< 10$   \\
Dead time \\
\hspace{2mm} in acquisition mode (ms) & 200        & 1       & 600      \\
\hspace{2mm} in calibration mode (ms) & 50         & 1       & 120      \\
ADC resolution (mV/ch)                & 10         & 1       & 1        \\
Energy offset (ch)                    & 0          & 0       & 0        \\
ADP offset of 4096 ch (ch)            & $-$45 to +25 & 0     & 70--120  \\
\end{tabular}
\end{table}

     The counting systems reside in a specially designed air-conditioned room
in the underground facility.  To minimize the fast neutron and gamma ray
flux, the walls are made from low-radioactivity concrete with an outer steel
shell.  The entire room is lined with sheets of 1 mm zinc-galvanized steel to
reduce radio-frequency noise.  Power to the counting electronics is supplied
by a filtered uninterruptible power supply, with signal and power cables laid
inside independent steel conduits.  The data acquisition computers, which are
in the counting room, are networked so that the systems can be monitored
outside the underground laboratory.  The counting room is kept locked and
access is restricted to counting personnel.

     System 3 was moved to BNO and installed in the underground counting room
in 1988.  It can record events from up to eight counters which are placed
inside the well of a NaI crystal that serves as an active shield (crystal, 23
cm diameter by 23 cm height; well, 9 cm diameter by 15 cm height).  There are
two layers of passive shielding.  An inner layer of square tungsten rods (10
mm$\times$10 mm) encloses the NaI and the photomultipliers are shielded by
Pb.  All components are made from low-radioactivity materials which were
assayed prior to construction by a low-background solid-state Ge detector.
The preamplifiers are mounted as close to the counters as possible, but are
separated by a thick layer of copper.  The counters are sealed nearly air
tight inside the apparatus.  Dry nitrogen gas from evaporation of liquid
nitrogen flows continuously through the NaI well to remove Rn.  The entire
apparatus may be lowered with a hoist into an outer shield whose bottom and
sides consist of 24--32 mm of copper, 210 mm of lead, and 55 mm of steel, and
whose top has 34 mm of Cu and 250 mm of steel.  The cavity between the inner
and outer shields is also purged continuously by gas from evaporating liquid
nitrogen; to preclude backstreaming the flow rate is kept below that in the
inner shield.

\begin{figure*}
\begin{center}
\includegraphics[width=6.500in]{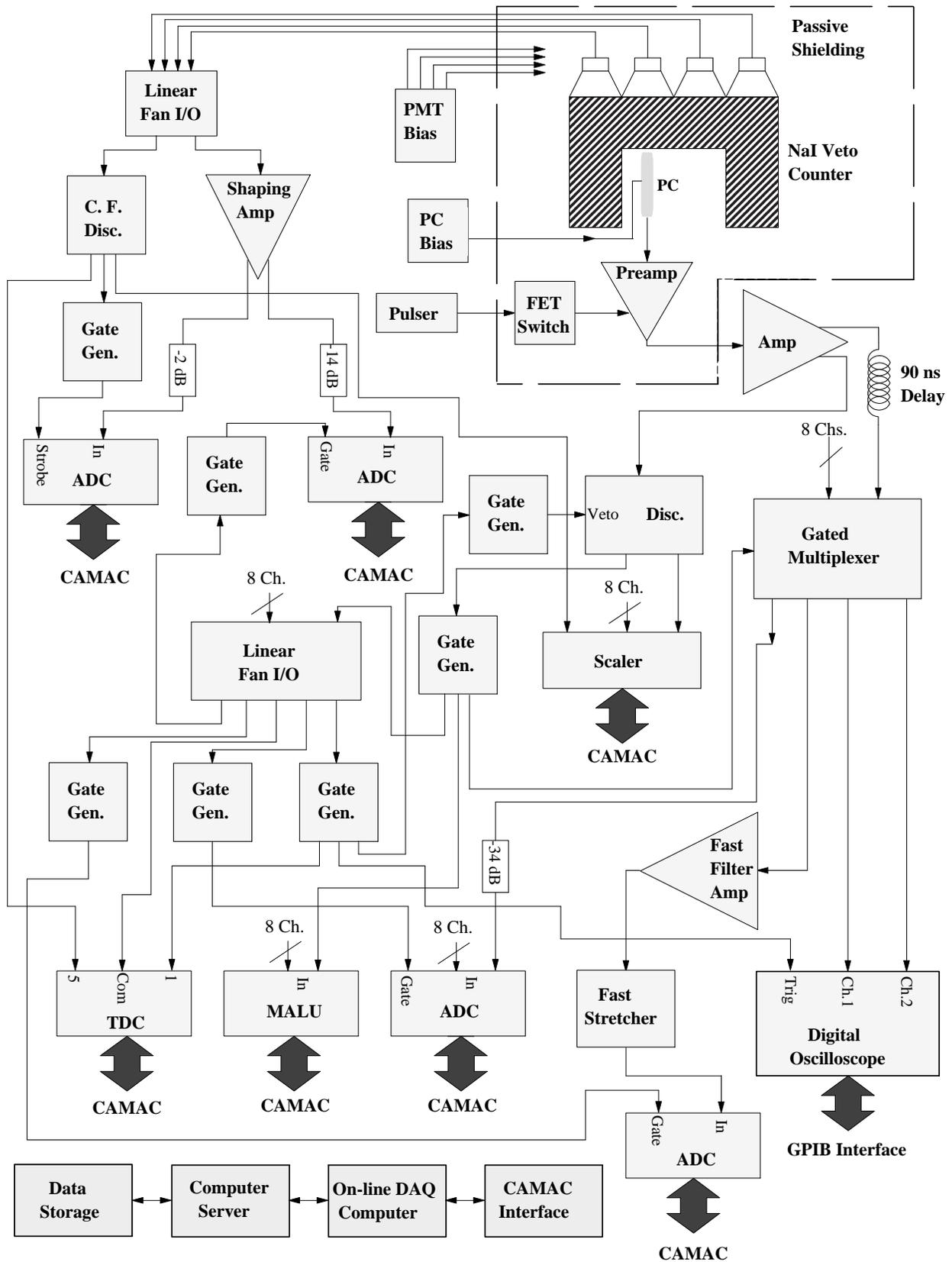}
\end{center}
\caption{Block diagram of one channel of the eight-channel system 3 counting
electronics.  Abbreviations: PC, proportional counter; DAQ, data
acquisition.}
\label{sage_daq}
\end{figure*}

     To minimize the length of the signal cables, the rack of counting
electronics is immediately adjacent to the outer passive shield.  The
electronics is in a single rack designed to reduce rf interference.  The
block diagram of a single channel of system 3 is illustrated in
Fig.~\ref{sage_daq}.  Briefly, the analog signal processing proceeds as
follows: the proportional counter anode is directly connected to a
charge-sensitive preamplifier.  After further amplification the signal is
split, with one channel going to the digital logic to determine that an event
from that counter has occurred, and a second channel going to a 90 ns cable
delay and then to a gated multiplexer.  The signals from all 8 counters are
input to separate gates of this multiplexer and the appropriate gate is
opened by the digital logic for whichever counter has seen an event.  The
multiplexed output is split into four channels: two go to a digital
oscilloscope which records the counter wave form with 8-bit resolution for
800 ns after pulse onset at two different amplification ranges, one
appropriate for the \nuc{71}{Ge} $L$ peak and the other appropriate for the
$K$ peak.  One of the other two signals goes to an integrating ADC to measure
the total pulse energy; the second signal is differentiated with a time
constant of 10 ns, stretched, and input to a peak-sensing ADC.  This second
ADC measures the amplitude of the differentiated pulse, called ``ADP.''
Acquisition can be run in calibration or event acquisition modes.  For each
event in acquisition mode, the energy, ADP value, time of event, NaI time and
energy, and the two digitized wave forms (high- and low-gain channels) are
written to disk.

\section{Selection of Candidate $^{\bf71}$G\lowercase{e} Events}

     The counting data consist of a set of events for each of which there is
a set of measured parameters, such as wave form, energy, NaI coincidence,
etc.  The first step of analysis is to sort through these events and apply
various selection criteria to choose those events that may be from
\nuc{71}{Ge}.  We will describe here the selection procedure for events
measured in counting system 3; the procedure for system 6 is identical except
there are no measured wave forms, so the energy is measured by an ADC and the
ADP method is used for rise time determination.

\subsection {Standard analysis description}\label{stand_anal}

     The various steps to select potential \nuc{71}{Ge} events are the
following:

     (1) The first step of event selection is to examine the event wave form
and identify two specific types of events: those that saturate the wave form
recorder and those that originate from high-voltage breakdown.  Saturated
events are mostly produced by alpha particles from natural radioactivity in
the counter construction materials or from the decay of \nuc{222}{Rn} that
has entered the counter during filling.  Such events are easily identified
and labeled by looking at the pulse amplitude at the end of the wave form.
Saturated pulses have amplitude greater than 16 keV and occur in an average
run at a rate of approximately 0.5/day.  Since most such pulses are seen
after any initial \nuc{222}{Rn} has decayed, they are mainly from internal
counter radioactivity.  Events from high-voltage breakdown have a
characteristic wave form which rises very steeply and then plateaus.  A true
pulse from \nuc{71}{Ge} decay, in contrast, rises more slowly and after this
initial rise, has a slow, but steady, increase in amplitude as the positive
ions are collected.  Breakdown pulses are identified by determining the slope
of the wave form between 500 and 1000 ns after pulse digitization begins.

     (2) To minimize the concentration of Rn, the air in the vicinity of the
counters is continuously purged with evaporating liquid nitrogen.  Counter
calibrations, however, are done with the counter exposed to counting room air
which contains an average of 2 pCi of Rn per liter.  When the shield is
closed and counting begins, a small fraction of the decays of the daughters
of \nuc{222}{Rn} can make pulses inside the counter that mimic those of
\nuc{71}{Ge}.  To remove these false \nuc{71}{Ge} events, we delete 2.6 h of
counting time after any opening of the passive shield, and estimate the
background removal efficiency of this time cut to be nearly 100\%.  See
Sec.~\ref{ext_rn} for further details.

     (3) It is possible that the Xe-GeH$_4$ counter filling may have a small
admixture of \nuc{222}{Rn} that enters the counter when it is filled.  Most
of the decays of Rn give slow pulses at an energy outside the \nuc{71}{Ge}
peaks, but approximately 8\% of the pulses from Rn and its daughters make
fast pulses in the $K$ peak that are indistinguishable from those of
\nuc{71}{Ge}.  Since Rn has a half-life of only 3.8 days, these events will
occur early in the counting and be falsely interpreted as \nuc{71}{Ge}
events.  Each \nuc{222}{Rn} decay is, however, accompanied by three $\alpha$
particles, which are detected with high efficiency and usually produce a
saturated pulse in the counter.  Since the radon decay chain takes on average
only about 1 h from the initiating decay of \nuc{222}{Rn} to reach $^{210}$Pb
with a 22-yr half-life, deleting all data for a few hours around each
saturated event removes most of these false \nuc{71}{Ge} events.  We choose
to delete from 15 min prior to 3 h after each saturated pulse.  The
efficiency of this cut in time is 95\%.  Further details are given in
Sec.~\ref{int_rn}.

     (4) All events whose pulse is coincident with a NaI detector response
are then eliminated.  Since \nuc{71}{Ge} has no $\gamma$ rays associated with
its decay, this veto reduces background from natural radioactivity.

\begin{table}[t]
\caption{Effect of cuts on the experimental live time and events for all runs
of SAGE~II and SAGE~III that were counted in both $L$ and $K$ peaks (except
May 1996).  The results of the cut on each row include the effect of all cuts
on preceding rows.  Because the rise time cut varies with energy, no entry
can be given for the ``2-15 keV'' column.}
\label{Live_Time_Cut_table}
\begin{tabular}{ccccc}

                         & Live   & \multicolumn{3}{c}{Number of events} \\
\cline{3-5}
                         & time   & 2--15& $L$  & $K$  \\
Cut description          & (days) & keV  & peak & peak \\
\hline
None                     & 4129   & 4209 & 1990 & 821  \\
Shield open time cut     & 4040   & 3962 & 1864 & 785  \\
Saturated event time cut & 3862   & 3641 & 1733 & 728  \\
NaI coincidence cut      & 3862   & 1275 & 1106 & 519  \\
Rise time cut            & 3862   &  NA  &  408 & 314
\end{tabular}
\end{table}

     (5) The next step is to set the energy windows for the Ge $L$ and $K$
peaks.  The measure of energy is the integral of the pulse wave form for 800
ns after pulse onset.  The peak position for each window is based on the
calibration with \nuc{55}{Fe}, with appropriate corrections for
polymerization, as described in Sec.~\ref{ext_cals}, and for nonlinearity, as
described in Sec.~\ref{gain_linearity}.  If the peak position changes from
one calibration to the next, then the energy window for event selection is
slid linearly in time between the two calibrations.  The resolution at each
peak is held constant and is set to be the average of the resolutions with
\nuc{55}{Fe} for all counter calibrations, scaled to the $L$- or $K$-peak
energy as described in Sec.~\ref{ext_cals}.  (In the rare cases that the
resolution of the first \nuc{55}{Fe} calibration is larger than the average,
the resolution of the first calibration is used throughout the counting.)
Events are then accepted as candidates only if their energy is within $\pm1$
FWHM of the central peak energy.

     (6) Finally, events are eliminated unless their rise time is in the
range of what is expected for \nuc{71}{Ge} decays.  For runs with wave form
recording, the rise time is derived from a fit to the pulse shape with an
analytical function, as described below in Sec.~\ref{rise_time_techs}.  For
those runs without wave form recording, the $L$ peak is not analyzed and the
ADP measure of rise time is used to set the acceptance window for $K$-peak
events.

\begin{figure}
\begin{center}
\includegraphics[width=3.375in]{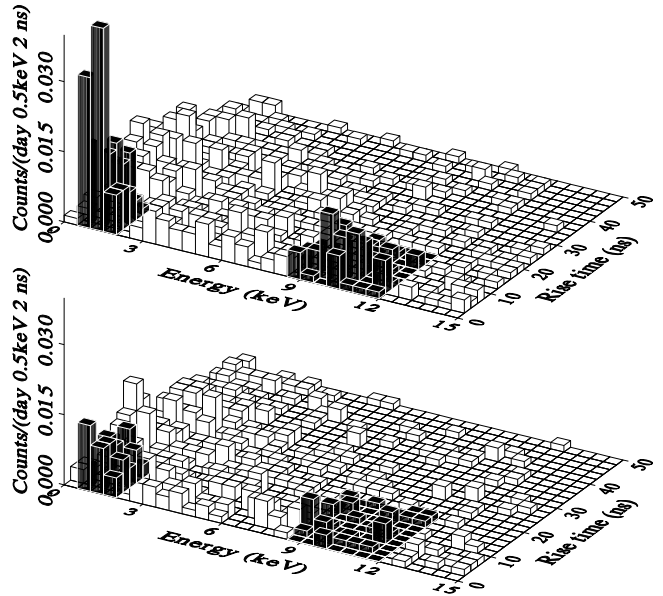}
\end{center}
\caption{Upper panel shows the energy rise time histogram of all events
observed during the first 30 days after extraction for all runs that could be
counted in both $L$ and $K$ peaks (except May 1996).  The live time is 711.1
days.  The expected location of the \nuc{71}{Ge} $L$ and $K$ peaks as
predicted by the \nuc{55}{Fe} and \nuc{109}{Cd} calibrations is shown
darkened.  Lower panel shows the same histogram for all events that occurred
during an equal live time interval at the end of counting.}
\label{2d_hist}
\end{figure}

     For the 30 runs of SAGE~II and III that could be counted in both the $L$
and $K$ peaks, the effect on the live time of each successive cut and the
total number of candidate \nuc{71}{Ge} events that survive is given in
Table~\ref{Live_Time_Cut_table}.  (The run of May 1996 is excluded because
the counter was slightly contaminated with residual \nuc{37}{Ar} which had
been used to measure this counter's efficiency.)  Figure~\ref{2d_hist} shows
all events from these same runs that survive the first four cuts.  Events
that occurred early in the counting are shown in the upper panel and at the
end of counting in the lower panel.

     Several runs were compromised and some were completely lost due to
operational failures.  Failure of an electronic component made it impossible
to use the $L$ peak in the extractions of April 1993, May 1993, July 1993,
and October 1994.  Similar problems made it impossible to make a rise time
cut in the $K$ peak for the runs of June 1991, July 1993, October 1994, and
October 1997.  These runs thus have a larger than normal number of events.
If an electronic component fails that deteriorates the rise time response and
the failure occurs early in the counting, while the \nuc{71}{Ge} is decaying,
our policy is to not use any rise time cut in the $K$ peak and to reject this
run in the $L$ peak.  If the failure occurs later, the rise time cut is
retained and the interval of failure is removed from the data.  Extractions
in March 1993, January 1995, May 1995, and March 1996 were entirely lost due
to counter failure.  The extractions of September 1993, September 1994-2, and
July 1996 were lost because either the counter stopcock failed or some other
gas fill difficulty occurred.  Electronic failures caused the loss of the
extractions of September 1993-1, May 1994-2, and April 1995.  Extractions in
June 1995 were lost due to radioactive contamination of the counters with
isotopes that were being used at this time for counter efficiency
measurement.  Finally, we exclude several extractions from one reactor that
were systematic studies in preparation for the Cr source experiment.  Since
their mass was no more than 7.5 tons of gallium, less than one atom of
\nuc{71}{Ge} is detected on the average in such runs in the combination of
both the $L$ and $K$ peaks.  Two-reactor extractions, however, whose mass is
approximately 15 tons, give on the average 1.5 \nuc{71}{Ge} events,
sufficient to determine the solar neutrino capture rate, albeit with a large
error \cite{Bahcall85}.

\subsection {Rise time analysis techniques}\label{rise_time_techs}

\begin{figure}
\begin{center}
\includegraphics[width=3.375in]{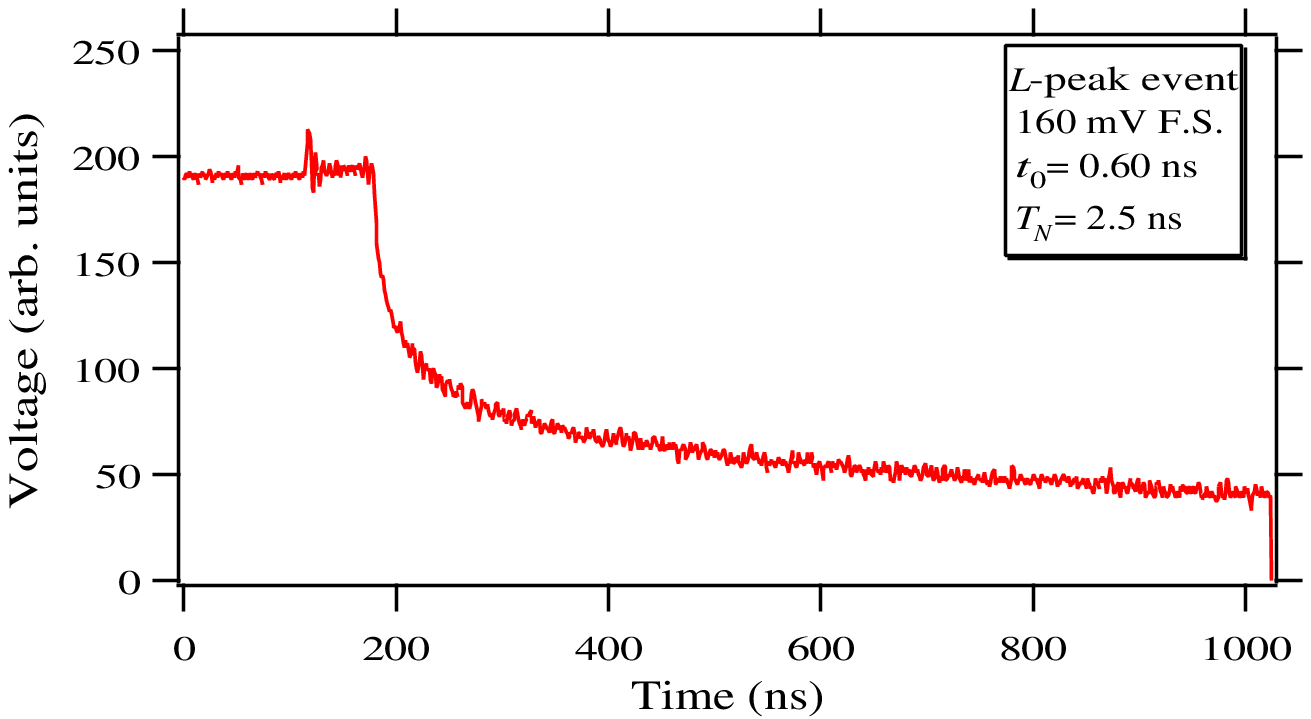}
\includegraphics[width=3.375in]{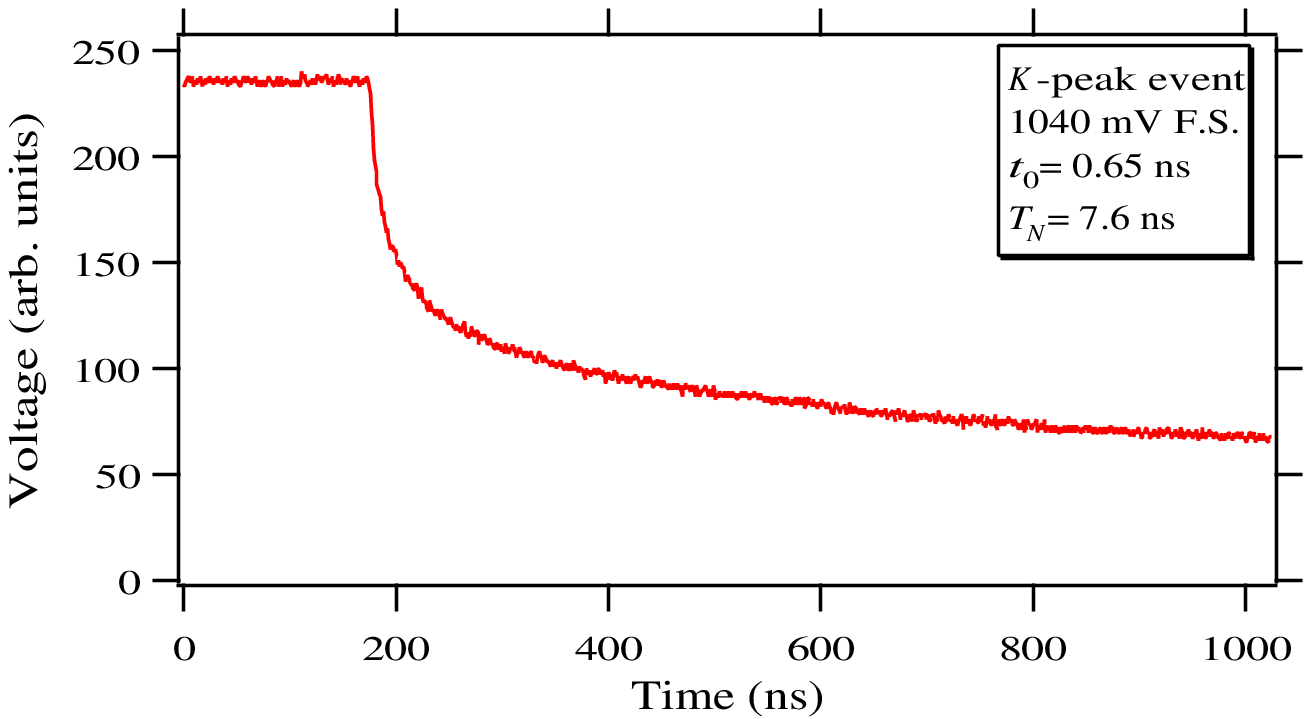}
\end{center}
\caption{\nuc{71}{Ge} events in $L$ and $K$ peaks.}
\label{example_LK_candidate_waveforms}
\end{figure}

     As described in Secs.\ \ref{extrac_hist} and \ref{elec_sys}, the data
acquisition system electronics has evolved over the course of SAGE.  The data
from SAGE~I relied entirely on a hardware measurement of the rise time.  This
ADP technique suffices well in studies of the $K$-peak counter response, but
is not capable of adequately differentiating rare $L$-peak events from noise.

     Wherever possible for SAGE~II, and throughout SAGE~III, we derive a
parameter that characterizes the rise time from the wave form, and are thus
able to present both $L$- and $K$-peak results.  For those runs with only ADP
data, the $L$ peak cannot be analyzed and we present only $K$-peak data.  All
wave form data come from counting system 3.

\subsubsection{Waveform rise time determination -- $T_N$}\label{T_N}

     Figure~\ref{example_LK_candidate_waveforms} shows typical pulses in the
$L$ and $K$ peaks from a \nuc{71}{Ge}-filled counter as captured by the
digitizing oscilloscope in system 3.  There are 256 channels full scale on
the y axis corresponding to 1.040 V (130 mV/div) for digitizer channel 1 and
0.160 V (20 mV/div) for channel 2.  The x axis has 1024 digitization points
each with 1 ns duration.  The relevant features of the pulses are the base
line from $t = 0$ to roughly 120 ns, the dc offset that occurs when the gate
opens at 120 ns, and the fast onset of the pulse at about 180 ns.  The exact
values of these times and offsets vary depending on the counting channel and
the run; they even vary slightly from pulse to pulse within a given run.
When determining the energy and rise time of the pulse, it is therefore
necessary to determine accurately the onset of the pulse both in time and dc
voltage level.

\begin{figure}
\begin{center}
\includegraphics[width=3.375in]{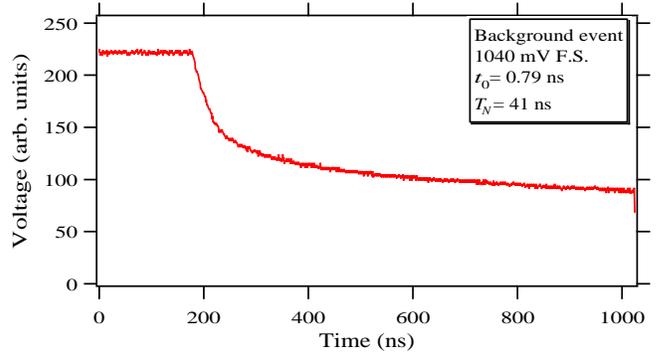}
\end{center}
\caption{Background candidate event in $K$ peak.  Note the much slower fall
when the pulse begins at $\sim 200$ ns than for the true \nuc{71}{Ge}
$K$-peak event in Fig. \ref{example_LK_candidate_waveforms}.}
\label{example_K_background_waveform}
\end{figure}

     By treating the trail of ionization in the proportional counter as a
collection of point ionizations and integrating over their arrival time at
the anode, it can be shown \cite{ELL90} that the voltage output $V$ of an
infinite bandwidth preamplifier as a function of time $t$ after pulse onset
has the form

\begin{equation}
\begin{array}{rcll}
\label{Tn_Formula}
V(0<t<T_N) & = & V_0[ & \frac{t + t_0}{T_N}\ln(1 + \frac{t}{t_0})
                                           -  \frac{t}{T_N}], \\
V(t>T_N)   & = & V_0[ & \ln(1 + \frac{t - T_N}{t_0}) - 1 \\
           &   &      & -\frac{t + t_0}{T_N}\ln(1 - \frac{T_N}{t+t_0})],
\end{array}
\end{equation}

\noindent with $V(t<0) = 0$, where $T_N$ is the time duration over which the
ionization arrives at the anode, $t_0$ is a time inversely proportional to
the ion mobility, and $V_0$ is proportional to the total amount of ionization
deposited in the counter.  The parameter $T_N$ characterizes the rise time of
the wave form.  For the case of true point ionization, $T_N$ should be near
zero.  When $T_N$ is zero, the function reduces to the Wilkinson form
$V(t:T_N=0) = V_0 \ln(1 + t/t_0)$.  When $T_N$ is large, the event is
characteristic of extended ionization, and is most likely a background event
from a high-energy $\beta$ particle traversing the counter.
Figure~\ref{example_K_background_waveform} is an example of such a slow pulse
in the $K$ peak.

     Because this form for the pulse shape has a sound physical basis and
reasonable mathematical simplicity, we fit every pulse that is not identified
as saturation or breakdown to Eq.\ (\ref{Tn_Formula}).  To account for the
fact that the pulse onset time $t_{\text{onset}}$ is not at time zero, we
replace $t$ by $t-t_{\text{onset}}$, and since the pulse begins at a finite
voltage $V_{\text{offset}}$, we replace $V$ by $V-V_{\text{offset}}$.  The
fit is made from 40 ns before the time of pulse onset to 400 ns after onset.
Five parameters are determined by the fit: $t_{\text{onset}}$,
$V_{\text{offset}}$, $V_0$ (a measure of the energy deposited during the
event which is not used in analysis), $t_0$ (whose value of slightly less
than 1 ns is approximately constant for all pulses), and $T_N$ (the rise
time).

\subsubsection{Alternative wave form analysis methods}

     Although we use fits to $T_N$ as our standard analysis technique, we
also developed two alternative methods to discriminate pointlike ionization
from extended-track ionization in the proportional counter pulses.  These
serve as checks on the event selection based on $T_N$.  One technique is
based on a fast Fourier transform (FFT) of the digitized wave form.  No
specific functional form for the pulse is assumed and hence this method has
the advantage that it is sensitive to potential alterations in the pulse
shape.  See Appendix~\ref{fft} for further information concerning the FFT
method.  The second method of wave form analysis that was investigated also
assumes no particular form for the pulse.  This method, called the ``RST
method,'' deconvolutes the observed wave form to find the initial ionization
pattern in the counter.  See Appendix~\ref{RST} for further details.

     Since these three techniques are sensitive to different characteristics
of the wave form, their selection of events is, not unexpectedly, different.
Nonetheless, when many data sets are considered in combination, their results
for the overall production rate are in good agreement, which provides strong
support for the validity of our wave form analysis procedure.

\subsubsection{Hardware rise rime measurement: ADP}\label{adp}

     The amplitude of the differentiated pulse is proportional to the product
of the original pulse amplitude and the inverse rise time.  The quantity
ADP/energy is thus proportional to the inverse rise time.  Events due to
low-energy Auger electrons and x rays that produce point ionization in the
counter all have a fast rise time.  Events with a slower rise time (small
ADP) are due to background pulses that produce extended ionization.  Events
with a very fast rise time (large ADP) are due to electronic noise or
high-voltage breakdown.

     Inherent in an ADP analysis is the uncertainty that arises from an
imprecise knowledge of the offset for a given run.  Nonzero offset occurs
when the gate is opened after an event trigger.  The electronic components
which process the pulse are subject to small drifts in their offsets that are
functions of external parameters, such as temperature.  These nonzero offsets
contribute to the dc offset on which the event pulse rides.  Our approach has
been to extrapolate ADP vs energy plots from the \nuc{55}{Fe} calibrations
using the 5.9-keV peak and the escape peak to obtain an offset for each
calibration.  Since the offsets are typically distributed in a Gaussian
manner with a sigma of 1 or 2 channels, the average is a good approximation
when determining the $K$-peak selection window.  For the $L$ peak, however,
uncertainties of a few channels lead to significant variations in event
selection.  Utilizing the digitized pulses, it is possible to eliminate this
uncertainty by determining every offset on a pulse-by-pulse basis.

     A further disadvantage of the ADP method is that it is only responsive
to the initial rise of the pulse.  Occasional small pulses from high-voltage
breakdown have rise time the same as for true $L$-peak \nuc{71}{Ge} pulses,
but after their initial rise they turn flat, rather than gently rise as the
positive ions are collected as with a real \nuc{71}{Ge} event.  A breakdown
event of this type is not distinguished from a \nuc{71}{Ge} event by the ADP
method, but is easily recognized by examining the recorded wave form long
after pulse onset.

\subsection{Calibration of rise time response}\label{rise_tim_cal}

     To determine the values of $T_N$ for true \nuc{71}{Ge} pulses, we have
filled counters with typical gas mixtures (20\% GeH$_4$ and 80\% Xe at a
pressure of 620 mm Hg), added a trace of active \nuc{71}{Ge}H$_4$, and
measured the pulses in each of the system 3 counting slots.  All events
inside 2 FWHM of the $L$ and $K$ peaks are then selected and the rise time
$T_N$ of each event calculated with Eq.~(\ref{Tn_Formula}).  The rise time
values are arranged in ascending order and an upper rise time limit set such
that 5\% of the events are excluded.  This leads to event selection limits on
$T_N$ of 0.0--10.0 ns in the $L$ peak and 0.0--18.4 ns in the $K$ peak.  The
variation with electronics channel and with counter filling, over the range
of our usual gas mixtures, was measured to be approximately 1.2 ns.  We
choose to fix the event selection limits at the values given above, and
include in the systematic error an uncertainty in the efficiency of $\pm1$\%
due to channel and filling variations.  A major advantage of using $T_N$ is
that the rise time limits are fixed and are the same for all extractions.
The purpose of the calibrations with \nuc{55}{Fe} and other sources is solely
to determine the energy scale.

     For those runs in which the ADP method of rise time discrimination is
used, the limits for the ADP cut are determined separately for each run from
the \nuc{55}{Fe} calibrations.  Histograms of the values of ADP/energy for
the events within 2 FWHM of the 5.9-keV energy peak are analyzed to determine
the cut point for 1\% from the fast region (to eliminate noise) and 4\% from
the slow region (to eliminate background).  All calibrations from a run are
analyzed and the ADP window for event selection is slid linearly in time from
one calibration to the next.

\section{Statistical Analysis and Results of Single
Runs}\label{results_stat_anal}

     In this section we describe how the data are analyzed to determine the
\nuc{71}{Ge} production rate.  We then give the results for individual runs
and for all runs in the $L$- and $K$-peak regions.

\subsection{Single-run results}
     The above selection criteria result in a group of events from each
extraction in both the $L$- and $K$-peak regions which are candidate
\nuc{71}{Ge} decays.  To determine the rate at which \nuc{71}{Ge} was
produced during the exposure time, it is assumed in each peak region that
these events originate from two sources: the exponential decay of a fixed
number of \nuc{71}{Ge} atoms and a constant-rate background (different for
each peak).  Under this assumption the likelihood function \cite{CLE83} for
each peak region is

\begin{equation}
\label{likelihood_function}
{\cal L} = e^{-m} \prod_{i=1}^{N} (b + a e^{-\lambda t_{i}}),
\end{equation}
\noindent where
\begin{eqnarray}
m      & = & bT + a \Delta/\lambda,  \nonumber \\
T      & = & \sum_{k=1}^{n} (t_{ek} - t_{bk}), \nonumber \\
\Delta & = & \sum_{k=1}^{n} (e^{-\lambda t_{bk}}
                           - e^{-\lambda t_{ek}}). \nonumber
\end{eqnarray}

\noindent Here $b$ is the background rate, $\lambda$ is the decay constant of
\nuc{71}{Ge}, $t_i$ is the time of occurrence of each event with $t = 0$ at
the time of extraction, and $N$ is the total number of candidate events.  The
production rate $p$ of \nuc{71}{Ge} is related to the parameter $a$ by

\begin{equation}
\label{prod_rate_def}
a =  \epsilon p (1 - e^{-\lambda \Theta}),
\nonumber
\end{equation}

\noindent where $\Theta = t_E - t_B$ is the exposure time (i.e., the time of
end of exposure $t_E$ minus the time of beginning of exposure $t_B$), and
$\epsilon$ is the total efficiency for the extraction (i.e., the product of
extraction and counting efficiencies).  The total counting live time is given
by $T$ and is a sum over the $n$ counting intervals, each of which has a
starting time $t_{bk}$ and ending time $t_{ek}$.  The parameter $\Delta$ is
the live time weighted by the exponential decay of \nuc{71}{Ge}.  Its value
would be unity if counting began at the end of extraction and continued
indefinitely.  We convert the production rate (in \nuc{71}{Ge} atoms produced
per day) to the solar neutrino capture rate (in SNU) using the conversion
factor $2.977 \times 10^{-4}$ atoms of \nuc{71}{Ge} produced/(SNU day ton of
gallium), where the mass of gallium exposed in each extraction is given in
Table \ref{run_parameter_table}.

     Because of the eccentricity of the Earth's orbit, the Earth-Sun
distance, and thus the production rate, varies slightly during the year.  We
correct the production rate for this effect by multiplying $\epsilon$ by the
factor $1 + C$ where $C$ is given by

\begin{eqnarray}
\label{earth_sun_corr_factor}
C = \bigg(\frac{2e}{S[1+r^2]}\bigg)[& \cos X_E & + r \sin X_E \\
                            - (1-S)(& \cos X_B & + r \sin X_B)],  \nonumber
\end{eqnarray}

\noindent with
\begin{eqnarray}
r   & = & \omega/\lambda, \nonumber  \\
X_E & = & \omega(t_E - t_p), \nonumber  \\
X_B & = & \omega(t_B - t_p), \nonumber  \\
S   & = & 1 - e^{-\lambda \Theta}. \nonumber
\end{eqnarray}

\noindent Here $e$ is the eccentricity of the orbit ($e = 0.0167$), $\omega$
is the angular frequency ($\omega = 2\pi/365.25 \text{ day}^{-1}$), and $t_p$
is the moment of perihelion passage, which has been 2--5 January for the past
number of years.  We use $t_p = 3.5$ days.

     The best estimate of the solar neutrino capture rate in each peak region
is determined by finding the values of $a$ and $b$ which maximize $\cal L$.
In doing so we exclude unphysical regions; i.e., we require $a>0$ and $b>0$.
The uncertainty in the capture rate is found by integrating the likelihood
function over the background rate to provide a likelihood function of signal
only, and then locating the minimum range in signal which includes 68\% of
the area under that curve.  This procedure is done separately for the $L$ and
$K$ peaks and the results are given in Tables \ref{L_peak_table} and
\ref{K_peak_table}.  We call the set of events in each peak region a ``data
set.''

\begin{table*}[t]
\squeezetable
\caption{Results of analysis of $L$-peak events.  $Nw^2$ is a measure of the
goodness of fit.  See \protect \cite{CLE98} for its calculation and use.  The
probability values are derived from 1000 simulations and have an uncertainty
of $\sim 1.5$\%.  The total counting live time is 10.379 yr.  The entries in
columns 2--4 include all analysis time cuts.}
\label{L_peak_table}
\begin{tabular}{l d d d d d d r @{--} d d d}

           & Lead     & Live     &       & Number of & Number & Best &
\multicolumn{2}{c}{68\% conf.} \\
Exposure   & time     & time     &       & candidate & fit to & fit  &
\multicolumn{2}{c}{range}             &        & Probability \\
 date      & (h)      & (days)     & Delta  & events    & $^{71}$Ge  &
(SNU)    & \multicolumn{2}{c}{(SNU)}  & $Nw^2$ &       (\%)  \\
\hline
 Sep. 92   & 29.0 & 103.8 & 0.811 &  7. &  4.0 & 109. &   46 & 174. & 0.039 &
82. \\
 Oct. 92   & 27.3 &  96.3 & 0.839 & 10. &  0.0 &   0. &    0 &  61. & 0.179 &
19. \\
 Nov. 92   & 30.7 &  66.7 & 0.688 & 12. &  0.0 &   0. &    0 &  62. & 0.238 &
12. \\
 Dec. 92   & 26.7 &  47.5 & 0.835 & 10. &  7.4 & 153. &   51 & 208. & 0.055 &
65. \\
 Jan. 93   & 29.9 &  23.4 & 0.518 &  4. &  4.0 & 135. &   35 & 181. & 0.092 &
67. \\
 June 93   & 33.3 & 120.7 & 0.699 &  9. &  1.1 &  29. &    0 & 107. & 0.490 &
2. \\
 Oct. 93-2 & 51.9 &  71.6 & 0.686 &  2. &  2.0 & 193. &   19 & 297. & 0.097 &
55. \\
 Oct. 93-3 & 31.6 & 102.7 & 0.772 &  3. &  3.0 & 287. &   88 & 428. & 0.078 &
68. \\
 July 94   & 45.6 & 136.0 & 0.782 & 10. &  2.2 &  65. &   11 & 131. & 0.026 &
95. \\
 Aug. 94   & 32.6 & 116.4 & 0.838 & 20. &  0.0 &   0. &    0 &  67. & 0.056 &
73. \\
 Sep. 94-1 & 40.5 & 120.0 & 0.729 & 20. &  4.7 & 171. &   54 & 300. & 0.087 &
32. \\
 Nov. 94   & 30.4 & 112.3 & 0.660 & 10. &  2.7 &  76. &   18 & 143. & 0.041 &
79. \\
 July 95   & 35.5 & 110.6 & 0.776 & 16. &  1.2 &  35. &    0 & 104. & 0.336 &
3. \\
 Aug. 95   & 35.2 & 108.9 & 0.698 & 16. &  3.9 & 113. &   42 & 200. & 0.095 &
28. \\
 Sep. 95   &124.3 &  80.4 & 0.561 & 23. &  0.2 &   8. &    0 & 179. & 0.160 &
23. \\
 Oct. 95   & 39.3 & 120.7 & 0.793 & 17. &  3.2 & 169. &   33 & 319. & 0.041 &
78. \\
 Nov. 95   & 37.2 & 149.9 & 0.759 & 19. &  8.4 & 214. &  124 & 310. & 0.064 &
45. \\
 Dec. 95-2 & 78.3 & 119.9 & 0.530 & 22. &  0.8 &  40. &    0 & 174. & 0.102 &
42. \\
 Jan. 96   & 33.9 & 141.2 & 0.767 & 21. &  0.0 &   0. &    0 &  61. & 0.065 &
66. \\
 May  96   & 35.2 & 117.8 & 0.628 & 25. &  3.5 & 104. &   23 & 200. & 0.038 &
82. \\
 Aug. 96   & 32.9 & 148.7 & 0.790 & 20. &  5.6 & 126. &   58 & 204. & 0.048 &
68. \\
 Oct. 96   & 33.6 & 155.5 & 0.785 & 11. &  0.0 &   0. &    0 &  48. & 0.119 &
39. \\
 Nov. 96   & 35.0 & 162.5 & 0.795 & 13. &  0.2 &   5. &    0 &  58. & 0.042 &
85. \\
 Jan. 97   & 34.5 & 160.0 & 0.816 & 16. &  1.2 &  24. &    0 &  68. & 0.581 &
1. \\
 Mar. 97   & 36.3 & 160.9 & 0.814 & 10. &  2.6 &  45. &    9 &  89. & 0.126 &
17. \\
 Apr. 97   & 35.0 & 167.4 & 0.791 & 12. &  0.0 &   0. &    0 &  27. & 0.108 &
45. \\
 June 97   & 35.8 & 173.4 & 0.797 & 16. &  4.5 &  95. &   40 & 161. & 0.089 &
30. \\
 July 97   & 37.0 & 140.0 & 0.752 & 13. &  0.7 &  14. &    0 &  61. & 0.238 &
10. \\
 Sep. 97   & 33.6 & 166.6 & 0.826 & 12. &  1.1 &  24. &    0 &  77. & 0.059 &
64. \\
 Oct. 97   & 34.2 & 149.5 & 0.780 & 19. &  4.7 &  99. &   42 & 167. & 0.041 &
76. \\
 Dec. 97   & 34.4 & 137.1 & 0.726 & 15. &  3.1 &  69. &   18 & 131. & 0.045 &
73. \\
\hline
\multicolumn{4}{l}{Combined (31 data sets)} & 433. &   64.3 &  55. &   43 &
68. & 0.020 & $>99$. \\
\end{tabular}
\normalsize
\end{table*}

\begin{table*}
\squeezetable
\caption{Results of analysis of $K$-peak events.  The uncertainty in the
probability is $\sim 1.5$\%.  No probability can be given for exposure Oct.
93-1 because no counts were detected.  The total counting live time is 18.342
yr.}
\label{K_peak_table}
\begin{tabular}{l d d d d d d r @{--} d d d}

           & Lead     & Live     &       & Number of & Number & Best &
\multicolumn{2}{c}{68\% conf.} \\
Exposure   & time     & time     &       & candidate & fit to & fit  &
\multicolumn{2}{c}{range}             &        & Probability \\
 date      & (hours)  & (days)     & Delta  & events    & $^{71}$Ge  &
(SNU)    & \multicolumn{2}{c}{(SNU)}  & $Nw^2$ &       (\%)  \\
\hline
 Jan. 90   & 25.0 &  57.4 & 0.849 &  8. &  0.0 &   0. &    0 &  64. & 0.367 &
4. \\
 Feb. 90   & 25.0 &  57.3 & 0.886 &  2. &  2.0 &  95. &   18 & 159. & 0.164 &
26. \\
 Mar. 90   & 25.0 &  47.5 & 0.839 &  9. &  2.8 & 107. &    0 & 224. & 0.053 &
66. \\
 Apr. 90   & 29.8 &  90.4 & 0.881 &  9. &  0.0 &   0. &    0 & 112. & 0.104 &
40. \\
 July 90   & 22.6 &  59.3 & 0.870 & 15. &  0.0 &   0. &    0 & 213. & 0.142 &
28. \\
 June 91   & 20.5 & 108.3 & 0.904 & 10. &  0.4 &  13. &    0 & 119. & 0.211 &
14. \\
 July 91   & 26.1 &  59.2 & 0.877 &  1. &  1.0 &  55. &    0 & 115. & 0.159 &
26. \\
 Aug. 91   & 73.8 &  94.4 & 0.651 & 16. &  9.8 & 412. &  243 & 577. & 0.036 &
83. \\
 Sep. 91   & 35.3 &  68.9 & 0.827 &  8. &  3.5 &  73. &   20 & 126. & 0.041 &
79. \\
 Nov. 91   & 40.8 & 112.6 & 0.822 & 14. &  2.4 &  48. &    0 & 102. & 0.095 &
30. \\
 Dec. 91   & 26.2 & 111.8 & 0.917 & 10. & 10.0 & 180. &   99 & 217. & 0.063 &
77. \\
 Feb. 92-1 & 21.5 & 192.7 & 0.900 & 14. &  0.0 &   0. &    0 &  43. & 0.057 &
74. \\
 Feb. 92-2 & 43.0 &  43.2 & 0.800 &  1. &  1.0 & 101. &    0 & 192. & 0.085 &
88. \\
 Mar. 92   & 26.0 & 167.8 & 0.840 & 21. & 10.1 & 245. &  155 & 342. & 0.043 &
72. \\
 Apr. 92   & 21.5 & 144.9 & 0.717 & 15. &  2.3 &  55. &   13 & 111. & 0.143 &
18. \\
 May  92   & 54.0 & 114.9 & 0.843 &  4. &  0.0 &   0. &    0 &  74. & 0.134 &
30. \\
 Sep. 92   & 29.0 & 103.8 & 0.811 &  6. &  2.1 &  55. &   12 & 104. & 0.108 &
25. \\
 Oct. 92   & 27.3 & 134.2 & 0.840 & 11. &  2.7 &  52. &   13 &  98. & 0.046 &
71. \\
 Nov. 92   & 30.7 & 123.4 & 0.695 & 16. &  5.1 & 130. &   57 & 210. & 0.046 &
68. \\
 Dec. 92   & 26.7 & 140.7 & 0.871 & 18. &  9.1 & 176. &  107 & 250. & 0.075 &
36. \\
 Jan. 93   & 29.9 & 119.1 & 0.816 & 13. &  5.6 & 111. &   45 & 181. & 0.130 &
14. \\
 Feb. 93   & 26.2 & 169.6 & 0.839 &  3. &  0.0 &   0. &    0 &  48. & 0.116 &
41. \\
 Apr. 93   & 25.0 & 155.3 & 0.820 &  7. &  2.9 &  71. &   25 & 124. & 0.041 &
82. \\
 May  93   & 33.4 & 126.8 & 0.411 &  8. &  1.4 &  64. &    5 & 153. & 0.073 &
51. \\
 June 93   & 33.3 & 120.7 & 0.699 &  9. &  2.1 &  51. &    3 & 111. & 0.154 &
11. \\
 July 93   & 27.5 & 124.5 & 0.761 & 28. &  7.6 & 224. &  114 & 348. & 0.040 &
78. \\
 Aug. 93-1 & 26.8 & 129.0 & 0.877 &  4. &  2.5 &  66. &   20 & 116. & 0.048 &
79. \\
 Aug. 93-2 & 53.8 &  53.0 & 0.769 &  1. &  1.0 & 120. &    0 & 227. & 0.093 &
67. \\
 Oct. 93-1 & 26.7 &  54.5 & 0.733 &  0. &  0.0 &   0. &    0 & 158. &
\multicolumn{1}{c}{NA}   &  \multicolumn{1}{c}{NA}  \\
 Oct. 93-2 & 51.9 &  72.6 & 0.694 &  2. &  0.8 &  69. &    0 & 198. & 0.048 &
86. \\
 Oct. 93-3 & 31.6 & 103.7 & 0.782 &  4. &  0.3 &  27. &    0 & 192. & 0.024 &
99. \\
 July 94   & 45.6 & 136.7 & 0.783 & 12. &  1.1 &  30. &    0 &  88. & 0.056 &
68. \\
 Aug. 94   & 32.6 & 117.2 & 0.841 &  7. &  3.0 &  71. &   25 & 123. & 0.042 &
78. \\
 Sep. 94-1 & 40.5 & 120.8 & 0.751 & 10. &  2.6 &  87. &   22 & 165. & 0.043 &
76. \\
 Oct. 94   & 55.4 & 120.3 & 0.681 & 44. &  4.8 & 136. &   27 & 257. & 0.075 &
45. \\
 Nov. 94   & 30.4 & 112.3 & 0.660 & 13. &  5.6 & 164. &   79 & 259. & 0.035 &
84. \\
 Dec. 94   & 29.3 & 100.0 & 0.803 &  9. &  0.0 &   0. &    0 & 236. & 0.184 &
19. \\
 Mar. 95   & 29.3 & 151.4 & 0.772 & 23. &  3.7 & 147. &   47 & 266. & 0.042 &
77. \\
 July 95   & 35.5 & 110.6 & 0.776 & 17. &  4.3 & 128. &   39 & 229. & 0.114 &
19. \\
 Aug. 95   & 35.2 & 108.9 & 0.698 &  8. &  3.6 & 100. &   38 & 168. & 0.058 &
59. \\
 Sep. 95   &124.3 &  80.4 & 0.561 & 10. &  1.0 &  48. &    0 & 201. & 0.144 &
19. \\
 Oct. 95   & 39.3 & 120.7 & 0.793 &  9. &  3.3 & 160. &   51 & 286. & 0.060 &
54. \\
 Nov. 95   & 37.2 & 149.9 & 0.759 & 13. &  2.7 &  66. &   18 & 125. & 0.039 &
83. \\
 Dec. 95-2 & 78.3 & 119.9 & 0.530 & 18. &  0.0 &   0. &    0 & 127. & 0.044 &
85. \\
 Jan. 96   & 33.9 & 141.2 & 0.767 & 14. &  4.6 & 117. &   45 & 193. & 0.091 &
29. \\
 May  96   & 35.2 & 117.8 & 0.628 &  6. &  2.3 &  66. &   13 & 126. & 0.028 &
95. \\
 Aug. 96   & 32.9 & 148.9 & 0.800 &  6. &  0.0 &   0. &    0 &  51. & 0.102 &
45. \\
 Oct. 96   & 33.6 & 155.5 & 0.785 & 10. &  5.0 & 107. &   55 & 165. & 0.066 &
47. \\
 Nov. 96   & 35.0 & 162.5 & 0.795 & 15. &  1.9 &  40. &    0 &  88. & 0.110 &
29. \\
 Jan. 97   & 34.5 & 160.0 & 0.816 &  8. &  1.4 &  29. &    0 &  70. & 0.123 &
23. \\
 Mar. 97   & 36.3 & 160.9 & 0.814 & 14. &  3.6 &  64. &   20 & 116. & 0.058 &
52. \\
 Apr. 97   & 35.0 & 167.4 & 0.791 & 10. &  4.2 &  84. &   38 & 137. & 0.052 &
63. \\
 June 97   & 35.8 & 160.2 & 0.797 & 11. &  5.8 & 121. &   65 & 183. & 0.033 &
86. \\
 July 97   & 37.0 & 127.3 & 0.751 &  9. &  0.0 &   0. &    0 &  37. & 0.204 &
16. \\
 Sep. 97   & 33.6 & 124.5 & 0.794 &  5. &  2.6 &  61. &   22 & 107. & 0.109 &
29. \\
 Oct. 97   & 34.2 & 149.5 & 0.780 &  7. &  0.3 &   6. &    0 &  42. & 0.429 &
3. \\
 Dec. 97   & 34.4 & 108.4 & 0.519 &  9. &  3.0 &  90. &   31 & 159. & 0.044 &
77. \\
\hline
\multicolumn{4}{l}{Combined (57 data sets)} &  604. &  143.7 &  73. &   64 &
82. & 0.110 &  25. \\
\end{tabular}
\normalsize
\end{table*}

\begin{table}
\squeezetable
\caption{Results of combined analysis of $L$-peak and $K$-peak events for all
31 runs that could be analyzed in both peaks.  The uncertainty in the
probability is $\sim 1.5$\%.  Treating the $L$- and $K$-peak regions as two
separate data sets, the total counting live time is 21.282 yr.}
\label{K_plus_L_peak_table}
\begin{tabular}{l d d d r @{--} d l d}

             & Number of & Number     & Best & \multicolumn{2}{c}{68\% conf.}
\\
Exposure     & candidate & fit to     & fit  & \multicolumn{2}{c}{range} &  &
\multicolumn{1}{c}{Prob.} \\
 date        & events    & $^{71}$Ge  & (SNU)& \multicolumn{2}{c}{(SNU)} &
$Nw^2$ & \multicolumn{1}{c}{(\%)} \\
\hline
 Sep. 92             &   13. &    6.0 &  79. &   44 & 123. & 0.097 &  25. \\
 Oct. 92             &   21. &    3.3 &  32. &    4 &  67. & 0.105 &  26. \\
 Nov. 92             &   28. &    4.3 &  56. &   10 & 111. & 0.047 &  70. \\
 Dec. 92             &   28. &   16.8 & 168. &  115 & 229. & 0.057 &  53. \\
 Jan. 93             &   17. &   10.0 & 124. &   81 & 177. & 0.089 &  32. \\
 June 93             &   18. &    3.3 &  42. &    4 &  92. & 0.557 & $<1$.\\
 Oct. 93-2           &    4. &    3.0 & 141. &   60 & 245. & 0.049 &  83. \\
 Oct. 93-3           &    7. &    4.0 & 185. &   80 & 303. & 0.052 &  77. \\
 July 94             &   22. &    3.4 &  47. &    9 &  94. & 0.027 &  95. \\
 Aug. 94             &   27. &    3.9 &  46. &   15 &  85. & 0.075 &  52. \\
 Sep. 94-1           &   30. &    6.5 & 112. &   50 & 188. & 0.082 &  39. \\
 Nov. 94             &   23. &    8.0 & 116. &   66 & 176. & 0.015 &$>99$.\\
 July 95             &   33. &    5.0 &  74. &   19 & 138. & 0.063 &  55. \\
 Aug. 95             &   24. &    7.4 & 105. &   60 & 161. & 0.061 &  56. \\
 Sep. 95             &   33. &    1.2 &  28. &    0 & 142. & 0.058 &  73. \\
 Oct. 95             &   26. &    6.5 & 163. &   75 & 270. & 0.019 &$>99$.\\
 Nov. 95             &   32. &   10.2 & 127. &   78 & 185. & 0.032 &  88. \\
 Dec. 95-2           &   40. &    0.5 &  12. &    0 &  95. & 0.068 &  62. \\
 Jan. 96             &   35. &    3.5 &  45. &    0 & 101. & 0.047 &  76. \\
 May  96             &   31. &    5.3 &  78. &   31 & 136. & 0.039 &  90. \\
 Aug. 96             &   26. &    4.5 &  51. &   14 &  96. & 0.089 &  35. \\
 Oct. 96             &   21. &    5.4 &  58. &   28 &  95. & 0.046 &  74. \\
 Nov. 96             &   28. &    1.9 &  21. &    0 &  57. & 0.103 &  37. \\
 Jan. 97             &   24. &    2.6 &  26. &    0 &  60. & 0.190 &  13. \\
 Mar. 97             &   24. &    6.1 &  54. &   24 &  90. & 0.134 &  15. \\
 Apr. 97             &   22. &    2.7 &  27. &    3 &  57. & 0.037 &  86. \\
 June 97             &   27. &   10.4 & 109. &   71 & 155. & 0.078 &  35. \\
 July 97             &   22. &    0.0 &   0. &    0 &  24. & 0.333 &   7. \\
 Sep. 97             &   17. &    4.3 &  49. &   22 &  84. & 0.043 &  80. \\
 Oct. 97             &   26. &    3.4 &  36. &    9 &  72. & 0.083 &  49. \\
 Dec. 97             &   24. &    6.2 &  80. &   40 & 128. & 0.031 &  89. \\
\hline
Combined             &  753. &  152.1 &  64. &   56 &  72. & 0.033 &  93. \\
\end{tabular}
\normalsize
\end{table}

     The overall likelihood function for a single extraction is the product
of the separate likelihood functions for the $L$- and $K$-peak regions.  The
best fit capture rate is found by maximizing this function, allowing the
independent background rates in the $L$ and $K$ peaks to be free variables.
The uncertainty in this result is determined by finding the values of the
capture rate at which the logarithm of the likelihood function decreases by
0.5, again choosing the background rates at these two points to be those
which maximize the likelihood function.  The results for all extractions that
could be analyzed in both peaks are given in Table \ref{K_plus_L_peak_table}.

\begin{figure*}
\begin{center}
\includegraphics[width=5.000in]{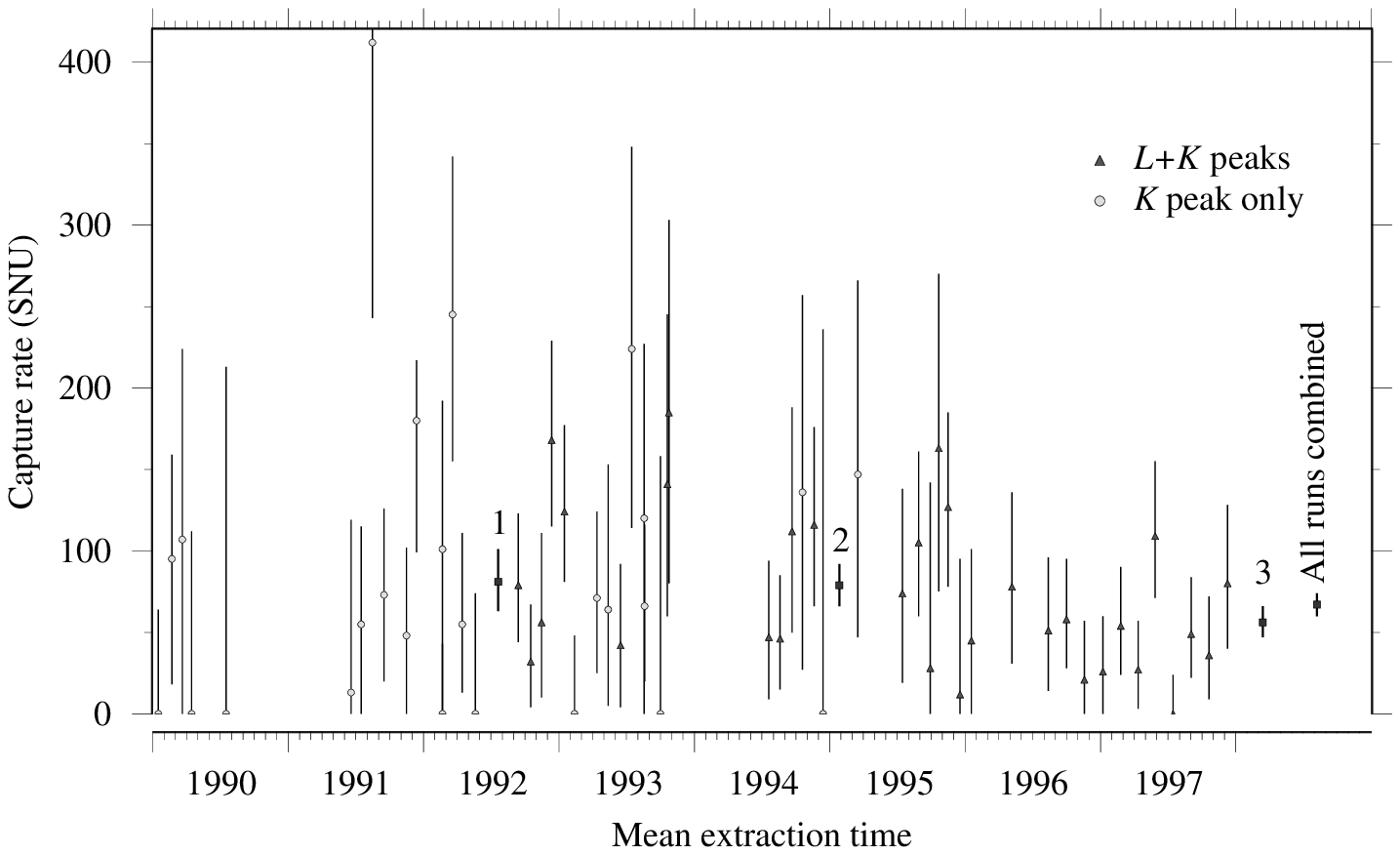}
\end{center}
\caption{Capture rate for each extraction as a function of time.  All
uncertainties are only statistical.  The symbols 1, 2, and 3 show the
combined result for SAGE~I, II, and III, respectively.}
\label{All_extraction_results}
\end{figure*}

     The capture rate for each extraction of all runs of SAGE is plotted in
Fig.\ \ref{All_extraction_results}.  These results are derived from the $K$
peak plus $L$ peak wherever possible, otherwise from the $K$ peak alone.  For
those readers who may be interested in looking for temporal phenomena, the
beginning time $t_B$ and ending time $t_E$ for each run can be inferred from
the mean exposure date $t_m$ and total exposure time $\Theta$ given in
Table~\ref{run_parameter_table} by the relationships
\begin{eqnarray}
t_B & = & t_m - \frac{1}{\lambda}\ln\bigg(\frac{1 +
                                            e^{\lambda\Theta}}{2}\bigg), \\
t_E & = & \Theta - t_B. \nonumber
\end{eqnarray}

\subsection{Global fits}\label{global_fits}

     The combined likelihood function for any set of extractions is the
product of the overall likelihood functions for each extraction.  The best
fit capture rate for the set of extractions is determined by maximizing this
function, requiring the production rate per unit mass of Ga to be the same
for each extraction, and allowing the background rates in both the $L$ and
$K$ peaks to be different for each extraction.  The uncertainty is found in
the same way as for the $L$- and $K$-peak combination for a single
extraction.  There are a number of other techniques for estimating the
uncertainties in addition to the two described and used here.  When many runs
are combined, the likelihood as a function of capture rate approaches a
Gaussian, and the difference between the results of these techniques becomes
slight.

     The results of global fits to our data are given in Sec.~\ref{Results}.

\section{Systematic Uncertainties}\label{systematics}

     There are four basic sources of systematic error in SAGE: uncertainty in
the chemical extraction efficiency, uncertainty in the counting efficiency,
uncertainty due to nonsolar neutrino production of \nuc{71}{Ge} (such as by
cosmic rays), and uncertainty due to nonconstant events which mimic
\nuc{71}{Ge} (such as may be made by \nuc{222}{Rn}).
Table~\ref{uncertainty_table} summarizes the results of our consideration of
all these effects and additional information regarding each of these items
follows.

\begin{table}[t]
\caption{Summary of the systematic uncertainties.  The SNU values for
extraction and counting efficiency are based on a rate of 67.2 SNU.}
\label{uncertainty_table}
\begin{tabular}{l d @{\hspace{-9.0mm}} d}

                                       & \multicolumn{2}{c}{Uncertainty}\\
Origin of uncertainty                    & in percent   & in SNU        \\
\hline
\multicolumn{3}{l}{Extraction efficiency}                            \\
 \hspace{ 5mm} Ge carrier mass           & $\pm$2.1\% & $\pm$ 1.4    \\
 \hspace{ 5mm} Mass of extracted Ge      & $\pm$2.5\% & $\pm$ 1.7    \\
 \hspace{ 5mm} Residual Ge carrier       & $\pm$0.8\% & $\pm$ 0.5    \\
 \hspace{ 5mm} Ga mass                   & $\pm$0.3\% & $\pm$ 0.2    \\
 \hspace{2.5mm} Total (extraction)       & $\pm$3.4\% & $\pm$ 2.3    \\
\multicolumn{3}{l}{Counting efficiency}                              \\
 \hspace{ 5mm} Volume efficiency         & $\pm$1.4\% & $\pm$ 0.9    \\
 \hspace{ 5mm} End losses                & $\pm$0.5\% & $\pm$ 0.3    \\
 \hspace{ 5mm} Monte Carlo interpolation & $\pm$1.0\% & $\pm$ 0.7    \\
 \hspace{ 5mm} Shifts of gain            & $-$3.1\%   & $+$2.1       \\
 \hspace{ 5mm} Resolution                & $+$0.5\%,$-$0.7\% & $-$0.3,+0.5 \\
 \hspace{ 5mm} Rise time limits          & $\pm$1.0\% & $\pm$ 0.7    \\
 \hspace{ 5mm} Lead and exposure times   & $\pm$0.8\% & $\pm$ 0.5    \\
 \hspace{2.5mm} Total (counting)         & $+$2.3\%,$-$3.9\% & $-$1.5,+2.6 \\
\multicolumn{3}{l}{Nonsolar neutrino production of \nuc{71}{Ge}}     \\
 \hspace{ 5mm} Fast neutrons             &               & $<-$ 0.02 \\
 \hspace{ 5mm} \nuc{232}{Th}             &               & $<-$ 0.04 \\
 \hspace{ 5mm} \nuc{226}{Ra}             &               & $<-$ 0.7  \\
 \hspace{ 5mm} Cosmic-ray muons          &               & $<-$ 0.7  \\
 \hspace{2.5mm} Total (nonsolar)         &               & $<-$ 1.0  \\
\multicolumn{3}{l}{Background events that mimic \nuc{71}{Ge}}        \\
 \hspace{ 5mm} Internal \nuc{222}{Rn}    &               & $<-$ 0.2  \\
 \hspace{ 5mm} External \nuc{222}{Rn}    &               & $  $ 0.0  \\
 \hspace{ 5mm} Internal \nuc{69}{Ge}     &               & $<-$ 0.6  \\
 \hspace{2.5mm} Total (background events)&               & $<-$ 0.6  \\
\hline
 Total                                   &               & $-$3.0,+3.5
\end{tabular}
\end{table}

\subsection{Chemical extraction efficiency}\label{chem_extrac_eff}

     The way in which the chemical extraction efficiency is determined was
described in Sec.~\ref{Chemical_extraction_efficiency}.  There are four
sources of uncertainty.

\subsubsection{Mass of Ge carrier}
     The extraction efficiency is measured by adding to the Ga metal several
slugs of Ga-Ge alloy which contain a known mass of Ge.  This alloy is
produced in large batches by reduction of Ge by Ga metal from chloride
solution, and then divided into several hundred small slugs, each of which
weighs 18--20 g and contains about 40 $\mu$g of Ge.  The equality of Ge
content was measured by extracting the Ge from a few dozen slugs.  The
standard deviation of these measurements was 2.1\%, which we take as the
uncertainty in the mass of added Ge carrier.

\subsubsection{Mass of extracted Ge}
     There is also an uncertainty in how much carrier has been synthesized
into GeH$_4$.  This is determined by the accuracy to which the GeH$_4$ volume
can be determined and is estimated to be 2.5\%.

\subsubsection{Residual Ge carrier}
     Since the extraction of carrier Ge is not complete, residual Ge carrier
from preceding extractions will contribute to the extraction efficiency
measurement.  Each extraction for solar neutrino data is followed by a second
extraction to remove this surplus carrier.  The amount removed during the
first two extractions is at least 95\%, but is uncertain as described above,
which leads to an uncertainty in the amount of remaining carrier.  The
extraction efficiency uncertainty due to the uncertainty in the residual
carrier is $\pm 0.8$\%.

\subsubsection{Mass of Ga}
     The total mass of Ga has been weighed periodically with a precision of
0.3\%.  The amount removed during each extraction is small (typically 0.1\%)
and is known well (2\%).  We take the uncertainty in the Ga mass for all runs
to be $\pm 0.3$\%.

\subsection{Counting efficiency}\label{cnting_effs}

     The counter efficiency is calculated using Eq.~\ref{efficiency_formula}.
There are thus three sources of uncertainty: the volume efficiency
$\epsilon_v$, the end effects (or equivalently the fraction of degraded
events), and the gas efficiency.

\subsubsection{Volume efficiency}\label{vol_eff}
     As described in Sec.\ \ref{counter_efficiency}, the volume efficiency of
seven counters of the ``LA'' type has been directly measured with an
uncertainty of 0.6\%.  These seven counters were used for 48 of our 88 data
sets.  The uncertainty in $\epsilon_v$ for counters of this same type used in
36 other data sets is estimated from the spread in the measured $\epsilon_v$
for the measured counters, which is $\pm 2.3$\% relative uncertainty.  The
uncertainty in $\epsilon_v$ for counters of the ``Ni'' and ``RD'' types,
which were used for four data sets, is taken as $\pm 3$\%.  Averaging over
the different types of counters used for all extractions, the uncertainty
assigned to volume efficiency is taken as $\pm 1.4$\%.

\subsubsection{End effects}
     The reduced electric field near the ends of the counter cathode results
in a fraction of events that lie outside the $\pm1$ FWHM energy windows.
Uncertainties in these end effects are due to variations in the physical
dimensions of the counters.  Based on measurements of various ``LA''
counters, these dimensional differences lead to an uncertainty of $\pm 4.1$\%
in the end effect.  This gives $\pm 0.5$\% relative uncertainty in the factor
$(1 - f_D)$.  This should be valid for measurements made with the ``LA''
counters, which were used for most of our data, and this value is taken for
the entire data set.

\subsubsection{Monte Carlo interpolation of measured gas
parameters}\label{mc_gas}
     The uncertainties in the gas efficiency consist of three components:
uncertainty in the Monte Carlo calculations, uncertainty in the measured gas
pressure, and uncertainty in the measured percentage of GeH$_4$.  The limited
statistics used in the Monte Carlo calculations to determine the constants in
the gas efficiency formula leads to an uncertainty of 1.0\% in the
determination of the gas efficiency.  The uncertainty in the gas pressure
measurements is $\pm 5$ Torr, which corresponds to an uncertainty in the gas
efficiency for a typical counter filling (710 Torr at 24\% GeH$_4$) of $\pm
0.2$\% relative change.  The uncertainty in the measured percentage of
GeH$_4$ is taken to be $\pm 1$\%, which corresponds to an uncertainty in the
gas efficiency for an average counter filling of $\pm0.2$\% relative change.
Adding these three contributions in quadrature yields a relative total
uncertainty in the gas efficiency of $\pm 1.0$\%.

\subsubsection{Gain shifts}\label{gain_shifts}
     If the calibration mean shifts between two calibrations, there is an
error made in the efficiency estimate.  This error has been minimized by two
features of our standard analysis.  First, we use a two FWHM wide energy
window.  Since the peak is relatively Gaussian and the window limits are far
out on the tail, uncertainties in the location of the centroid of the peak do
not greatly affect the efficiency.  Second, by sliding the energy window
between calibrations we hope to minimize any error in estimating the centroid
due to the observed gain shifts.  Although the correction for nonlinearity of
the counter response [Eq.\ (\ref{Gain_Scaling_Formula})] results in an
additional uncertainty in the gain of 0.7\%, the total uncertainty in the
gain is dominated by the shifts.

     To estimate the error generated by using an incorrect centroid, we
computed the area under a Gaussian between two integration limits which are
shifted by an amount $\delta$.  We then compared this number to the 0.9815
number expected from integration limits of $\pm 2$ FWHM.  Using a typical
$K$-peak resolution of 20\%--23\% we calculated the true efficiency for
various values of $\delta$ expressed as a fraction of the true mean.

     Typical gain shifts are of the order of a few percent.  This results in
an uncertainty of approximately $-3.1$\% in the efficiency.  Note that this
effect can only decrease our efficiency so it is a one sided systematic
uncertainty.

\subsubsection{Energy resolution}
     As a result of the statistics of our calibration spectra, the resolution
is known to about 2.1\%.  For the $K$ peak, there is an additional
uncertainty due to the counter nonlinearity
[Eq.\ (\ref{Resolution_Scaling_Formula})] of $\pm 4.5$\%.  Adding these in
quadrature, the uncertainty in the resolution results in an uncertainty in
the efficiency of about +0.5\%, $-0.7$\%.  Again, because the energy window
is so wide, the uncertainty in the efficiency due to the resolution
uncertainty is not large.

\subsubsection{Rise time limits}\label{adp_lims}
     As described in Section \ref{rise_tim_cal}, when the wave form method of
rise time determination is used, there is an uncertainty in the efficiency of
$\pm1$\% that arises from changes in the rise time limits due to counting
channel and filling variations.  For those runs that used the ADP method of
rise time determination, we can find the uncertainty of the ADP cut as
follows:  Usually a calibration has between 1000 and 5000 events in the peak.
We base our lower ADP threshold on 4\% of those or 40--200 events.  This
small number of events is subject to statistical fluctuations.  For the most
extreme case, we take the square root of 40 (6.4) and notice that the
efficiency due to the ADP cut could actually be between 94.4\% and 95.6\%
instead of the 95\% we believe it to be.  Thus this is a $\pm 0.6$\%
uncertainty.  Since the vast majority of our data are based on wave form
analysis, we use $\pm1$\% for all runs.

\subsubsection{Lead and exposure times}\label{lead_exp_tims}
     Because extraction usually occurs from several reactors over the course
of 6--10 h, there is an uncertainty in the exposure time and in the time from
extraction to the start of counting (which we call the ``lead time'') of
roughly 3 h.  The lead time is typically 36 h and the exposure time is
typically 34 days.  These small uncertainties make a small contribution to
the uncertainty associated with the solar neutrino flux.  By
Eq.\ (\ref{prod_rate_def}), the solar neutrino production rate $p$ is
proportional to the quantity $[e^{-\lambda t_{\text{lead}}}(1 - e^{-\lambda
\Theta})]^{-1}$, where $\lambda$ is the \nuc{71}{Ge} decay constant,
$t_{\text{lead}}$ is the lead time, and $\Theta$ is the exposure time.  By
differentiation one finds that $\delta p/p$ due to $t_{\text{lead}}$ is about
$\pm 0.8$\% and due to $\Theta$ is about $\pm 0.11$\%.

\subsection{Nonsolar neutrino contributions to the $^{\bf71}$G\lowercase{e}
signal} \label{bkgds}

     In addition to solar neutrinos, \nuc{71}{Ge} can also be produced from
Ga by the reaction \nuc{71}{Ga}$(p,n)$\nuc{71}{Ge}.  The protons that
initiate this reaction can be secondaries made by the $(n,p)$ reaction of
fast neutrons or by the $(\alpha,p)$ reaction where the $\alpha$'s are from
radioactive decay or may arise from photonuclear reactions initiated by
cosmic-ray muons.  The yields of these reactions have been measured with
neutrons from radioactive sources, $\alpha$'s from a Van de Graaff generator,
and high-energy muons from accelerators (see, e.g., \cite{BAH78}).  Based on
these results, great care was taken in the design and construction of SAGE to
minimize these potential background sources.  A major advantage of using Ga
metal as the solar neutrino target (as opposed to an aqueous solution, such
as the GaCl$_3$ target of GALLEX) is that the target contains no free
protons, and thus the production rates of all these reactions are low.

     Any one of these processes could produce a background effect that must
be subtracted from our measured solar neutrino signal, but as will be seen
below, our best estimates for all of these effects are very small and have
large errors.  Thus, rather than making a background subtraction, we include
these effects here as systematic uncertainties.

     Other Ge isotopes that may be misidentified as \nuc{71}{Ge} can be
produced in similar reactions: \nuc{69}{Ge} can be made by
\nuc{69}{Ga}$(p,n)$\nuc{69}{Ge} and the spallation reaction on Ga by
throughgoing cosmic-ray muons can make \nuc{68}{Ge} and \nuc{69}{Ge}.  Since
the production rate of \nuc{68}{Ge} by the spallation reaction is comparable
to that of \nuc{71}{Ge} and its half-life is long (271 days), the
\nuc{68}{Ge} decay rate is much less than that of \nuc{71}{Ge} and can be
neglected.  The short-lived isotope \nuc{69}{Ge} has a greater potential to
give events that mimic \nuc{71}{Ge} and is considered below in
Sec.~\ref{69Ge_background}.

\subsubsection{External neutrons}\label{ext_neutrons}
     Fast neutrons mainly arise from the walls of the Ga chamber by
$(\alpha,n)$ reactions where the $\alpha$ particles are from \nuc{238}{U}
decay.  This background is expected to be small due to the low cross sections
for $(n,p)$ reactions on Ga isotopes and to the low fast neutron background,
which is because the Ga chamber is lined with low-radioactivity concrete and
steel.  The fast neutron flux in the gallium chamber was measured by
extracting \nuc{37}{Ar} from a tank with 187 kg of dry CaC$_2$O$_4$ and
counting in a proportional counter.  Fast neutrons above 3 MeV produce
\nuc{37}{Ar} through the \nuc{40}{Ca}$(n,\alpha)$\nuc{37}{Ar} reaction.  The
flux over 3 MeV is $(4.6 \pm 1.6) \times 10^{-3}$ neutrons/(cm$^2$ day)
\cite{Gavrin91b}.  The number of \nuc{71}{Ge} atoms produced by this flux in
60 tons of Ga metal is $<2.9 \times 10^{-4}$/day \cite{BAR79,BAR87,Korn98},
which, using the conversion factor 56 SNU/(\nuc{71}{Ge} atom produced per day
in 60 tons of Ga), corresponds to $< 0.016$ SNU.

\begin{table*}[t]
\caption{Rows 1 and 2: probability of a false \nuc{71}{Ge} event in the $L$
and $K$ peaks produced by the decay of each nucleus in the Rn decay chain per
decay of \nuc{222}{Rn}.  Row 3: probability that each nucleus in the chain
will decay later than 180 min after a \nuc{222}{Rn} decay.}
\label{intrntable}
\begin{tabular}{l d d d d d}
& \multicolumn{5}{c}{Decaying nucleus in the \nuc{222}{Rn} chain} \\
\cline{2-6}
\vspace{-0.8em} \\  
& \nuc{222}{Rn} & \nuc{218}{Po} & \nuc{214}Pb & \nuc{214}{Bi} & \nuc{214}{Po}
\\
\hline
Prob.\ false $L$ event  & 0.      & 0.0042 & 0.0110 & 0.0072 & 0.0016 \\
Prob.\ false $K$ event  & 0.00004 & 0.0490 & 0.0061 & 0.0018 & 0.0186 \\
Survival probability    & 0.      & 0.     & 0.011  & 0.035  & 0.035  \\
\end{tabular}
\end{table*}

\subsubsection{Internal radioactivity}\label{int_radio}
     The second possible background source is due to $\alpha$ radioactivity
in the gallium.  The only appreciable sources of high-energy $\alpha$'s are
from decays in the U and Th chains, mainly the 8.8-MeV $\alpha$ from
\nuc{212}{Po} at the end of the Th chain and the 7.8-MeV $\alpha$ from
\nuc{214}{Po} near the end of the U chain.  The concentrations of radioactive
impurities in the Ga have been measured in two ways: by direct counting in a
low-background Ge detector by the Institute for Nuclear Research (INR)
\cite{GAV86} and by glow discharge mass spectrometry by both Charles Evans
Associates \cite{Evans} and Shiva Technologies \cite{Shiva}.  No U, Th, or Ra
was detected in any of these measurements.  Expressed in grams of impurity
per gram of Ga, the limits are U $<2.0\times 10^{-10}$ (Evans) and
$<1.2\times 10^{-10}$ (Shiva); Th $<8.0\times 10^{-10}$ (INR), $<1.7\times
10^{-10}$ (Evans), and $<1.2\times 10^{-10}$ (Shiva); \nuc{226}{Ra} $<1.1
\times 10^{-16}$ (INR).  We take the INR limit for \nuc{226}{Ra} and the
Shiva limit for Th.  Using the measured \nuc{71}{Ge} yields \cite{BAH78} in
metallic Ga, the number of \nuc{71}{Ge} atoms produced per day in 60 tons of
Ga is $<0.001$/day from \nuc{232}{Th} and $< 0.013$/day from \nuc{226}{Ra},
which correspond to $<0.04$ SNU and $<0.7$ SNU, respectively.

\subsubsection{Cosmic-ray muons}\label{cr_muons}
     The third possible background source is production of Ge isotopes by
cosmic-ray muons.  The global cosmic-ray muon flux in the gallium laboratory
at BNO has been measured \cite{Gavrin91} to be $(3.03 \pm 0.10) \times 10^{-
9}$ muons/(cm$^2$ s).  This flux can be converted to a \nuc{71}{Ge}
production rate in 60 tons of Ga metal in two ways: (1) using cross sections
measured at accelerators \cite{BAH78,Cribier97} and scaling by the average
muon energy (which is $\sim 381$ GeV) to the 0.73 power \cite{Rya65}, one
predicts 0.012/day \cite{Korn98}, or (2) using cross sections for production
of \nuc{71}{Ge} calculated in \cite{GAV87}, one predicts 0.013/day.  Both of
these convert to rates of 0.7 SNU.  Since the error of these estimations is
about 100\% \cite{Korn98}, we consider this background as a systematic
uncertainty for the muon background rate.

\subsection{Background events that mimic $^{\bf71}$G\lowercase{e}}

     As is evident from Tables~\ref{L_peak_table} and \ref{K_peak_table}, a
large fraction of the events that we select as \nuc{71}{Ge} candidates are
not \nuc{71}{Ge}.  In the $L$ ($K$) peak we select 433 (604) events as
candidates, but most of these events occur late in the counting, and thus the
best fit to \nuc{71}{Ge} plus constant background assigns only 64.3 (143.7)
of them to be \nuc{71}{Ge}.  These late events that we cannot discriminate
from \nuc{71}{Ge} are produced by background processes, such as $\beta$ rays
whose path through the counter either is very short or is parallel to the
anode wire.  As long as these background events occur at a constant rate,
they only deteriorate our signal-to-background ratio, but do not change our
extracted \nuc{71}{Ge} signal rate.  If these background events mainly occur
early or late in the counting, however, the extracted \nuc{71}{Ge} signal
rate will be incorrect, too high or too low, respectively.  Particularly
insidious in this regard is the ubiquitous naturally occurring isotope
\nuc{222}{Rn}.

\subsubsection{Internal radon}\label{int_rn}

     Because it has a short half-life of only 3.8 days, and can produce
events that mimic \nuc{71}{Ge}, any \nuc{222}{Rn} that enters the counter at
the time it is filled will produce events early in the counting period that
may be falsely interpreted as \nuc{71}{Ge}, and thus give an incorrectly high
\nuc{71}{Ge} signal rate.  To understand this process, let us first consider
the principal \nuc{222}{Rn} decay sequence, which is

\def\mapright#1{\smash{\mathop{\longrightarrow}\limits_{#1}}}
\def\decayright#1#2{\kern#1em\raise1.1ex\hbox{$|$}\kern-.60em\mapright{#2}}
{\raggedright
\hspace{0.5em}
\nopagebreak[4]
$^{222}\text{Rn} \mapright{3.82\text{ d}} ^{218}\text{Po} + \alpha_1$ \\
\nopagebreak[4]
\hspace{0.5em}
\nopagebreak[4]
$ \decayright{5.9}{3.05\text{ m}} ^{214}\text{Pb} + \alpha_2 $ \\
\nopagebreak[4]
\hspace{0.5em}
\nopagebreak[4]
$ \decayright{9.4}{26.8\text{ m}} ^{214}\text{Bi} + \beta_1  $ \\
\nopagebreak[4]
\hspace{0.5em}
\nopagebreak[4]
$ \decayright{12.9}{19.7\text{ m}} ^{214}\text{Po} + \beta_2  $ \\
\nopagebreak[4]
\hspace{0.5em}
\nopagebreak[4]
$ \decayright{16.4}{10^{-4}\text{ s}} ^{210}\text{Pb} + \alpha_3. $
\vspace{0.6em}
}  

     Events that are falsely identified as \nuc{71}{Ge} can be produced by
one of the $\alpha$'s (this happens rarely as they are heavily ionizing), by
one of the $\beta$'s (this occurs more frequently), by the recoil nucleus
from the $\alpha$ decay if the initial nucleus is on the counter wall, or by
a low-energy x ray emitted by one of the heavy elements in the chain.
Fortunately, the start of this chain is easily recognized as at least one of
the first two heavily ionizing $\alpha$ particles usually produces a pulse
that saturates the energy scale.  Thus, since this chain takes on average
about 1 hour from the initiating decay of \nuc{222}{Rn} to reach 22-yr
\nuc{210}{Pb}, if one makes a time cut of a few hours after each saturated
event, then most of these false \nuc{71}{Ge} events will be removed.

     We choose to eliminate all events that occur from 15 min before to 180
min after each detected saturation event (energy greater than 16 keV).  To
determine the effect of this time cut, we filled a counter with a typical
mixture of Xe and GeH$_4$ to which \nuc{222}{Rn} had been added and measured
it in system 3 under conditions identical to those of solar runs.  Based on
these measurements and Monte Carlo modeling, the spectrum of pulses in the
counter was determined by each element in the Rn chain.  The probability of a
false \nuc{71}{Ge} event can then be directly calculated and the results are
given in Table~\ref{intrntable}.

     Folding the probability of a false event with the probability of
survival after the time cut (Table~\ref{intrntable}), we obtain the
probability of observing a false \nuc{71}{Ge} event after the time cut to be
0.00043 in the $L$ peak and 0.00078 in the $K$ peak.  The resolution of this
counter was better than for the average solar neutrino extraction.  If we use
the average resolution, the number of false \nuc{71}{Ge} events in a typical
solar neutrino run that satisfy all our event selection criteria divided by
the number of detected saturated events due to \nuc{222}{Rn} is calculated to
be 0.0006 and 0.0012 for the $L$ and $K$ peaks, respectively.

     To estimate the number of false \nuc{71}{Ge} events that survive the
time cut, we next calculate how many saturated events are present in our data
that can be attributed to Rn.  This is done by taking the data for each run,
making the usual time cut after shield openings, and selecting all saturated
events.  The time sequence of these events is then fit to a decaying
component with the 3.82-day half-life of \nuc{222}{Rn} plus a constant
background.  For the periods of SAGE~II and III, this yields 294 saturated
events initiated by \nuc{222}{Rn}, with 192.1 (294.0) events in the 31 (57)
data sets that give the $L$- ($K$-) peak results.  Since SAGE~I did not have
the capability to detect saturated events, we scale the number for the $K$
peak by the number of additional extractions (16) to make the $K$-peak total
376.5.

     Combining these results, the number of false \nuc{71}{Ge} events that
remain after the time cut is then given by $0.0006 \times 192.1 = 0.11$ in
the $L$ peak and $0.0012 \times 376.5 = 0.44$ in the $K$ peak.  Since we have
observed a total of 64.3 (143.7) events in the $L$ ($K$) peaks, the fraction
of false events is 0.2\% (0.3\%), which translates to a false signal rate of
0.1 (0.2) SNU.  Combining the $L$- and $K$-peak results gives a total false
signal rate of 0.2 SNU.  Because the uncertainty in this correction is
comparable with the magnitude of the correction itself, we choose to treat
this effect as a systematic uncertainty, rather than as a background to be
subtracted from the signal.

\subsubsection{External radon}\label{ext_rn}

     \nuc{222}{Rn} that is external to a proportional counter can also
produce false \nuc{71}{Ge} events.  The radon levels in the counting room
vary with external conditions, and usually fall within the range of $(2.0 \pm
1.0)$ pCi/l.  To reduce the level of Rn in the vicinity of the counters, all
passive shields are equipped with purge lines from evaporating liquid
nitrogen and the shields have been made fairly hermetic.  Each time a counter
is calibrated, however, the shield must be opened, and some mine air will
enter the volume around the counters.  Under normal circumstances
calibrations occur regularly every 2 weeks.  Any false \nuc{71}{Ge} events
that are produced by external Rn will thus occur more or less constantly in
time, and will be treated by the maximum likelihood analysis as a constant
background.  Nevertheless, we minimize the effect of external Rn by making a
time cut on the data for 2.6 h after any shield opening.

     A special counter was constructed to give information on the false
\nuc{71}{Ge} events that are produced by external \nuc{222}{Rn}.  This
consisted of one of our usual counters enclosed within a cylindrical quartz
capsule (20~mm diameter).  The sealed volume of the capsule was filled with
air to which \nuc{222}{Rn} was added to make the total activity 3.5 nCi.
This counter was measured in counting system 6 which uses the ADP method of
rise time determination.  In the $K$ peak of \nuc{71}{Ge}, after cuts for
energy, ADP, and NaI, the measured count rate was $0.81 \pm 0.11$ events/min.

     Because the internal volume of the proportional counter is shielded by
the Fe cathode of 1/3 mm thickness, the false \nuc{71}{Ge} events are mainly
produced by the $\beta$ particles from the decay of \nuc{214}{Bi} which have
sufficient energy to penetrate to the active volume of the counter.  We use
the measurements from the internal Rn section, our Monte Carlo model for the
counter response, make some reasonable assumptions regarding the location of
Rn-daughter products, take into account the reduction of Rn in the vicinity
of the counters due to the N$_2$ purge, and calculate the number of false
\nuc{71}{Ge} events to be 0.005 in the sum of the $L$ and $K$ peaks per run.
These calculations were made without taking into account the effect of the
2.6-h time cut after each shield opening, which reduces the number of events
even further.  Thus the number of false \nuc{71}{Ge} events produced by
external \nuc{222}{Rn} is negligible.

     This conclusion is verified by analyzing our full $L$ + $K$ data set
without making the shield opening time cut.  The result is 67.9 SNU, nearly
equal to the result of 67.2 SNU when the time cut is used.

\subsubsection{Internal \nuc{69}{Ge}}\label{69Ge_background}

     Because it can be produced by the same background reactions that make
\nuc{71}{Ge}, has a short half-life of 39 h, and 64\% of its decays are by
electron capture, \nuc{69}{Ge} can produce events that will be misidentified
as \nuc{71}{Ge}.  We can estimate the production rate of \nuc{69}{Ge} from
known data.  The cosmic-ray production rate of \nuc{69}{Ge} can be determined
in the same way as was done in Sec.~\ref{cr_muons} for \nuc{71}{Ge}.  Using
the measured muon flux in the laboratory, the cross section for production of
\nuc{69}{Ge} of 100 $\mu$b measured in GaCl$_3$ at CERN with 280 GeV muons
\cite{Cribier97}, the factor of 2 greater production rate of \nuc{69}{Ge} in
Ga metal compared to GaCl$_3$ measured at FNAL with 225 GeV muons
\cite{BAH78}, and scaling as the muon energy to the 0.73 power, we estimate a
production rate of 0.036 atoms of \nuc{69}{Ge} per day in 60 tons of Ga.  The
production rate of \nuc{69}{Ge} by $\alpha$ particles and neutrons is
comparable to that of \nuc{71}{Ge}, viz., 0.015 atoms/day.  We must add to
this the production rate of \nuc{69}{Ge} by \nuc{8}{B} neutrinos, which we
estimate as being comparable to that of \nuc{71}{Ge}, i.e., 5.8 SNU, thus
making a total estimated production rate of 0.21 \nuc{69}{Ge}/day in 60 tons
of Ga.  Since the usual exposure interval is at least 30 days, \nuc{69}{Ge}
will be fully saturated, and the total number of atoms at the end of exposure
will be approximately 0.5.  When counting starts, on the average 36 h after
extraction, nearly half of these atoms will have decayed.  Fortunately 86\%
of the decays of \nuc{69}{Ge} have a coincident $\beta^+$ or $\gamma$ and
will be vetoed by the surrounding NaI detector with approximately 90\%
efficiency.  Including the 14\% of \nuc{69}{Ge} decays that occur by electron
capture to the ground state, the total efficiency for \nuc{69}{Ge} detection
will be no more than 25\%.  Approximately 70\% of these decays will appear in
the $L$ and $K$ peaks, leaving a total of 0.045 observed \nuc{69}{Ge} decays
per run, or 1 event in every 44 data sets.  Since the 211 events that we have
assigned to \nuc{71}{Ge} in our 88 data sets correspond to 67 SNU, this
implies that the false \nuc{69}{Ge} background is approximately 0.6 SNU.
This very small value illustrates the desirability of siting the SAGE
detector at great depth and the advantage of a NaI veto on all channels
during counting.

\section{Results}\label{Results}

     If we combine SAGE~I with SAGE II (minus part 2) and SAGE III, the
global best fit capture rate for the 88 separate counting sets is $67.2
^{+7.2}_{-7.0}$ SNU, where the uncertainty is statistical only.  In the
windows that define the $L$ and $K$ peaks there are 1037 counts with 211.15
assigned to \nuc{71}{Ge} (the total counting live time is 28.7 yr).  If we
were to include the data from SAGE II part 2, the overall capture rate would
decrease by 7.1 SNU.  The systematic control of the experiment was suspect
during the period of the gallium theft (see Sec.~\ref{extrac_hist}), and thus
we exclude that data interval from our result.

     The total systematic uncertainty is determined by adding in quadrature
all the contributions given in Table~\ref{uncertainty_table} and is $-$3.0,
+3.5 SNU.  Our overall result is thus $67.2 ^{+7.2 +3.5}_{-7.0 -3.0}$ SNU.
If we combine the statistical and systematic uncertainties in quadrature, the
result is $67.2 ^{+8.0}_{-7.6}$ SNU.

     This section continues with the evidence that we are truly counting
\nuc{71}{Ge}, considers how well the observed data fit the models that are
assumed in analysis, and concludes with consideration of the internal
consistency of the SAGE results.

\subsection{Evidence for $^{71}$G\lowercase{e}}

     The most direct visual evidence that we are really observing
\nuc{71}{Ge} is in Fig.\ \ref{2d_hist}.  The expected location of the
\nuc{71}{Ge} $L$ and $K$ peaks is shown darkened in this figure.  These peaks
are apparent in the upper panel, but missing in the lower panel because the
\nuc{71}{Ge} has decayed away.  Events outside the two peak regions occur at
about the same rate in both panels because they are mainly produced by
background processes.

     A quantitative indication that \nuc{71}{Ge} is being counted can be
obtained by allowing the decay constant during counting to be a free variable
in the maximum likelihood fit, along with the combined production rate and
all the background rates.  The best fit half-life to all selected events in
both $L$ and $K$ peaks is then $10.5_{-1.9}^{+2.3}$ days, in good agreement
with the measured value \cite{HAM85} of 11.43 days.

\subsection{Consistency of the data with analysis hypotheses}

\subsubsection{Energy and rise time window positions}
     To test whether or not the energy and rise time windows are properly
set, the windows can be made wider and the data reanalyzed.  If the rise time
window for accepted events is increased by 30\%, i.e., from 0--10 ns to 0--13
ns in the $L$ peak and from 0--18.4 ns to 0--24.0 ns in the $K$ peak, then
the overall result of all runs of SAGE~II and III that were counted in system
3 is 68.3 SNU.  This change is entirely consistent with the $\sim 3$\%
increase in counting efficiency due to the increased size of the rise time
acceptance window.  Similarly, if the energy window in both $L$ and $K$ peaks
is opened from the usual 2 FWHM to 3 FWHM, then the overall result of all
runs of SAGE~II and III becomes 69.1 SNU.  This increase from the value of
67.2 SNU in the 2 FWHM energy window is because some of the \nuc{71}{Ge}
decays occur at the ends of the counter and their detected energy is reduced
from the full peak value.  This results in an increase in the counting
efficiency in the wider energy window of 2\%--3\%.  If this efficiency
increase is included in the analysis, then the results in the two energy
windows agree to better than 1\%.

\subsubsection{Time sequence}
     A major analysis hypothesis is that the time sequence of observed events
for each run consists of the superposition of events from the decay of a
fixed number of \nuc{71}{Ge} atoms plus background events which occur at a
constant rate.  The quantity $Nw^2$ and the goodness of fit probability
inferred from it provide a quantitative measure of how well the data fit this
hypothesis (see \cite{CLE98} for the definition and interpretation of
$Nw^2$).  These numbers are evaluated for each data set and are given in
Tables~\ref{L_peak_table}, \ref{K_peak_table}, and \ref{K_plus_L_peak_table}.
There are occasional runs with rather low probability of occurrence, but no
more of these are observed than are expected due to normal statistical
variation.

\begin{figure}[t]
\begin{center}
\includegraphics[width=3.375in]{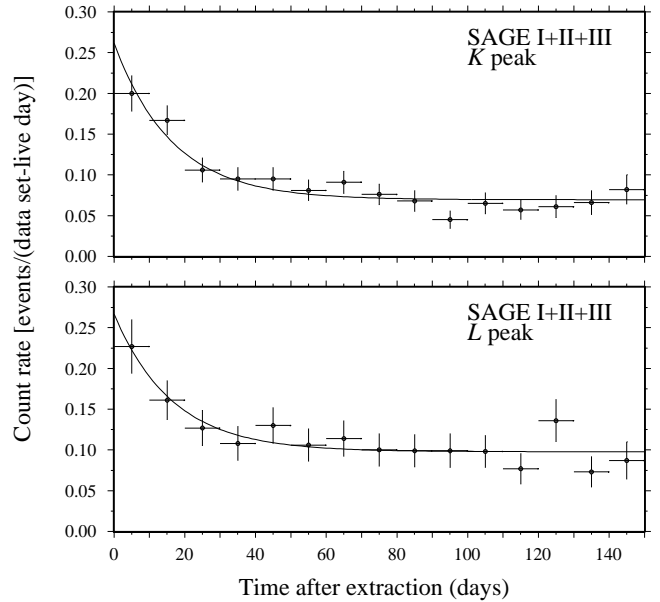}
\end{center}
\caption{Count rate for all runs in $L$ and $K$ peaks.  The solid line is a
fit to the data points with the 11.4-day half-life of \nuc{71}{Ge} plus a
constant background.  The vertical error bar on each point is proportional to
the square root of the number of counts and is shown only to give the scale
of the error.  The horizontal error bar is $\pm 5$ days, equal to the 10-day
bin size.}
\label{cntrate}
\end{figure}

     This method can also be used to determine the goodness of fit of the
time sequence for any combination of runs.  These numbers are given in the
various tables; for the combined time sequence of all $L$ plus $K$ events
from all runs, this test yields $Nw^2 = 0.074$, with a goodness-of-fit
probability of $(58 \pm 5)$\%.  A visual indication of the quality of this
fit is provided in Fig.\ \ref{cntrate} which shows the count rate for all
events in the $L$ and $K$ peaks vs time after extraction.  As is apparent,
the observed rate fits the hypothesis quite well.

\begin{table*}
\caption{Results of combined analysis of various segments of SAGE data.  The
time intervals for each segment are defined in
Table~\ref{Data_Assignment_Table}.  The uncertainty in the probability is
$\sim 4$\%.}
\label{Segment_table}
\begin{tabular}{l c d d d d r @{--} d d d}

        &      &           & Number of & Number     &            &
\multicolumn{2}{c}{}             &        &                           \\
Data    &      & Number of & candidate & fit to     & Best fit   &
\multicolumn{2}{c}{68\% conf.}   &        & Probability               \\
segment & Peak & data sets & events    & $^{71}$Ge  &   (SNU)    &
\multicolumn{2}{c}{range (SNU)}  & $Nw^2$ & (\%)                      \\
\hline
SAGE I   & $K$   & 16. &  157. &  41.2 & 81. & 63 & 101. & 0.097 &  24. \\
SAGE II  & $L+K$ & 33. &  342. &  85.5 & 79. & 66 &  92. & 0.105 &  32. \\
SAGE III & $L+K$ & 39. &  538. &  87.0 & 56. & 47 &  66. & 0.040 &  90. \\
\\
All      & $L$   & 31. &  433. &  64.3 & 55. & 43 &  68. & 0.020 & $>99$. \\
All      & $K$   & 57. &  604. & 143.7 & 73. & 64 &  82. & 0.110 &  25. \\
All      & $L+K$ & 88. & 1037. & 211.1 & 67. & 60 &  74. & 0.074 &  58. \\
\end{tabular}
\end{table*}

\subsubsection{Production rate sequence}
     Another analysis hypothesis is that the rate of \nuc{71}{Ge} production
is constant in time.  By examination of Fig.~\ref{All_extraction_results}, it
is apparent that, within the large statistical uncertainty for each run,
there are no substantial long-term deviations from constancy.

\begin{figure}[t]
\begin{center}
\includegraphics[width=3.375in]{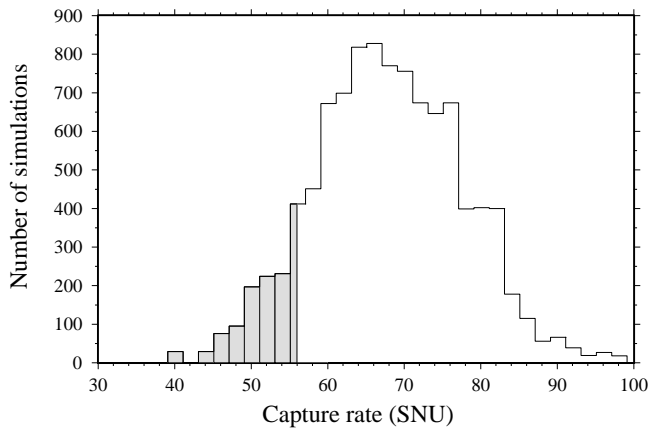}
\end{center}
\caption{Distribution of capture rate from 10\~000 simulations of SAGE III
assuming true production rate of 67.2 SNU.  The probability of a rate less
than or equal to the observed rate of 56 SNU is 11\% and is shown shaded.}
\label{sage3kl}
\end{figure}

     To quantitatively test whether or not it is reasonable to assume that
the production rate is constant, we can consider the three segments of SAGE
data, whose results are given in Table~\ref{Segment_table}.  A test of the
consistency of any data segment with the overall result of 67 SNU can be made
by Monte Carlo simulation.  For the purposes of illustration, we choose the
most deviant segment, SAGE III, whose overall result is 56 SNU.  We then
simulate all 39 data sets of SAGE III assuming that the true production rate
is 67 SNU.  To ensure that these simulations parallel the real data as
closely as possible, all parameters of the simulation, such as background
rates, efficiencies, exposure times, and counting times, are chosen to be the
same as for the real data.  From the sequence of simulated event times, the
combined production rate is calculated in exactly the same manner as for the
real data.  This process is repeated 10\~000 times and a histogram of the
combined rate is produced.  From the position of the observed rate for the
real data in this histogram, we can calculate the probability that the real
data are produced by the assumed initial production rate.  As shown in
Fig.\ \ref{sage3kl}, we find that $(11 \pm 0.3)$\% of the 10\~000 simulations
of SAGE~III have a value that is lower than the observed value of 56 SNU.
Since this probability is one tailed (maximum of 50\%), this is the most
aberrant of the three sections of SAGE data, and no systematic uncertainties
were included in the simulations, a value of 11\% is not extremely unusual,
and there is thus no statistically significant evidence for production rate
variation.  The same analysis applied to SAGE~I and SAGE~II yields
probabilities of 35\% and 38\%, respectively, highly consistent with the
assumption of constant production rate.

\begin{figure}
\begin{center}
\includegraphics[width=3.375in]{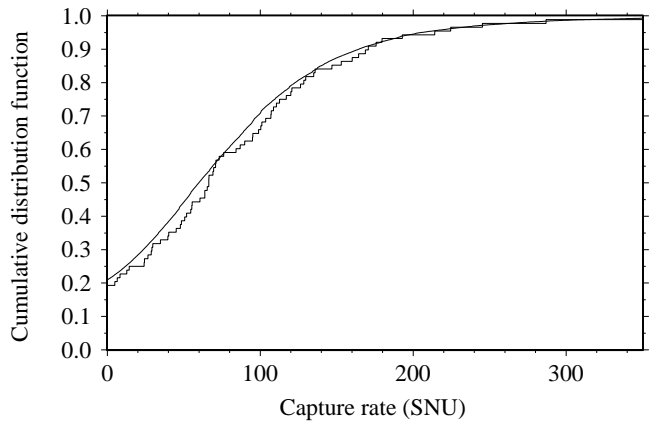}
\end{center}
\caption{Measured capture rate for all SAGE data sets (jagged curve) and the
expected distribution derived by 1000 Monte Carlo simulations of each set
(smooth curve).  The capture rate in the simulations was assumed to be 67.2
SNU.}
\label{prod}
\end{figure}

     Another way to consider this question is to use the cumulative
distribution function of the production rate $C(p)$, defined as the fraction
of data sets whose production rate is less than $p$.  Figure \ref{prod} shows
this distribution for all data sets and the expected distribution from
simulation, assuming a constant production rate of 67 SNU.  The two spectra
parallel each other closely and can be compared by calculating the $Nw^2$
test statistic \cite{CLE98}.  This gives $Nw^2$ = 0.343 whose probability is
10\%.

\begin{table}[t]
\squeezetable
\caption{Capture rate results for yearly, monthly, and bimonthly combinations
of SAGE data.  Runs are assigned to each time period by their mean exposure
time.}
\label{combinations}
\begin{tabular}{c d d r @{--} d}
             &           & Best  & \multicolumn{2}{c}{68\% conf.} \\
Exposure     & Number of & fit   & \multicolumn{2}{c}{range}      \\
interval     & data sets & (SNU) & \multicolumn{2}{c}{(SNU)}      \\
\hline
1990    &   5. &   43. &   2 &  78. \\
1991    &   6. &  112. &  82 & 145. \\
1992    &  13. &   76. &  59 &  95. \\
1993    &  15. &   84. &  65 & 105. \\
1994    &  10. &   73. &  51 &  98. \\
1995    &  13. &  101. &  77 & 128. \\
1996    &  10. &   49. &  32 &  68. \\
1997    &  16. &   46. &  35 &  58. \\
\\
Jan     &   7. &   47. &  24 &  74. \\
Feb     &   6. &   41. &  20 &  63. \\
Mar     &   3. &  198. & 137 & 266. \\
Apr     &   5. &   41. &  22 &  63. \\
May     &   6. &   83. &  58 & 111. \\
Jun     &   3. &   37. &   3 &  80. \\
Jul     &   9. &   40. &  22 &  62. \\
Aug     &   9. &   79. &  57 & 102. \\
Sep     &  12. &   63. &  47 &  82. \\
Oct     &  11. &   64. &  42 &  90. \\
Nov     &   9. &   73. &  52 &  96. \\
Dec     &   8. &  123. &  95 & 153. \\
\\
Jan+Feb &  13. &   44. &  28 &  60. \\
Mar+Apr &   8. &   70. &  48 &  94. \\
May+Jun &   9. &   71. &  50 &  95. \\
Jul+Aug &  18. &   60. &  45 &  77. \\
Sep+Oct &  23. &   64. &  50 &  79. \\
Nov+Dec &  17. &   95. &  77 & 113. \\
\\
Feb+Mar &   9. &   69. &  48 &  92. \\
Apr+May &  11. &   60. &  44 &  78. \\
Jun+Jul &  12. &   39. &  23 &  59. \\
Aug+Sep &  21. &   70. &  57 &  84. \\
Oct+Nov &  20. &   69. &  54 &  86. \\
Dec+Jan &  15. &   88. &  70 & 106.
\end{tabular}
\normalsize
\end{table}

     Although these statistical tests are consistent with a constant
production rate, they can never exclude the possibility of a cyclic time
variation whose magnitude is comparable with the statistical uncertainty.  We
thus give in Table~\ref{combinations} the capture rate result for several of
the possible temporal combinations of SAGE data.  Each of these data
divisions fits well to the constant rate of 67 SNU, as is verified by
$\chi^2$/degree of freedom = 8.2/7 (yearly), 14.6/11 (monthly), 4.9/5
(January + February bimonthly), and 3.9/5 (February + March bimonthly), which
have probabilities of 32\%, 20\%, 43\%, and 56\%, respectively.  We remind
those readers who are interested in short-term periodicity that the known
variation due to the change in Earth-Sun distance has been removed from our
reported capture rate [see Eq.\ (\ref{earth_sun_corr_factor})].

\subsection{Internal consistency of SAGE results}

     The combined results for all runs in the $L$ and $K$ peaks are given in
Table~\ref{Segment_table}.  The $L$-peak result is 12 SNU below the overall
value of 67 SNU and the $K$-peak result is 6 SNU above.  The statistical
$1\sigma$ error of these results, however, extends upward to 68 SNU in the
$L$ peak and downward to 64 SNU in the $K$ peak.  Both $L$- and $K$-peak
results thus overlap the overall value, and there is no evidence for
inconsistency between the results in the $L$ and $K$ peaks.

     As noted in Sec.~\ref{Chemical_extraction_efficiency}, so as to remove
most of the residual Ge carrier from the Ga metal, it is customary to make a
second extraction 2 or 3 days after each solar neutrino extraction.  Although
these second extractions are usually counted, until recently they were often
measured in counters which did not have the lowest background rates, and were
rarely counted in electronic system 3 with the wave form recorder.  Further,
these runs were seldom counted for a long time.  As a consequence, it was not
possible for us to give a result for the production rate from these second
extractions.  This situation changed at the beginning of 1996, however,
because SAGE then switched to a 6-week extraction schedule, which freed some
better low background counters and made it possible to measure these samples
from second extractions in system 3.  Ten such extractions have been measured
since 1996.  Taking into account the delay between the first and second
extractions and the extraction and counting efficiencies, in these ten
extractions we expect to detect three \nuc{71}{Ge} atoms that are leftover
from the first extraction, and seven \nuc{71}{Ge} atoms that are produced by
solar neutrinos during the interval between extractions.  The total number of
\nuc{71}{Ge} atoms detected in these ten extractions was 1.1 with a 68\%
confidence range from 0.0 to 8.7.  The number observed is statistically
consistent with the number expected, thus confirming our extraction
efficiency.  Further, it establishes that the \nuc{71}{Ge} we detect is not
an artifact of the extraction process and that our counting and data analysis
do not find a significant quantity of \nuc{71}{Ge} if it is not present.

\section{Summary and Conclusions}\label{conclusions}

     We have presented the methods and procedures of the SAGE experiment: the
extraction of Ge from Ga, the subsequent Ge purification, the counting of
\nuc{71}{Ge}, the identification of candidate \nuc{71}{Ge} events, and the
analysis of the counting data to obtain the solar neutrino production rate.
Eight years of measurement of the solar neutrino flux give the capture rate
result $67.2 ^{+7.2}_{-7.0}$ SNU, where the uncertainty is statistical only.
Analysis of all known systematic effects indicates that the total systematic
uncertainty is $^{+3.5}_{-3.0}$ SNU, considerably smaller than the
statistical uncertainty.  Finally, we have examined the counting data and
shown that there is good evidence that \nuc{71}{Ge} is being counted, that
the counting data fit the analysis hypotheses, and that the counting data are
self-consistent.

     The SAGE result of 67.2 SNU represents from 52\% \cite{BAH98} to 53\%
\cite{TUR98} of SSM predictions.  Given the extensive systematic checks and
auxiliary measurements that have been performed, especially the \nuc{51}{Cr}
neutrino source experiment \cite{ABD96,ABD98}, this $7\sigma$ reduction in
the solar neutrino flux compared to SSM predictions is very strong evidence
that the solar neutrino spectrum below 2 MeV is significantly depleted, as
was previously shown for the \nuc{8}{B} flux by the Cl and Kamiokande
experiments.  If we take into account the results of all experiments,
astrophysical solutions to the solar neutrino deficit can now nearly be
excluded \cite{Berezinsky96,Berezinsky97,Dar98}.  This conclusion is indeed
implied by the SAGE result itself, as it lies $2.5 \sigma$ below the capture
rate prediction of $88.1 _{-2.4} ^{+3.2}$ SNU obtained by artificially
setting the rate of the \nuc{3}{He}$(\alpha,\gamma)$\nuc{7}{Be} reaction to
zero and $1.5 \sigma$ below the astrophysical minimum capture rate of $79.5
_{-2.0} ^{+2.3}$ SNU \cite{BAH97}.  The solar neutrino problem is now a
model-independent discrepancy \cite{BAH982,HEE97} that does not depend on the
details of solar models or their inputs.

\begin{figure}[t]
\begin{center}
\includegraphics[width=3.375in]{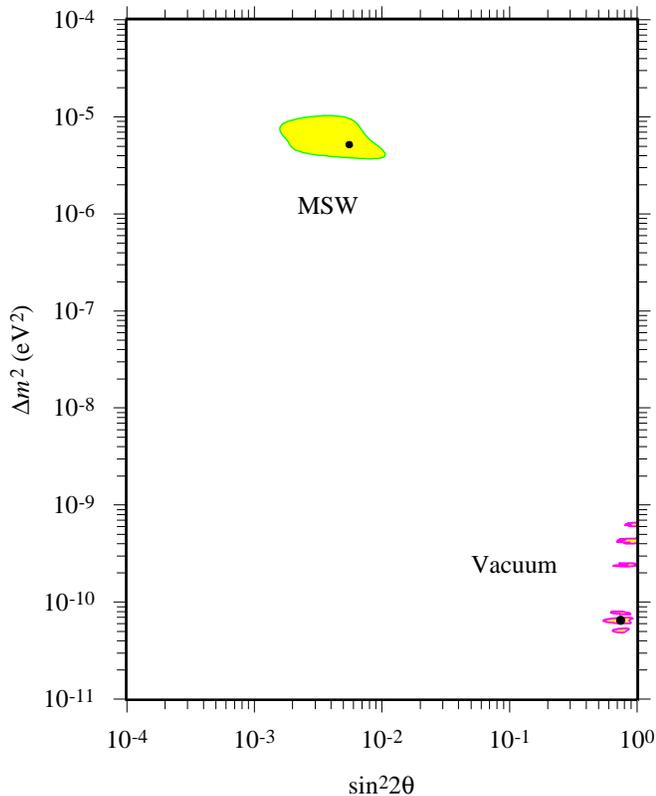}
\end{center}
\caption{Allowed regions of neutrino parameter space for two-flavor
oscillations into active neutrino species.  The analysis uses the results of
all solar neutrino experiments, including the constraints from the energy
spectrum and zenith-angle dependence measured by Super-Kamiokande.  The black
circles are the best fit points and the shading shows the allowed regions at
99\% confidence.  Thee figure is based on calculations in
Ref.~\protect\cite{BAH982}.}
\label{neutosc}
\end{figure}

     More credible explanations for the solar neutrino deficit involve either
matter-enhanced Mikheyev-Smirnov-Wolfenstein (MSW) neutrino oscillations, in
which the solar $\nu_e$ oscillates into other flavor neutrinos or a sterile
neutrino \cite{BAH982,CAL98,Bilenky99,HAT97}, or vacuum oscillations
\cite{Krastev-Petcov,BAH982,GelbRosen98}.  For both of these possibilities,
the allowed regions of $\Delta m^2 - \sin^2 2 \theta$ parameter space
determined from solar neutrino experiments for two-flavor oscillations into
active neutrino species are shown in Fig.\ \ref{neutosc}.  The fit quality is
about the same in both regions.  There is also a fit with similar quality for
MSW oscillations into sterile neutrinos, whose allowed region approximately
coincides with the region shown for MSW oscillations with active neutrinos.

     There are now very strong indications that the solar neutrino deficit
has a particle physics explanation and is a consequence of neutrino mass.  To
fully unravel the solar neutrino story, however, will require more
experiments, especially those with sensitivity to low-energy neutrinos or to
neutrino flavor.  SAGE continues to perform regular solar neutrino
extractions every 6 weeks with $\sim 50$ tons of Ga and will continue to
reduce its statistical and systematic uncertainties, thus further limiting
possible solutions to the solar neutrino problem.

\section*{Acknowledgments}\label{acks}

We thank J.~N.~Bahcall, M.~Baldo-Ceolin, P.~Barnes, L.~B.~Bez\-ru\-kov,
S.~Brice, L.~Callis, A.~E.~Chudakov, A.~Dar, G.~T.~Garvey, W.~Haxton,
V.~N.~Kornoukhov, V.~A.~Kuzmin, V.~A. Matveev, L.~B.~Okun, V.~A.~Rubakov,
R.~G.~H.~Robertson, N.~Sapporo, A.~Yu.~Smirnov, A.~A.~Smolnikov,
A.~N.~Tavkhelidze, and many members of GALLEX for their continued interest
and for fruitful and stimulating discussions.  We acknowledge the support of
the Russian Academy of Sciences, the Institute for Nuclear Research of the
Russian Academy of Sciences, the Ministry of Science and Technology of the
Russian Federation, the Russian Foundation of Fundamental Research under
Grant No.\ 96-02-18399, the Division of Nuclear Physics of the U.S.
Department of Energy, the U.S. National Science Foundation, and the U.S.
Civilian Research and Development Foundation under award No.\ RP2-159.  This
research was made possible in part by Grant No.\ M7F000 from the
International Science Foundation and Grant No.\ M7F300 from the International
Science Foundation and the Russian Government.

\appendix

\section{Other Counting Systems}
\label{other_counting_systems}

     The counting systems have been designated by the numbers 1--6.  The
initial developmental work on system 1 \cite{BAR83,GOG83}, which used the
amplitude of the differentiated pulse (ADP) method \cite{DAV72} to separate
\nuc{71}{Ge} events from background, was done in Russia during the early
1980s.  Based on this work, system 2 was developed at BNO during the years
1985--1988.  System 2 was completed in 1989 and counted all but two first
extractions through May 1992 (SAGE~I).  Counting system 5, which used the ADP
method of rise time measurement, was used to count the other two first
extractions during 1990 and 1991.  During the summer of 1992, system 3, which
has the capability to record the counter wave form, was brought on line;
since that time, it has been used to count almost all first extractions.
After the implementation of system 3 as the primary counting system,
extensive upgrades to reduce backgrounds were performed on system 2 to enable
SAGE to have low-noise counting capability in more than eight channels.  The
upgraded system is referred to as system 6.  It has counted seven first
extractions during SAGE~II and III, mostly from low-mass samples of Ga, and
has been used mainly for developmental work, such as testing proportional
counters and counting cleanup extractions of gallium.

\subsubsection{Counting system 2}\label{sys_2}
     System 2 was a seven-channel system where each PC was counted in an
independent passive shield; five of those channels had active shielding with
NaI crystals.  The passive shield consisted of an internal wall of tungsten
(40--80 mm thick) or copper (20--30 mm thick) surrounded by lead (150 mm
thick).  The NaI events were recorded in coincidence mode with events from
the proportional counters.  Several of the performance characteristics of
system 2 are given in Table~\ref{sys_specs}.

\subsubsection{Counting system 6}\label{sys_6}
     Modifications to system 2 began during 1992 when system 3 became the
primary acquisition system; the improvements were so extensive that it was
redesignated as system 6.  The counting system has seven channels of
acquisition with independent passive shields for each proportional counter.
Six channels have an active shield, which operates in coincidence mode with
events in the proportional counter.  A modified ADP method with the
application of several differentiation time constants is used for rejection
of point ionization events from backgrounds.  System 6 became fully
operational in early 1993.  To give this system wave form recording
capability a digitizing oscilloscope was added, but this improvement has
never been fully implemented.

\section{Fourier transform of the wave form -- $\bbox{R}$}\label{fft}

\begin{figure}
\begin{center}
\includegraphics[width=3.375in]{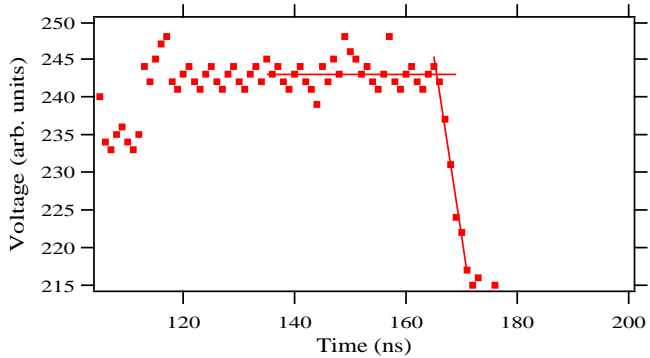}
\end{center}
\caption{Determination of the time and dc offset of a candidate event.}
\label{offset_det}
\end{figure}

     In contrast to the $T_N$ method, the pulse offset is determined
independently from the wave form.  One uses the intersection point of two
lines, the zero-slope line of the offset and the initial slope of the pulse,
to obtain the onset position in time and voltage.  The initial slope is
defined as a certain number of points before and after the point at 20\% of
the maximum pulse height.  The exact number of points to fit is determined
individually for each pulse since the number of points available will depend
on the pulse height.  This region is chosen so that the points are
sufficiently linear.  Figure \ref{offset_det} illustrates graphically how the
onset point is determined.

     Two data runs with counters filled with \nuc{71}{Ge} were used to check
the energy offset from this wave form analysis determination.  The runs were
separated by three years and used different digitizer settings.  The Gaussian
centroid of each $L$ peak and $K$ peak was calculated, with each peak
containing a few thousand counts.  The extrapolated intercepts in energy are
0.005(6) keV and 0.022(8) keV using $L$- and $K$-peak energies of 1.17 keV
and 10.37 keV.  Given the energy resolution of our counters, the energy
offset is effectively zero.

     The algorithm for determining the pulse onset was checked using
computer-simulated pulses, both with and without Gaussian noise.  It
correctly identifies the time offset to within 1 ns and the dc offset to
within one channel.  Those limits are, of course, dependent on the noise
levels, but the levels used were approximately the same as for typical data.
Thus, if each pulse is properly normalized to both zero time and dc offset,
there is no need to apply an energy offset correction.

     This technique uses the zero- and lowest-frequency values from a FFT to
obtain measures of the energy and rise time of a pulse.  The determination of
the energy is straightforward from the definition of the Fourier transform,

\begin{equation}
\label{Fourier}
F(\omega) = \int_{-\infty}^{+\infty}{f(t)e^{-i \omega t} dt}.
\end{equation}

\noindent At $\omega = 0$, $F(\omega)$ equals the area under the curve
$f(t)$, which in this case is the digitized wave form of the event convoluted
with a Hanning windowing function.  We select an integration time of 800 ns,
which is the maximum time allowable given the variation in the time of pulse
onset.  In effect, this technique is equivalent to summing channels used with
$T_N$ and is analogous to an ADC that integrates for 800 ns.

     In a Fourier analysis, the rise time behavior of a typical pulse clearly
will be a very-low-frequency component.  Studies with actual \nuc{71}{Ge}
pulses and computer-simulated pulses generated with Eqs.\ (\ref{Tn_Formula})
show that one can accurately identify several distinct features of the wave
form.  As expected, the dominant components are the lowest frequencies along
with the random noise that spans all frequencies.  One can identify structure
as well; most of it originates from the intrinsic properties of the
oscilloscope, such as dithering and the finite digitization size.  One of the
advantages of a Fourier analysis is that such structure appears at high
frequencies and is well separated from the rise time information.  The
lowest, nonzero, real component $F(1)$ scales similarly to an ADP value but
is independent of electronic offsets and high-frequency noise contributions
to the pulse.  Dividing it by the energy $F(0)$ of the pulse produces a
parameter $R$ that is proportional to the inverse rise time.  Thus, one can
perform a complementary analysis of the data that is analogous to the ADP
method but is based solely on the digitized pulse and is independent of any
underlying assumptions of its functional form.

\section{RST method}
\label{RST}

     In the standard analysis of our data we use the $T_N$ method and fit the
observed pulse to Eq.\ \ref{Tn_Formula}.  This function gives the correct
description of the shape of the voltage pulse as recorded by the digital
oscilloscope when the ionization produced in the proportional counter
consists of a set of point ionizations evenly distributed along a straight
track.  Since \nuc{71}{Ge} events are usually a single cluster of ionization,
this method works satisfactorily to select \nuc{71}{Ge} candidate events.  It
is, however, restricted to the particular form of ionization that is assumed,
and gives a poor fit to other types of charge deposit in the counter, such as
the combination of a point event from \nuc{71}{Ge} $K$-electron capture
followed by capture of the 9.3-keV x ray at some other location in the
counter.  To give us the capability to investigate all possible events that
may occur in the counter, we have also developed a more general method which
can analyze an event produced by ionization with an arbitrary distribution of
charge.  We call this the ``restored pulse method'' or ``RST method'' for
short.

     We begin with the measured voltage pulse $V(t)$ as recorded by the
digitizer.  For an ideal point charge that arrives at the counter anode wire,
$V(t)$ has the Wilkinson form $V(t) = W(t) = V_0 \ln(1 + t/t_0)$, provided
the counter is ideal and the pulse processing electronics has infinite
bandwidth.  For a real event from the counter, with unknown charge
distribution, $V(t)$ can in general be expressed as the convolution of the
Wilkinson function with a charge collection function $G(t)$:
\begin{equation}
\label{G_definition}
V(t) = W(t) \otimes G(t).
\end{equation}
The function $G(t)$ contains within it the desired information about the
arrival of charge at the counter anode, coupled with any deviations of the
counter or electronics from ideal response.  Equation~(\ref{G_definition})
can be considered as the definition of $G(t)$.

     To get the desired function $G(t)$, one must deconvolute
Eq.\ (\ref{G_definition}).   To perform this deconvolution, we have found it
mathematically convenient to use the current pulse $I(t)$, which is obtained
by numerical differentiation of $V(t)$:
\begin{eqnarray}
I(t) & = & \frac{dV}{dt} = \frac{d}{dt} [W(t) \otimes G(t)] \\ \nonumber
     & = & \frac{dW}{dt} \otimes G(t) = W^{'}(t) \otimes G(t),
\end{eqnarray}
where $W^{'}(t)$ is normalized over the observed time of pulse measurement,
$T_{\text{obs}},$ such that $\int_0^{T_{\text{obs}}} W^{'}(t)dt = 1$.

     To deconvolute, we Fourier transform to the frequency domain and then
use the theorem that convolution in the time domain becomes multiplication in
the frequency domain \cite{NUMREP}.  This simply gives $I(f) = W^{'}(f)
G(f)$, which can be solved for $G(f)$.  We then Fourier transform $G(f)$ back
to the time domain to get the desired function $G(t)$.  The energy of the
event is given by $\int_0^{T_{\text{obs}}} G(t) dt$.  The duration of the
collection of ionization is given by the width of $G(t)$, which can be used
as a measure of the rise time.

\begin{figure}
\begin{center}
\includegraphics[width=3.375in]{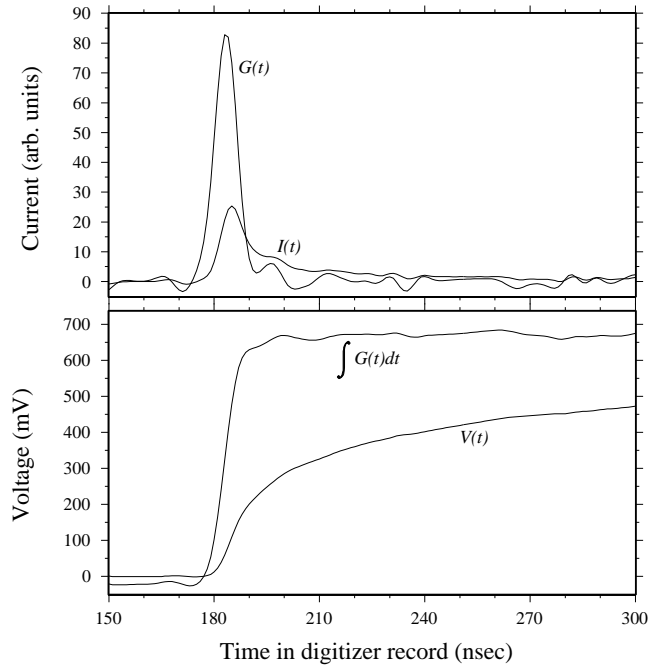}
\end{center}
\caption{Analysis of typical \nuc{71}{Ge} pulse by the RST method.  See text
for explanation.}
\label{RST_example}
\end{figure}

     An example of this procedure as applied to a typical \nuc{71}{Ge}
$K$-peak event is given in Fig.\ \ref{RST_example}.  This pulse has $T_N =
3.9$ ns.  The recorded voltage pulse after inversion and smoothing is given
by $V(T)$ in the lower panel.  The current pulse, obtained by numerical
differentiation of the voltage pulse, is given by $I(t)$ in the upper panel.
The deduced function $G(t)$ is also shown in the upper panel.  It has a FWHM
of about 15 ns, found to be typical for true \nuc{71}{Ge} $K$-peak events.
The integrated current pulse, which records the pulse energy, is given by
$\int G(t)dt$ in the lower panel.

     This method has the advantage that it can reveal the basic nature of the
ionization in the counter for an arbitrary pulse.  It is also capable of
determining the pulse energy over a wider range than the $T_N$ method.  A
problem that has been found with this method in practice, however, is that
when \nuc{71}{Ge} data are analyzed one obtains multiple collection functions
[i.e., $G(t)$ has several distinct peaks separated in time] more often than
is expected from the known physical processes that take place in the counter.
These multiple peaks are due to noise on the pulse and cutoff of the system
frequency response at about 100 MHz.  Attempts have been made to remove these
extraneous peaks by filtering and smoothing the original pulse, but they have
not been fully successful.  Evidently we need faster electronics and a
reduction in the noise level to be able to fully exploit this pulse shape
analysis technique.  As a result, we have only been able to use this method
to select events on the basis of energy.

\end{document}